\documentclass[referee, pdflatex]{sn-jnl}

\usepackage{url}
\usepackage{hyperref}
\usepackage{float}
\usepackage{booktabs}
\usepackage{multirow}
\usepackage{fancyvrb}
\usepackage[table]{xcolor}
\usepackage{tikz}
\usepackage{mathtools}
\usepackage{graphicx} 
\usepackage{rotating} 
\usepackage{appendix} 
\usepackage{textcomp} 
\usepackage{manyfoot} 
\usepackage{etoolbox} 
\usepackage{algorithm} 
\usepackage{algorithmicx} 
\usepackage{algpseudocode} 
\usepackage{listings} 
\usepackage{lmodern}
\usepackage[T1]{fontenc}
\usepackage[square,numbers]{natbib}
\usepackage{longtable}

\usepackage{array}
\newcolumntype{P}[1]{>{\raggedright\arraybackslash}p{#1}}

\usetikzlibrary{arrows.meta, positioning,shapes.geometric, arrows}
\tikzstyle{startstop} = [rectangle, rounded corners, minimum width=3cm, minimum height=1cm,text centered, draw=black, fill=none]
\tikzstyle{process} = [rectangle, minimum width=3cm, minimum height=1cm, text centered, draw=black, fill=none]
\tikzstyle{arrow} = [thick,->,>=stealth]

\usepackage{amsmath,amsfonts,amssymb}
\graphicspath{{Fig/}}

\begin{document}

\title[BayesianFitForecast]{\texttt{BayesianFitForecast}: A User-Friendly R Toolbox for Parameter Estimation and Forecasting with Ordinary Differential Equations}

\author[1]{Hamed Karami}
\author[2]{Amanda Bleichrodt}
\author[2]{Ruiyan Luo}
\author*[2,3]{Gerardo Chowell} \email{gchowell@gsu.edu}

\affil[1]{\orgdiv{Department of Mathematics and Statistics}, \orgname{Georgia State University}, \orgaddress{\street{Atlanta},\state{Georgia}, \country{USA}}}
\affil[2]{\orgdiv{Department of Population Health Sciences}, \orgname{Georgia State University}, \orgaddress{\street{Atlanta},\state{Georgia}, \country{USA}}}
\affil[3]{\orgdiv{Department of Applied Mathematics}, \orgname{Kyung Hee University}, \orgaddress{\street{Yongin}, 17104, \country{Korea}}}

\abstract{

\textbf{Background:} Mathematical models based on ordinary differential equations (ODEs) are essential tools across various scientific disciplines, including biology, ecology, epidemic modeling, and healthcare informatics, where they are used to simulate complex dynamic systems and inform decision-making. However, implementing Bayesian calibration and forecasting typically requires substantial coding in Stan or similar tools. To support Bayesian parameter estimation and forecasting for such systems, we introduce \texttt{BayesianFitForecast}, a user-friendly R toolbox specifically developed to streamline Bayesian parameter estimation and forecasting in ODE models, making it particularly relevant to health informatics and public health decision-making (\texttt{https://github.com/gchowell/BayesianFitForecast/}).

\textbf{Results:} This toolbox enables automatic generation of Stan files, allowing users to configure models, define priors, and analyze results with minimal programming expertise. By eliminating manual coding, \texttt{BayesianFitForecast} significantly lowers the technical barrier to Bayesian inference with dynamical systems. We demonstrate its flexibility and usability through applications to historical epidemic datasets (e.g., the 1918 influenza pandemic in San Francisco and the 1896–1897 Bombay plague) and simulated data, showing robust parameter estimation and forecasting performance under Poisson and negative binomial observation error structures. The toolbox also provides robust tools for evaluating model performance, including convergence diagnostics, posterior distributions, credible intervals, and performance metrics.

\textbf{Conclusion:} By improving the accessibility of advanced Bayesian methods, \texttt{BayesianFitForecast} broadens the application of Bayesian inference in time-series modeling, healthcare forecasting, and epidemiological applications. In addition to the R scripting interface, a built-in Shiny web application is included, enabling interactive model configuration, visualization, and forecasting. A tutorial video demonstrating the toolbox’s functionality is also available (\texttt{https://youtu.be/jnxMjz3V3n8}).
}


\maketitle

\newpage

\section{Background}\label{sec:intro}

Mathematical models using ordinary differential equations (ODEs) are widely used across scientific fields including biological and social sciences \citep{strogatz2024nonlinear,brauer2012mathematical,yan2019quantitative}. These models, characterized by sets of ODEs and their associated parameters, are essential for analyzing the complex dynamics that arise across different regions of the parameter space. They enable researchers to simulate intricate processes, such as the spread of infectious diseases and population dynamics. However, a key challenge in utilizing ODE models is parameter estimation, where parameters are inferred from observed time-series data. This process, often referred to as the inverse problem, involves estimating model parameters while accounting for uncertainty. Advanced techniques, such as Bayesian inference and maximum likelihood estimation, are frequently employed to address this challenge, enabling the calibration of ODE models for more accurate predictions and policy development.

Bayesian estimation methods have gained significant traction in the calibration of epidemic models based on ODEs. These methods integrate prior knowledge with new data to refine parameter estimates, yielding posterior distributions that explicitly account for uncertainty and incorporate expert knowledge into the modeling process \citep{grinsztajn2021bayesian,bouman2024bayesian,gelman2020bayesian,belasso2023bayesian,girolami2008bayesian,kypraios2017tutorial,mckinley2014simulation,gelman1995bayesian}. Using Bayes’ theorem, these methods merge prior parameter distributions with the likelihood of observed data, resulting in posterior distributions. Markov Chain Monte Carlo (MCMC) algorithms are often employed to approximate these posterior distributions. MCMC works by generating a sequence of random samples from the posterior distribution, which we then use to estimate the distribution's characteristics, such as the mean and credible intervals. The flexibility and robustness of Bayesian methods make them particularly valuable in the context of emerging epidemics, where data may be limited or noisy \citep{alahmadi2020influencing,o1999bayesian,jackson2015calibration,trejo2022modified}. 

However, implementing Bayesian calibration and forecasting for ODE-based systems  often requires extensive programming knowledge and manual coding, presenting a significant barrier for many researchers. Tools like Stan \cite{stan2023} facilitate the implementation of Bayesian estimation and forecasting, allowing for rigorous uncertainty quantification and model validation \citep{annis2017bayesian,kelter2020bayesian,sennhenn2018bayesian,sorensen2015bayesian,monnahan2017faster,burkner2017brms,carpenter2017stan}.  Yet, the technical complexity involved in writing code and the need for specialized programming skills can deter many potential users.

In response to these challenges, we introduce \texttt{BayesianFitForecast}, a user-friendly R toolbox specifically designed to streamline the Bayesian calibration and forecasting process for real-world epidemic data and other dynamic systems, making it accessible to a broad audience, including graduate students in applied mathematical and statistical sciences. Traditionally, employing Bayesian estimation necessitates writing code in Stan \citep{annis2017bayesian,grinsztajn2021bayesian}, a probabilistic programming language. However, our toolbox eliminates this barrier by automatically generating the necessary Stan files based on user-defined options. \texttt{BayesianFitForecast} offers robust tools for evaluating model performance, including convergence diagnostics, posterior distributions, credible intervals, and performance metrics, all with minimal programming expertise required. We have included a detailed tutorial within this paper to guide users through the various features and functionalities of the toolbox, ensuring that even those with minimal programming experience can effectively leverage Bayesian methods with ODE models. The toolbox and accompanying tutorial materials are available at (\texttt{https://github.com/gchowell/BayesianFitForecast}) with a full video walkthrough at (\texttt{https://youtu.be/jnxMjz3V3n8}).

\section{Implementation\label{sec:imple}}

\subsection{Ordinary differential equation (ODE) models}
Mathematical models based on ODEs are fundamental tools for understanding dynamic processes across various scientific fields. These models describe how quantities evolve over time through differential equations and parameters. For instance, in an epidemiological context, an ODE model might track the number of susceptible, exposed, infected, and recovered individuals within a population. Accurately estimating the parameters of such a model, including transmission rates and the impact of interventions, is crucial for making precise predictions and guiding public health decisions. The model's reliability depends significantly on the effective calibration of these parameters to observed data and on comprehending how parameter variations influence the system’s behavior. A system of ODEs with $h$ differential equations is typically represented as follows:
\begin{align*}
    \dot{x}_1(t) &= g_1(x_1, x_2, \cdots, x_h, \Theta), \\
    \dot{x}_2(t) &= g_2(x_1, x_2, \cdots, x_h, \Theta), \\
    &\;\;\vdots \\
    \dot{x}_h(t) &= g_h(x_1, x_2, \cdots, x_h, \Theta),
\end{align*}
where \(\dot{x}_i\) represents the rate of change of the system state \(x_i\) for \(i = 1, 2, \ldots, h\), while \(\Theta = (\theta_1, \theta_2, \ldots, \theta_m)\) represents the set of model parameters. The system states can be observable or latent. The observed state refers to the specific state variable of the ODE system that has been recorded or measured in research or experiments. In contrast, latent states are those ODE states that are not directly observed but are inferred through mathematical modeling of the observed variables. In epidemic contexts, the observed state typically corresponds to the number of new cases over time.

Using the toolbox introduced in this tutorial, users can comprehensively define ODE models consisting of one or more differential equations. These models can be employed to simulate the system, estimate parameters by fitting the model to data, and forecast with quantified uncertainty based on the calibrated model. To achieve this, users must create a string variable that defines the ODE model, including the system of equations that describe the changes in state variables over time. Additionally, users should specify the characteristics of the model parameters, such as their names, ranges, prior distributions, initial guesses, and whether they are estimated or fixed based on prior information. Users should also provide details about the state variables, including their names and initial conditions.

\subsection{Model Calibration and Parameter Inference}
In this toolbox, we assume that there is a single observed state in the system. Let \( y(t_1), y(t_2), \ldots, y(t_n) \) represent the time series of the observed state used to calibrate the model. Here, \( t_j \) for \( j=1,2,\ldots,n \), are the time points for the time series data, and \( n \) is the total number of observations. Let \( f(t,\Theta) \) denote the expected temporal trajectory of the observed state. We estimate the set of model parameters, denoted by \( \Theta \) using Bayesian inference, which enables the integration of uncertainty into parameter estimates and the incorporation of expert knowledge through prior distributions.  Detailed information on the parameter estimation methodology, uncertainty quantification, and model assessment implemented in this toolbox is provided in the following sections.

\subsubsection{Bayesian inference}
Bayesian inference integrates prior knowledge about parameters with observed data to create a posterior distribution. This method allows for updating our understanding of parameter values as new data becomes available, providing a rigorous framework for uncertainty quantification. In Bayesian inference, the posterior distribution of the parameters is determined by Bayes’ rule:
\begin{equation}
    \label{eq:bayes}
    p(\theta \vert \mathbf{Y} ) \propto p(\boldsymbol{\theta})p(\mathbf{Y}\vert \theta),
\end{equation} 
where $p(\boldsymbol{\theta})$ denotes the prior distribution, $p(\mathbf{Y}\vert \boldsymbol{\theta})$ is the likelihood, and $p(\boldsymbol{\theta} \vert Y )$ represents the posterior distribution of the parameters $\boldsymbol{\theta}$.

\noindent \textbf{The likelihood}. For a given model, the likelihood represents the probability or density of observing the data given the parameters. We assume that the observations \( y(t_j)\), for $j=1, \ldots, n$, are independently obtained with mean \( \mu_{\Theta}(j) \) and a certain error structure,  where \( \mu_{\Theta}(j) = f(t_j, \Theta) \) represents the mean of \( y_j \) and \( f(t, \Theta) \) is the trajectory curve obtained by solving the differential equations given $\Theta$. 

Selecting the appropriate error structure is crucial for ensuring model accuracy. For instance, we can assume a Poisson error structure, where: 
\begin{equation}
    \label{eq:PoissonError}
     y_{t_j}|\boldsymbol{\Theta} \sim \text{Poisson}(\mu_{\Theta}(j)),
\end{equation} 
which has mean and variance equal to $\mu_{\Theta}(j)$. 
However, the Poisson distribution may struggle with overdispersed data, leading to biased estimates. To account for possible overdispersion in the data, we also consider the negative binomial distribution:
\begin{equation}
    \label{eq:NegBinError}
     y_{t_j}|\boldsymbol{\Theta},\phi \sim \text{NB}(\mu_{\Theta}(j), \phi),
\end{equation} 
which has mean $\mu_{\Theta}(j)$ and variance $\mu_{\Theta}(j)+\frac{\mu_{\Theta}(j)^2}{\phi}$. The negative binomial distribution introduces an additional parameter to account for overdispersion, making it better suited for real-world scenarios with high data variability. We also include the normal error structure:
\begin{equation}
    \label{eq:NormalError}
 y_{t_j}|\boldsymbol{\Theta},\phi \sim N(\mu_{\Theta}(j), \sigma^2)
\end{equation} 
with variance $\sigma^2$. Model parameters are estimated using a Bayesian approach, incorporating assumptions specific to the chosen error structure (e.g., Poisson, negative binomial, or normal). 

\noindent \textbf{Prior distributions}. In Bayesian inference, prior distributions represent existing knowledge or beliefs about parameters before observing the data. Within the toolbox, users have the flexibility to choose any distribution for their parameters. However, to simplify this process and provide examples, we offer predefined option files featuring normal and uniform prior distributions for ODE parameters. Below, we briefly explain these distributions.

\begin{unenumerate}
\item \textbf{a) Normal priors}. This type of prior concentrates the distribution closely around the anticipated value. We start by considering a range of potential values for a parameter, then use the midpoint of this range as the mean and choose a standard deviation of 10 to ensure the distribution is sufficiently broad. This prior is particularly suitable for case studies with substantial prior knowledge about the parameter value. 
\item \textbf{b) Uniform priors}. When using uniform priors, we assume minimal prior knowledge about the parameters. This involves establishing a range for parameter values without assuming they center around any specific value. Typically, a uniform distribution is used for this purpose. Uniform priors are particularly useful when reliable prior information about the parameters is lacking. By using uniform distributions, we allow the data to play a more significant role in shaping the posterior distributions of the parameters. This method is especially advantageous in the early stages of an outbreak when specific prior knowledge may not be available.
\end{unenumerate}
The error structures, such as the negative binomial distribution and normal distribution, have additional variance-related parameters, whose prior distributions will be specified in the toolbox illustration. 

\noindent \textbf{Posterior distributions}. In Bayesian inference, the posterior distribution of parameters is a fundamental concept that represents the updated beliefs about parameters after incorporating observed data. It combines prior information with the likelihood of the observed data, resulting in a comprehensive distribution that reflects both prior knowledge and empirical evidence. The posterior distribution serves as a comprehensive summary of the uncertainty surrounding parameter estimates, allowing for probabilistic interpretation and inference. It provides a basis for making predictions, estimating credible intervals, and conducting hypothesis testing within the Bayesian framework. As new data becomes available, the posterior distribution can be further updated, continually refining the understanding of the parameters of interest. In the toolbox, the posterior distribution is obtained by numerical simulation using Stan. % and a comprehensive analysis of posterior distributions is available, including convergence diagnostics and parameter estimation.\\

\subsubsection{Numerical Implementation}
\noindent\textbf{Stan and MCMC}. Stan is a probabilistic programming language designed for Bayesian statistical modeling and inference, utilizing the MCMC algorithm and the No-U-Turn Sampler (NUTS) to sample from complex posterior distributions \citep{grinsztajn2021bayesian}. Stan employs a sophisticated variant of Markov chain Monte Carlo (MCMC) called Hamiltonian Monte Carlo (HMC) \citep{betancourt2017conceptual}, which leverages gradient information for efficient sampling, and the No-U-Turn Sampler (NUTS), an adaptive algorithm that automatically adjusts the trajectory length to avoid revisiting previous states. This approach enhances convergence speed and sampling accuracy without manual tuning. Stan's automatic differentiation capability ensures precise computation of gradients, which is crucial for HMC performance. By using Stan, users can specify a wide variety of statistical models and obtain samples from the posterior distribution, facilitating Bayesian calibration, uncertainty quantification, and predictive inference, especially in forecasting ODE-based systems driven by real-world epidemic data. In this toolbox, the Stan file is generated automatically for parameter estimation, eliminating the need for users to write Stan code manually. 

\noindent\textbf{Iterations and convergence diagnostic}. 
The number of iterations, representing the number of times the MCMC algorithm samples from the posterior distribution, should be sufficiently large to ensure convergence. Half of these iterations are typically used as burn-in to allow the Markov chain to reach a stable state. Rhat values, a convergence diagnostic used in Bayesian estimation, should be close to 1, indicating good convergence and mixing of the Markov chains. Posterior distributions and forecasts are summarized using posterior medians and credible intervals, providing a comprehensive view of the parameter estimates and predictive uncertainties.

\noindent\textbf{Model Selection and Quality of Model Fit}. To compare the quality of model fit, we use several criteria that are widely adopted in Bayesian modeling, including the Deviance Information Criterion (DIC), Watanabe-Akaike Information Criterion (WAIC), and Leave-One-Out cross-validation (LOO)~\cite{spiegelhalter2002bayesian,watanabe2010asymptotic,vehtari2017practical}.

The DIC is a Bayesian generalization of the Akaike Information Criterion (AIC) and is given by:
\[
\text{DIC} = \bar{D} + p_D
\]
where $\bar{D}$ is the mean deviance and $p_D$ is the effective number of parameters. DIC provides a balance between model fit and complexity, penalizing over-parameterization.

The WAIC is a fully Bayesian alternative that approximates Bayesian cross-validation. It is calculated using the log pointwise predictive density (lppd) and includes a penalty term for model complexity based on the variance of the log-likelihood:
\[
\text{WAIC} = -2 \left( \text{lppd} - p_{\text{WAIC}} \right)
\]
where $p_{\text{WAIC}}$ is the estimated effective number of parameters. WAIC is asymptotically equivalent to Bayesian cross-validation and is especially useful for hierarchical models. \texttt{BayesianFitForecast} calculates WAIC automatically from the posterior samples, enabling users to compare models with different structures or prior assumptions based on out-of-sample predictive accuracy.

Lastly, LOO uses leave-one-out cross-validation by approximating the model’s predictive accuracy when each data point is held out in turn. We employ Pareto-smoothed importance sampling (PSIS-LOO), a computationally efficient approach to estimate LOO from a single posterior sample. LOO provides a robust, data-driven estimate of out-of-sample predictive performance, making it a valuable tool for model comparison, particularly when overfitting is a concern.

\subsubsection{Model-based forecasts with quantified uncertainty}

When fitting data to the model, two options can be specified: (a) the index of the model's variable to be fitted to the observed time series data, or (b) the index of the model's variable derivative to be fitted to the observed time series data. In each case, the predicted values are determined as follows:

\begin{itemize}
\item For fitting to the variable’s value: Assume that the given data should be fitted to $x_j(t)$,  where \( 1 \leq j \leq h \). In this case, the predicted values \( \hat{x}_j(t) \) are given by:
\begin{equation}
    \label{eq:fittingVal}
    \hat{x}_j(t) =  f(t,\Theta) \coloneqq \phi_j(t,\Theta),
\end{equation}
where \( \phi_j(t,\Theta) \) represents the \( j \)-th component of the trajectory curve obtained by solving the differential equations, parameterized by \(\Theta\), which can be sampled from the posterior distribution of the parameters using MCMC. Thus, the predicted values \( \hat{x}_j(t) \) represent the model's output given  \(\Theta\).

\item For fitting to the variable’s derivative: Assume that the given data should be fitted to \(\dot{x}_j(t)\), where \(1 \leq j \leq h\). In this case, the predicted derivatives \(\widehat{\dot{x}}_j(t)\) are given by:
\begin{equation}
    \label{eq:fittingDer}
    \widehat{\dot{x}}_j(t) = \dot{f}(t,\Theta) \coloneqq \psi_j (t,\Theta),
\end{equation}
where \(\psi_j (t,\Theta)\) is the \(j\)-th component of the rate of change of the trajectory curve, also parameterized by \(\Theta\). This function captures the model’s rate of change given  \(\Theta\).
\end{itemize}

Let \(\{\Theta^{(i)}:i = 1, 2, \ldots, N\}\) denote the \(N\) MCMC samples from the converged chains. We define the median trajectory curve \(\tilde{y}(t)\) as follows:
\begin{equation}
    \label{eq:medValues}
    \tilde{y}(t) = \text{median} \left\{ \phi_j(t,\Theta^{(i)}) \mid \Theta^{(i)} \sim P(\Theta \mid Y), \, i = 1, 2, \ldots, N \right\},
\end{equation}
or, if dealing with derivatives,
\begin{equation}
    \label{eq:medDer}
\tilde{y}(t) = \text{median} \left\{ \psi_j(t,\Theta^{(i)}) \mid \Theta^{(i)} \sim P(\Theta \mid Y), \, i = 1, 2, \ldots, N \right\}.
\end{equation}
Thus, the function \(\tilde{y}(t)\) represents the central tendency of the model's output or its derivatives, based on the posterior distribution of the parameters.
To quantify the uncertainty of the model predictions, we calculate the 95\% prediction interval (PI) for the predicted values or derivatives at each time point \(t\). This is achieved by sorting the predicted values \(\phi_j(t,\Theta^{(i)})\) or derivatives \(\psi_j(t,\Theta^{(i)})\) for \(i=1,2,\cdots, N\), and then identifying the 2.5th and 97.5th percentiles, which correspond to the lower and upper bounds of the 95\% PI:
\begin{equation}
    \label{eq:percValues}
    \text{PI}_{95\%}(t) = \left[ \text{Percentile}_{2.5\%} \left( \left\{ \phi_j(t,\Theta^{(i)}) \right\} \right), \, \text{Percentile}_{97.5\%} \left( \left\{ \phi_j(t,\Theta^{(i)}) \right\} \right) \right],
\end{equation}
or, for derivatives,
\begin{equation}
    \label{eq:percDer}
\text{PI}_{95\%}(t) = \left[ \text{Percentile}_{2.5\%} \left( \left\{ \psi_j(t,\Theta^{(i)}) \right\} \right), \, \text{Percentile}_{97.5\%} \left( \left\{ \psi_j(t,\Theta^{(i)}) \right\} \right) \right].
\end{equation}

This PI provides a range within which the model predictions or their rates of change are likely to fall with 95\% probability, thereby quantifying the uncertainty in the model's predictions.
\subsection{Performance metrics}
To assess forecasting performance, we employ four key performance metrics: the mean absolute error (MAE), the mean squared error (MSE), the coverage of the $95\%$ PIs, and the weighted interval score (WIS) \cite{gneiting2007strictly}. While it is possible to generate forecasts $H$ time units ahead for an evolving process, those forecasts cannot be evaluated until sufficient data for the $H$-time units ahead have been collected.

The \textit{mean absolute error} (MAE) is given by
\begin{equation}
\label{eq:mae}
    MAE = \frac{1}{N}\sum _{h=1}^{N}\left\vert f(t_h,\hat{\Theta})-y_{t_h}\right\vert,
\end{equation}
where $t_h$, $h=1,\cdots,N$ are the time points of the time series data \cite{kuhn2013applied}, $N$ is the number of observed data in the calibration period or forecasting period. 
Similarly, the $\textit{mean squared error}$ (MSE) is given by:
\begin{equation}
    \label{eq:mse}
     MSE = \frac{1}{N}\sum _{h=1}^{N}\left( f(t_h,\hat{\Theta})-y_{t_h}\right)^2.
\end{equation}

The coverage of the 95\% PI   corresponds to the fraction of data points that fall within the 95\% PI, calculated as:

\begin{equation}
    \label{eq:cov}
    \text{95\%PI coverage} = \frac{1}{N} \sum _{t=1}^N \mathbf{1}(Y_t>L_t\cap Y_t<U_t),
\end{equation}
where $L_t$ and $U_t$ are the lower and upper bounds of the 95\% PIs, respectively, $Y_t$ are the data and $\mathbf{1}$ is an indicator variable that equals 1 if $Y_t$ is in the specified interval and 0 otherwise.

The weighted interval score (WIS) \cite{gneiting2007strictly}, is a proper score that provides quantiles of predictive forecast distribution by
combining a set of Interval Scores (IS) for probabilistic forecasts. An IS is a simple proper score that requires only a central $(1-\alpha )\times 100\%PI$ \cite{gneiting2007strictly} and is described as:
\begin{equation}
    \label{eq:IS}
    IS_\alpha (F,y) = (u-l) +\frac{2}{\alpha}(l-y)\times \mathbf{1}(y<l) +\frac{2}{\alpha}(y-u)\times \mathbf{1}(y>u).
\end{equation}

In this equation, \textbf{1} refers to the indicator function, meaning that $\mathbf{1}(y<l)=1$ and 0 otherwise. The terms $l$ and $u$ represent the $\frac{\alpha}{2}$ and $(1-\frac{\alpha}{2})$ quantiles of the forecast $F$. The IS consists of three distinct quantities:
\begin{itemize}
    \item The sharpness of $F$, given by the width $u - l$ of the central $(1 - \alpha ) \times  100\% PI$.
    \item  A penalty term $\frac{2}{\alpha}(y-u)\times \mathbf{1}(y<l)$ for the observations that fall below the lower end point $l$ of the $(1 - \alpha ) \times  100\% PI$. This penalty term is directly proportional to the distance between $y$ and the lower end $l$ of the PI. The strength of the penalty depends on the level $\alpha$.
    \item An analogous penalty term $\frac{2}{\alpha}(y-u)\times \mathbf{1}(y>u)$ for the observations falling above the upper limit $u$ of the PI.
\end{itemize}

To provide more detailed and accurate information on the entire predictive distribution, we report several central PIs at different levels $(1-\alpha _1)<(1-\alpha _2)<\cdots(1-\alpha _k)$ along with the predictive median, $\tilde{y}$, which can be seen as a central prediction interval at level $(1-\alpha_0)\rightarrow 0$. This is referred to as the WIS, and it can be evaluated as follows:
\begin{equation}
    \label{eq:wis}
    WIS_{\alpha_{0:K}}(F,y) = \frac{1}{K+\frac{1}{2}}\left ( w_0\vert y-\tilde{y}\vert +\sum _{k=1}^K w_kIS_{\alpha_k}(F,y)\right),
\end{equation}
where $w_k=\frac{\alpha_k}{2}$ for $k=1,\cdots,K$ and $w_0=\frac{1}{2}$. Hence, WIS can be interpreted as a measure of how close the entire distribution is to the observation in units on the scale of the observed data \citep{bracher2021evaluating,cramer2022evaluation}.

\subsection{Overview of the Bayesian toolbox}

\begin{figure}[t] 
    \centering 
    \includegraphics[width=0.8\textwidth]{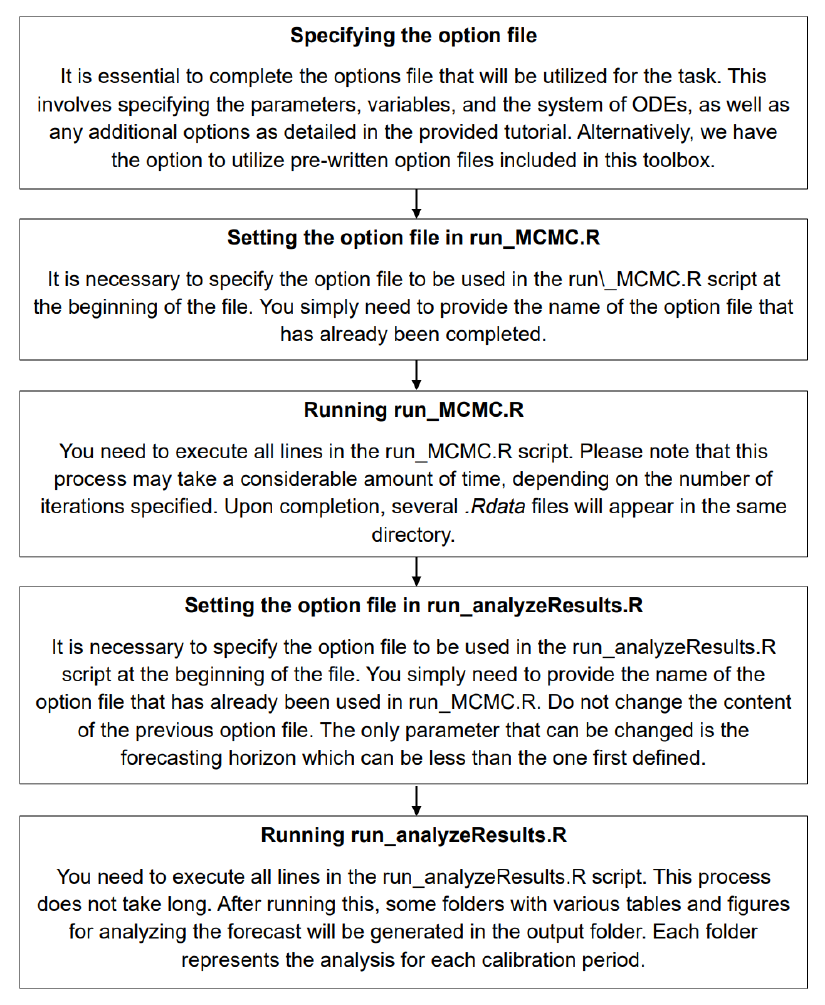}
    \caption{A comprehensive workflow diagram illustrating the steps involved in parameter estimation and forecasting using \texttt{BayesianFitForecast} with dynamical models based on ODEs. The figure outlines the sequential process of model specification, parameter estimation, and uncertainty quantification, providing users with a clear guide to effectively navigate the toolbox.}\label{fig:overview} 
\end{figure}

In this section, we demonstrate how to use the toolbox with the SEIR (Susceptible-Exposed-Infectious-Recovered) model \citep{Murray2003} to understand the early transmission dynamics of infectious disease spread, specifically in the context of the 1918 influenza pandemic in San Francisco. We provide step-by-step instructions for both general and tutorial-specific applications of the toolbox's functions and concise descriptions of the data and SEIR model employed in the tutorial. Figure~\ref{fig:overview} provides an overview of the steps involved in using \texttt{BayesianFitForecast}.
\subsubsection{Installation and Setup}\label{sec:Installation and setup}

\begin{itemize}
    \item \href{https://github.com/gchowell/BayesianFitForecast/}{Download the R code located in the folder from the GitHub repository.}
    
    \item Install \href{https://cran.r-project.org/}{R (recommended version <= 4.5)} from the Comprehensive R Archive Network (CRAN) and \href{https://posit.co/download/rstudio-desktop/}{RStudio} as the integrated development environment.
    
    \item Ensure that \href{https://cran.r-project.org/bin/windows/Rtools/}{Rtools} and \href{https://www.java.com/en/download/}{Java} are downloaded and up to date.
    
    \item Start an R session in RStudio.
    
    \item If running the code in RStudio for the first time, it is necessary to install all required packages before proceeding. To do so, navigate to the \textbf{Packages} tab in the upper-right panel of RStudio, and select \textbf{Install}. In the dialog box, enter the names of the following packages: \texttt{bayesplot}, \texttt{readxl}, \texttt{xlsx}, \texttt{openxlsx}, \texttt{rstan}, \texttt{ggplot2}, \texttt{loo}, \texttt{stringr}, and \texttt{gridExtra}.
    
    \item \textbf{Important:} This toolbox is compatible with R versions \textbf{prior to 4.5} and Stan versions \textbf{prior to 2.36}. Using later versions may lead to installation or runtime errors due to major changes in package dependencies and Stan behavior.
\end{itemize}

\subsubsection{File descriptions}
Upon downloading the relevant folder, you will find several files inside. Some of these files are meant to be used, while others should remain untouched. The untouched files are functions utilized in other files. Table~\ref{tab:file_roles} provides an overview of each file in the folder and its respective function or purpose. The main files that the user should work with are: \texttt{"options\_generaluse.R"}, \texttt{"run\_MCMC.R"} and \texttt{"run\_analyzeResults.R"}. In addition, the user can generate the prior solution of the model by running \texttt{prior\_sltn.R}. 
\begin{table}[ht]
\caption{Description of key toolbox files and their roles.}
\centering
\begin{tabular}{|p{3.5cm}|p{8cm}|}
\hline
\textbf{Files} & \textbf{Role} \\
\hline
\texttt{diff.R} & This function is designed to identify the derivative of the variable described in the model. It plays a crucial role in fitting models that involve derivatives of variables, ensuring accurate model representation and parameter estimation.\\
\hline
\texttt{Metric\_functions.R} & Contains all the functions necessary for calculating performance metrics such as Mean Absolute Error (MAE), Mean Squared Error (MSE), Weighted Interval Score (WIS), and the coverage of the $95\%$ prediction interval (PI). These metrics are essential for evaluating both model calibration and forecasting performance. \\
\hline
\texttt{ode\_rhs.R} & A function used to extract the right-hand side (RHS) of the ODEs. This file is particularly useful when fitting the model to the derivative of a variable, allowing users to focus on the rate of change within the model.\\
\hline
\texttt{options\_generaluse.R} & This file allows the user to define the model parameters, variables, and other key settings for running the toolbox. It serves as the primary configuration file, offering flexibility in defining the options used in each analysis. \\
\hline
\texttt{prior\_sltn.R}
& This file generates the forward solution of the model by sampling from the prior distribution.
\\
\hline
\texttt{run\_MCMC.R} & This file is responsible for fitting the model to the data by estimating parameters based on the options provided in the configuration file. It is the core component of the toolbox's parameter estimation functionality. \\
\hline
\texttt{run\_analyzeResults.R} & Generates Excel files containing the fitted model results, including forecasts, parameter estimation summaries, and performance metrics. These files are saved in a folder in the output directory, allowing users to easily access, visualize, and interpret the results. \\
\hline
\texttt{stancreator} & Generates a Stan file based on the specified ODE model and parameters, enabling Bayesian inference via Stan. This file automates the complex process of writing Stan code, making Bayesian analysis more accessible to users who may not be familiar with probabilistic programming languages. \\
\hline
\end{tabular}
\label{tab:file_roles}
\end{table}

{\subsubsection{options\_generaluse.R}\label{sec:option}} 
This file includes all setting parameters influencing the model fit and playing a key role. We will describe each of them step by step to ensure clarity.

\subsubsection{calibrationperiods}
These are the specific time intervals used to fit the model to the data. The selection of calibration periods affects the model's ability to accurately represent the observed data. These should be entered as an array of numbers, with each number selected from 1 to the total number of data points. Even if the user wants to consider only one calibration period, the value should still be entered as an array, with only one element. For example:

\begin{Verbatim}

                calibrationperiods <- c(10, 15, 20, 25, 30)

\end{Verbatim}
indicates that the user is interested in fitting the model at several calibration periods: 10, 15, 20, 25, 30. However, if the user is interested in calibrating the model only at 10, they should write:

\begin{Verbatim}

                    calibrationperiods <- c(10).

\end{Verbatim}
Please note that the numbers in \texttt{"calibrationperiods"} should be between 1 and the total number of data points. For instance, in the San Francisco data, the calibration periods can be any positive integer in the interval (1,63).

\subsubsection{forecastinghorizon} This parameter defines how far into the future predictions are made after the model is fit. The choice of forecasting horizon impacts the accuracy and uncertainty of the predictions. However, by initially choosing a large forecasting horizon, you gain the flexibility to generate metrics for a zoomed-in version later. When generating the Excel files and results, you can select a smaller forecasting horizon than the one initially defined. For example, if you set the \texttt{"forecastinghorizon"} to 10 when running \texttt{"run\_MCMC"}, you can later generate results for any horizon less than 10, such as 5, to focus on a specific part of the forecast and obtain updated metrics for that smaller segment. But note that if you initially define the \texttt{"forecastinghorizon"} as 5, the results might slightly differ from the zoomed version when the \texttt{"forecastinghorizon"} is set to 10 due to randomness in MCMC sampling. The zoomed version can be obtained by changing the \texttt{"forecastinghorizon"} in the options file immediately after running \texttt{"run\_MCMC.R"} and before running \texttt{"run\_analyzeResults.R"}. The \texttt{"forecastinghorizon"} can only be a positive integer. For example:
\begin{Verbatim}

                      forecastinghorizon <- 30.
\end{Verbatim}
\subsubsection{model\_name}
This parameter allows the user to assign a name to the project, helping to avoid any confusion when extracting the results. In this toolbox, efforts have been made to assign a specific name to the results folder based on the options. However, to ensure that folder names are concise and easily readable, users can choose their own project name. For example, if two projects have similar settings but different priors, the results of the second project could overwrite those of the first in the same folder. To avoid this, assigning distinct names for each set of parameters can help differentiate between projects. For example:
\begin{Verbatim}

                   model_name <- "Bayesian-normalpriors"
\end{Verbatim}
or,
\begin{Verbatim}
                   model_name <- "Bayesian-niter=1000".
\end{Verbatim}
\subsubsection{vars}
Here, we define the variables in the model. The variables in a model are represented by the letters corresponding to the derivatives taken with respect to $t$. For example, let's first define the simplest SEIR model. Let $S$ denote the susceptible population, $E$ the exposed, $I$ the infected, $R$ the recovered, and $C$ the cumulative number of infected individuals. The SEIR model, incorporating a reporting proportion parameter, is described by the following system of ODEs:
\begin{equation} \label{eq:SEIR}
%\begin{cases}
\frac{dS}{dt} = -\beta \frac{IS}{N}, \quad \frac{dE}{dt} = \beta \frac{IS}{N} - \kappa E, \quad \frac{dI}{dt} = \kappa E - \gamma I,\quad \frac{dR}{dt} = \gamma I,\quad \frac{dC}{dt} = \kappa \rho E.
%\end{cases}
\end{equation}
where $\beta > 0$ represents the transmission rate, $\kappa > 0$ the incubation rate, $\gamma > 0$ the recovery rate, $\rho \in [0, 1]$ the reporting proportion, and $N$ the total population size which is assumed to be known. A compartmental diagram of this model is given in Figure~\ref{fig:seir_model}. 
In this model, $S$, $E$, $I$, $R$, and $C$ are called the state variables and $\beta$, $\kappa$, $\gamma$, $\rho$, and $N$ are called parameters. When defining the variables (\texttt{"vars"}), they should be listed in the same order as their derivatives appear in the system of ODEs. Please note that each variable should be defined as a string (enclosed in double quotes). Therefore, for this model, we have:

\begin{Verbatim}

                   vars <- c("S", "E", "I", "R", "C")
                   
\end{Verbatim}

\begin{figure}[t]
    \centering
     \resizebox{0.7\textwidth}{!}{
    \begin{tikzpicture}[node distance=2.5cm, auto]
        % Nodes
        \node [draw, circle, minimum size=1.3cm] (S) {$S$};
        \node [draw, circle, minimum size=1.3cm, right of=S] (E) {$E$};
        \node [draw, circle, minimum size=1.3cm, right of=E] (I) {$I$};
        \node [draw, circle, minimum size=1.3cm, right of=I] (R) {$R$};
        
        % Observation Node
        \node [draw, rectangle, below of=E, node distance=2.5cm] (obs) {Observations are new cases; $y(t) = \kappa\rho E(t)$};
        
        % Arrows
        \draw [->, thick] (S) -- node[midway, above] {$\frac{\beta SI}{N}$} (E);
        \draw [->, thick] (E) -- node[midway, above] {$\kappa E$} (I);
        \draw [->, thick] (I) -- node[midway, above] {$\gamma I$} (R);
        \draw [->, dashed, thick] (E) -- (obs);
    \end{tikzpicture}
    }
    \caption{\footnotesize Compartmental diagram of the SEIR model with underreporting, illustrating the epidemiological compartments (Susceptible, Exposed, Infected, Recovered, and Cumulative cases) and the transitions between them. Solid arrows indicate the transitions, while dashed arrows show the sources of observed data, specifically newly reported infections.}
    \label{fig:seir_model}
\end{figure}

As another example, we can introduce the SEIRD model (Figure~\ref{fig:seird_model}). In addition to the states $S$, $E$, $I$, and $R$, the SEIRD model also tracks the number of disease-induced deaths (denoted as $D$) and is given by:
\begin{equation}
\begin{gathered}
\label{eq:SEIRD}
%\begin{cases}
\frac{dS}{dt} = -\beta \frac{IS}{N}, \quad 
\frac{dE}{dt} = \beta \frac{IS}{N} - \kappa E, \quad
\frac{dI}{dt} = \kappa E - \gamma I,\quad\\
\frac{dR}{dt} = \gamma (1-\rho) I,\quad
\frac{dD}{dt} = \gamma \rho I,
%\end{cases}
\end{gathered}
\end{equation}
where $\beta$, $\kappa$, and $N$ are defined as above, whereas $\rho$ denotes the proportion of deaths out of the total cases, and $\gamma$ captures the transition rate from infection to recovery ($R$) or death ($D$).

\begin{figure}[t]
    \centering
     \resizebox{0.8\textwidth}{!}{
    \begin{tikzpicture}[node distance=2.5cm, auto]
        % Nodes
        \node [draw, circle, minimum size=1.3cm] (S) {$S$};
        \node [draw, circle, minimum size=1.3cm, right of=S] (E) {$E$};
        \node [draw, circle, minimum size=1.3cm, right of=E] (I) {$I$};
        \node [draw, circle, minimum size=1.3cm, above of=I, node distance=2.5cm] (R) {$R$};
        \node [draw, circle, minimum size=1.3cm, below of=I, node distance=2.5cm] (D) {$D$};
        
        % Observation Node
        \node [draw, rectangle, right of=I, node distance=4cm, align=left] (obs) {Observations are new deaths; \\ $y(t) = \gamma\rho I(t)$};
        
        % Arrows
        \draw [->, thick] (S) -- node[midway, above] {$\frac{\beta SI}{N}$} (E);
        \draw [->, thick] (E) -- node[midway, above] {$\kappa E$} (I);
        \draw [->, thick] (I) -- node[midway, right] {$\gamma (1-\rho) I$} (R);
        \draw [->, thick] (I) -- node[midway, right] {$\gamma \rho I$} (D);
        \draw [->, dashed, thick] (I) -- (obs);
    \end{tikzpicture}
    }
    \caption{\footnotesize Compartmental diagram of the SEIRD model with underreporting, incorporating disease-induced deaths (D). The figure highlights transitions between compartments, with dashed arrows indicating the source of observed data, specifically daily reported deaths.}
    \label{fig:seird_model}
\end{figure}

In this model, we have

\begin{Verbatim}

                  vars <- c("S", "E", "I", "R", "D").
                   
\end{Verbatim}

The SEIUR model (Figure~\ref{fig:seiur_model}) with a reporting proportion parameter is defined as follows:
\begin{equation}
\begin{gathered}
\label{eq:SEIUR}
%\begin{cases}
\frac{dS}{dt} = -\beta \frac{(I+U)S}{N}, \quad
\frac{dE}{dt} = \beta \frac{(I+U)S}{N} - \kappa E, \quad
\frac{dI}{dt} = \kappa \rho E - \gamma I,\\
\frac{dU}{dt} = \kappa (1-\rho) E - \gamma U,\quad
\frac{dR}{dt} = \gamma (I+U), \quad
\frac{dC}{dt} = \kappa \rho E,
%\end{cases}
\end{gathered}
\end{equation}
where all the parameters are defined as above ($\rho$ is the reporting proportion and $\gamma$ captures the transition rate from infection to recovery ($R$)). Here, the infected population is divided into two groups: the reported infected ($I$) and the unreported infected ($U$).
\begin{figure}[t]
    \centering
     \resizebox{0.8\textwidth}{!}{
    \begin{tikzpicture}[node distance=3cm, auto]
        % Nodes
        \node [draw, circle, minimum size=1.3cm] (S) {$S$};
        \node [draw, circle, minimum size=1.3cm, right of=S] (E) {$E$};
        \node [draw, circle, minimum size=1.3cm, above right=of E] (I) {$I$};
        \node [draw, circle, minimum size=1.3cm, below right=of E] (U) {$U$};
        \node [draw, circle, minimum size=1.3cm, right of=E, node distance=12cm] (R) {$R$};
        
        % Observation Node
        \node [draw, rectangle, right of=E, node distance=6cm] (obs) {Observations are new cases; $y(t) = \kappa\rho E(t)$};

        % Arrows
        \draw [->, thick] (S) -- node[midway, above] {$\frac{\beta (I+U)S}{N}$} (E);
        \draw [->, thick] (E) -- node[midway, sloped, above] {$\kappa \rho E$} (I);
        \draw [->, thick] (E) -- node[midway, sloped, below] {$\kappa (1-\rho) E$} (U);
        \draw [->, thick] (I) -- node[midway, sloped, above] {$\gamma I$} (R);
        \draw [->, thick] (U) -- node[midway, sloped, below] {$\gamma U$} (R);
        \draw [->, dashed, thick] (E) -- (obs);
    \end{tikzpicture}
    }
    \caption{\footnotesize Compartmental diagram of the SEIURC model with underreporting, showing an extension of the SEIR model that includes unreported infections (U). The diagram illustrates the transitions between compartments, with dashed arrows denoting the source of observed data—newly reported infection cases.}
    \label{fig:seiur_model}
\end{figure}
In this model, we have
\begin{Verbatim}

                  vars <- c("S", "E", "I", "U", "R", "C").     
\end{Verbatim}

\subsubsection{params}
The parameters in the ODE model are all letters that are not variables. The parameters can either be constants or values that need to be estimated to fit the model to the data. All parameters, whether constants or to be estimated, must be included in the vector \texttt{"params"}. Please note that each parameter should be defined as a string (enclosed in double quotes). For example, the parameters in both the SEIR and SEIRD models should be entered as:

\begin{Verbatim}

            params <- c("beta", "gamma", "kappa", "rho","N").
                  
\end{Verbatim}
Note that the name used does not matter as long as it is a string.

\subsubsection{time\_dependent\_templates}

The \texttt{time\_dependent\_templates} object defines a set of time-dependent parameters, which are useful when the ODE system includes parameters that vary with time. 

To incorporate such parameters into the model, simply define them as string functions, as shown in the examples below. Each function should be expressed in terms of the time variable \texttt{t} and elements from the \texttt{params} vector. These strings can later be parsed and injected into the model code (see~\ref{sec:time-dependent}).

Here are some common examples:

\begin{Verbatim}
            time_dependent_param1 <- "
            return (params1 * exp(-params2 * t));
            "
\end{Verbatim}

\begin{Verbatim}
            time_dependent_param2 <- "
            if (t < params1) {
              return(params2);
            } else if (t < params3) {
              return(params2 + (params4 - params2) * (t - params1)
              / (params3 - params1));
            } else {
              return(params4);
            }
            "
\end{Verbatim}

\begin{Verbatim}
            time_dependent_param3 <- "
            if (t < params1) {
              return(params2);
            } else {
              return(params3);
            }
            "
\end{Verbatim}

\begin{Verbatim}
            time_dependent_param4 <- "
            return (params1 + params2 * sin(2 * pi() * t / params3));
            "
\end{Verbatim}

\begin{Verbatim}
            time_dependent_param5 <- "
            return (params1 * exp(-0.5 * square((t - params2) / params3)));
            "
\end{Verbatim}
In summary, a time-dependent parameter is simply a user-defined function of time and model parameters. These functions reference the parameter values already included in the \texttt{params} vector and are designed to be evaluated at runtime within the ODE solver or model function.

\subsubsection{ode\_system}
The \texttt{"ode\_system"} is a string that defines the system of ODEs you want to work with. It includes parameters, variables, and the derivatives of variables with respect to $t$. The way we define the parameters, variables, and derivatives is by coding \texttt{"params\large\textbf{i}\normalsize "}, \texttt{"vars\large\textbf{i}\normalsize "}, and \texttt{"diff\_var\large\textbf{i}\normalsize "}, where \texttt{"\large\textbf{i}\normalsize "}  represents the index of the parameter or variable in the \texttt{"params"} or \texttt{"vars"} lists. It is important to note that the first line in the string should be left empty. Additionally, the last line should end with the final differential equation, with the closing quotation mark placed immediately after the last character of the string. For example, assuming:

\begin{Verbatim}

                vars <- c("S", "E", "I", "R", "C"),          
         params <- c("beta", "gamma", "kappa", "rho","N"),
                  
\end{Verbatim}
we should write the SEIR model (\ref{eq:SEIR}) as

\begin{Verbatim}

ode_system <- '
        diff_var1 = -params1 * vars3 * vars1 / params5
        diff_var2 = params1 * vars3 * vars1 / params5 - params3 * vars2
        diff_var3 = params3 * vars2 - params2 * vars3
        diff_var4 = params2 * vars3
        diff_var5 = params4 * params3 * vars2'. 
                  
\end{Verbatim}
As another example, assuming

\begin{Verbatim}

                vars <- c("S", "E", "I", "R", "D"),
         params <- c("beta", "gamma", "kappa", "rho","N"),
                  
\end{Verbatim}
we should write the SEIRD model (\ref{eq:SEIRD}) as

\begin{Verbatim}

ode_system <- '
        diff_var1 = -params1 * vars3 * vars1 / params5
        diff_var2 = params1 * vars3 * vars1 / params5 - params3 * vars2
        diff_var3 = params3 * vars2 - params2 * vars3
        diff_var4 = params2 * (1-params4) * vars3
        diff_var5 = params4 * params2 * vars3'.      
\end{Verbatim}

If the ODE system includes any time-dependent parameter, simply write it as \texttt{time\_dependent\_param1} as defined in the \texttt{time\_dependent\_templates} list.

\begin{Verbatim}
params <- c("beta0","beta1", "q", "kappa", "gamma", "rho", "N", "t_int")

time_dependent_templates <- list(
  time_dependent_param1 = "if (t < params8) { return (params1); } 
  else { return (params2 + (params1 - params2) 
  * exp(-params3 * (t - params8))); }"
)

ode_system <- '
  diff_var1 = -time_dependent_param1 * vars1 * vars3 / params7
  diff_var2 = time_dependent_param1 * vars1 * vars3 / params7 
                                    - params4 * vars2
  diff_var3 = params4 * vars2 - params5 * vars3
  diff_var4 = params5 * vars3
  diff_var5 = params4 * params6 * vars2'
\end{Verbatim}

\subsubsection{paramsfix}
Users may sometimes want to keep certain parameters fixed. For example, in many models, the population size is treated as a fixed parameter and does not require estimation. In this toolbox, users have the flexibility to designate any parameter as constant. We have defined an array \texttt{"paramsfix"} with the same length as the \texttt{"params"} array. If an element in \texttt{"paramsfix"} is set to 1, it indicates that the corresponding parameter in \texttt{"params"} is constant. If it is set to 0, the parameter will be estimated. For example, assuming:
\begin{Verbatim}

            params <- c("beta", "gamma", "kappa", "rho","N"),
                
\end{Verbatim}
and 
\begin{Verbatim}

                    paramsfix <- c(1,0,0,0,1)
                
\end{Verbatim}
indicates that the parameters $\beta$ and $N$ are constants, and their values should be specified at a later stage.

\subsubsection{Composite expressions}
Sometimes, users need to estimate derived parameters, such as the basic reproduction number, recovery time, or other relevant metrics. To facilitate this, users should include these parameters in the \texttt{composite\_expressions} list. Each entry in the list should pair the parameter name on the left with its corresponding formula on the right, expressed as a string and as a function of the model's parameters. For example, in the SEIR model, the basic reproduction number and recovery time might be defined as follows:
\begin{Verbatim}

            composite_expressions <- list(
                            R0 = "beta / gamma",
                            recovery_time = "1 / gamma"
                            )
\end{Verbatim}
\subsubsection{fitting\_index and fitting\_diff}
These parameters are arrays that specify which variable the data points correspond to. For example, if the data represents the number of infected people in the SEIR model, we should set \texttt{"fitting\_index"} to \texttt{c(3)}, since $I$ is the third variable in the \texttt{"vars"} array. Similarly, if the data pertains to the number of deaths in the SEIRD model, we should set \texttt{"fitting\_index"} to \texttt{c(5)} because $D$ is the fifth variable in \texttt{"vars"}.

Additionally, the data can be associated with the derivative of these variables by setting \texttt{"fitting\_diff"} to \texttt{c(1)}. For instance, if \texttt{"fitting\_index"} is \texttt{c(5)} and \texttt{"fitting\_diff"} is \texttt{c(1)} in the SEIRD model, it means that we are fitting the data to $\frac{dD}{dt}$, representing the number of daily deaths. Similarly, setting \texttt{"fitting\_index"} to \texttt{c(3)} and \texttt{"fitting\_diff"} to \texttt{c(1)} in the SEIR model would indicate fitting the data to $\frac{dI}{dt}$, representing the number of new infections per day.

It is important to note that these values should be entered as arrays, since we might have multiple datasets. For example, specifying \texttt{"fitting\_index"} to \texttt{c(3,5)} and \texttt{"fitting\_diff"} to \texttt{c(0,1)} in the SEIR model means that we have two different time-series: the first corresponding to the number of infected people, $I$, and the second to the number of case incidences, $\frac{dC}{dt}$.

\subsubsection{errstrc}
This parameter is an integer that specifies the type of error structure incorporated when fitting the model to the data. These refer to the statistical models used to describe the variability or noise in the data. In this paper, we explore three error structures: 1. negative binomial, which accounts for overdispersion in count data; 2. normal, which assumes constant variance; and 3. Poisson, which assumes that the variance equals the mean. The error structure can be selected by specifying the corresponding number for \texttt{"errstrc"}. One additional parameter needs to be estimated in negative binomial and normal distributions: the dispersion $\phi$ in negative binomial and the standard deviation $\sigma$ in normal.

\subsubsection{cadfilename1}\label{sec:Excel}
This string represents the name of the Excel file in the working directory, excluding the extension ``.xlsx''. The format of the Excel file should have at least two columns, where the first column contains the sequential time points (e.g., 0, 1, 2, \dots), representing days, weeks, years, etc. The second column and any additional columns should contain the temporal incidence data related to the state variables or their derivatives. It is important to start the second column with ``cases1'', and if additional columns exist, they should follow the pattern ``cases2'', ``cases3'', etc.

\subsubsection{caddisease}
This is a string representing the name of the disease or process. This parameter is used to create a unique name for the results folder, distinguishing it from folders related to other diseases and preventing overwriting.

\subsubsection{series\_cases and datetype}
The user can set these string parameters based on their dataset. They are used only when plotting the fit and forecasting curves. For instance, if you set \texttt{"series\_cases"} to \texttt{"Cases"}, it indicates that there is only one time series, and the y-axis label of the plot will be \texttt{"Cases"}. If you set \texttt{"series\_cases"} to \texttt{"infected", "recovered"}, it means there are two time series (cases1 and cases2), and the y-axis labels of the plots will be \texttt{"infected"} and \texttt{"recovered"}. Additionally, if you set \texttt{"datetype"} to \texttt{"Days"}, the x-axis label of the plot will be \texttt{"Days"}.

\subsubsection{Priors}
This section allows the user to specify the prior distributions for the parameters. If a parameter is fixed (i.e., the corresponding location in \texttt{"paramsfix"} is 1), the prior should be defined as a constant value and will not be changed. A distribution must be entered as a string (within quotation marks) if a parameter needs to be estimated. Various types of distributions can be used. When specifying the prior distributions, we should pay attention to the range of parameters. For example, the parameters $\beta$, $\gamma$, and $\kappa$ are positive. If we use normal distributions as their priors, it is recommended to truncate the distribution at 0 (by using $T[0,]$ at the end of the normal distribution).

Here’s how to enter the priors: You should define the priors for all parameters listed in \texttt{"params"}. The order in which you enter the priors must match the order in \texttt{"params"}. For example, suppose we have the following:

\begin{Verbatim}

            params <- c("beta", "gamma", "kappa", "rho","N"),
                            
\end{Verbatim}
and the prior beliefs are as follows:

\begin{enumerate}[left=2em] 
    \item \texttt{beta} is a constant with a value of 2.
    \item \texttt{gamma} and \texttt{kappa} follow normal prior distributions with a mean of 1 and a standard deviation of 2.
    \item \texttt{rho} has a uniform prior distribution ranging from 0 to 1.
    \item The population size \texttt{N} is 10,000.
\end{enumerate}
The prior distribution would be defined as follows:

\begin{Verbatim}

                    params1_prior <- 2
                    params2_prior <- "normal(1,2)T[0,]"
                    params3_prior <- "normal(1,2)T[0,]"
                    params4_prior <- "uniform(0,1)"
                    params5_prior <- 10000
\end{Verbatim}

\subsubsection{Lower and Upper Bounds}

This section allows the user to define the lower and upper bounds for the parameters. The bounds restrict the range within which the parameters can vary during estimation.

The lower bound (\texttt{"LB"}) specifies the minimum value a parameter can take, while the upper bound (\texttt{"UB"}) specifies the maximum value. If no lower or upper bound is necessary for a particular parameter, it can be defined as \texttt{"NA"}, implying no minimum or maximum value restriction.

Similar to the priors, the order in which you enter the bounds must match the order of parameters in \texttt{"params"}. For example, if the parameters are as follows:
\begin{Verbatim}

             params <- c("beta", "gamma", "kappa", "rho","N"),
    
\end{Verbatim}
and the bounds are defined as:
\begin{enumerate}[left=2em] 
    \item \texttt{beta}, \texttt{gamma} and \texttt{kappa} have lower bounds of 0 and no upper bounds.
    \item \texttt{rho} has a lower bound of 0 and an upper bound of 1.
    \item The population size \texttt{N} has no lower or upper bounds specified.
\end{enumerate}
The lower and upper bounds would be defined as follows:
\begin{Verbatim}

                        params1_LB <- 0
                        params2_LB <- 0
                        params3_LB <- 0
                        params4_LB <- 0
                        params5_LB <- NA
                            
                        params1_UB <- NA
                        params2_UB <- NA
                        params3_UB <- NA
                        params4_UB <- 1
                        params5_UB <- NA
    
\end{Verbatim}
Setting these bounds ensures that the parameter values remain within the specified range during the estimation process.
\subsubsection{Prior distribution for additional parameters in error structure}

As explained earlier, we have additional parameters for the negative binomial and normal error structures: $\phi$ for the negative binomial and $\sigma$ for the normal distribution. For these parameters, we must define a prior distribution. Here’s how to specify these priors: The standard deviation $\sigma$ in the normal error structure is defined using \texttt{normalerror\_prior}, and the dispersion parameter $\phi$ in the negative binomial error structure is defined using \texttt{negbinerror\_prior}. For example:

\begin{Verbatim}

                normalerror1_prior <- "cauchy(0, 2.5)",
    
\end{Verbatim}
which assumes a Cauchy prior distribution with a location parameter of 0 and a scale parameter of 2.5 for $\sigma$. If there are multiple datasets, we should define additional error parameters. For example, if there are three different time series, we have:

\begin{Verbatim}
    
                normalerror1_prior <- "cauchy(0, 2.5)"
                normalerror2_prior <- "cauchy(0, 2.5)"
                normalerror3_prior <- "cauchy(0, 2.5)"
    
\end{Verbatim}

It is important to note that if \texttt{errstrc} is set to 1, then only \texttt{negbinerror\_prior} needs to be defined, and \texttt{normalerror\_prior} will not play any role in fitting the model. Similarly, if \texttt{errstrc} is set to 2, then only \texttt{normalerror\_prior} needs to be defined, and \texttt{negbinerror\_prior} will not be used.

\subsubsection{Ic}
This is a vector representing the initial conditions when using a Bayesian method through Stan. The initial conditions correspond to the initial values of the state variables, and their order should align with the order in \texttt{vars}. For example, suppose that in the SEIR model, we have $S(0) = N - i_0$, $E(0) = 0$, $I(0) = i_0$, $R(0) = 0$, and $C(0) = i_0$, where $i_0$ is the first data point observed in the temporal incidence data, and $N$ is the population size. If $i_0 = 1$, we can write:

\begin{Verbatim}

                Ic = c(params5_prior - 1, 0, 1, 0, 1),
    
\end{Verbatim}
since $N =$ \texttt{params5\_prior} as explained before. Alternatively, we can write the value of $N$ explicitly:

\begin{Verbatim}

                    Ic = c(10000 - 1, 0, 1, 0, 1).
\end{Verbatim}

\subsubsection{vars.init}
Occasionally, a user may need to estimate the initial conditions when the initial number of infected, exposed, or other initial values are unknown. In such cases, mathematical tools can be used for estimation. Within this toolbox, this can be achieved by setting \texttt{vars.init = 0}. If the user already knows the initial value, they should set \texttt{vars.init = 1}. When \texttt{"vars.init"} is set to 0, additional adjustments will be necessary, as outlined below:

\begin{enumerate}[left=2em] 
    \item \texttt{"params"}: Add the new parameter (related to the initial value) to the \texttt{"params"} list. For example, in the SEIR model, if the initial number of infected people needs to be estimated, we can add $i_0$ to \texttt{"params"}.
    \item \texttt{"paramsfix"}: Update \texttt{"paramsfix"} to include the new parameter. Set the corresponding entry to 0, indicating that the new parameter should be estimated.
    \item \texttt{"priors"}: Include a prior distribution for the new parameter in the priors section.
    \item \texttt{"Ic"}: Specify the initial condition as a string. For example, in the SEIR model, if the new parameter is named $i_0$, then set \texttt{Ic = ("N-$i_0$", 0,"$i_0$", 0, "$i_0$")}. Ensure all letters are enclosed in  double quotation marks (strings).
\end{enumerate}
If vars.init = 1, no changes are needed.

\subsubsection{niter and num\_chain}
These are two positive integers that are crucial when working with Stan. niter specifies the number of iterations for MCMC sampling and num\_chain specifies the number of MCMC chains. They are important parameters to ensure and check convergence in Bayesian analysis. 

\subsubsection{prior\_sltn.R} 
\texttt{prior\_sltn.R} performs prior predictive simulations by sampling parameter values from user-defined prior distributions and using them to solve the specified ODE model. For each draw from the prior, the function integrates the system of equations to produce a trajectory, and the collection of these trajectories forms an ensemble that reflects the uncertainty inherent in the prior assumptions. To run the function, the user must define the ODE model, including its parameters and state variables, and specify the prior distributions for the parameters in the options file. It is worth noting that the main functions of this toolbox, \texttt{run\_MCMC} and \texttt{run\_analyzeResults}, which generate the posterior distribution of the parameters, will be described later. The \texttt{prior\_sltn.R} function is used solely to generate the prior solution of the model using the prior distribution of parameters. 

As an example, we demonstrate the use of the \texttt{prior\_sltn.R} function with the SEIR model described earlier. Simulated data for the incidence cases ($\kappa E$) are generated by solving the SEIR model forward using the parameter values $\beta = 0.6$, $\gamma = 0.1$, $\kappa = 0.2$, $\rho = 0.9$, and $N = 550{,}000$, and are stored in \texttt{SEIRC.xlsx}. To reflect strong prior knowledge, we specify informative normal prior distributions centered at these true values, each with a small standard deviation ($\sigma = 0.01$): \texttt{beta <- "normal(0.6, 0.01)T[0,]"}, \texttt{gamma <- "normal(0.1, 0.01)T[0,]"}, \texttt{kappa <- "normal(0.2, 0.01)T[0,]"}, and \texttt{rho <- "normal(0.9, 0.01)T[0,]"}. These specifications, defined in \texttt{options\_SEIR\_sanfrancisco\_prior.R}, constrain the sampled parameter values to remain very close to the truth, resulting in simulated trajectories that closely align with the simulated dataset.

In order to use the \texttt{prior\_sltn.R} function, we need to update line 10 of the code to include the option file \texttt{options\_SEIR\_sanfrancisco\_prior.R}. 

\begin{figure}[H]
\centering
\begin{minipage}{0.48\linewidth}
\centering
\includegraphics[width=\linewidth]{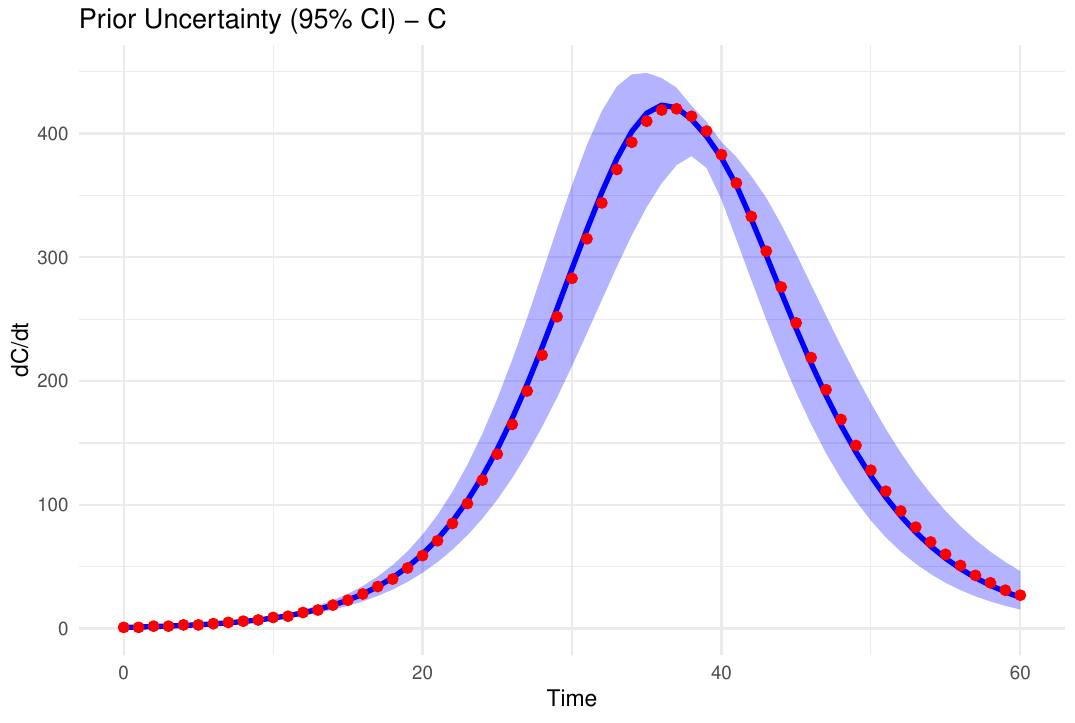}
\end{minipage}
\hfill
\begin{minipage}{0.48\linewidth}
\centering
\includegraphics[width=\linewidth]{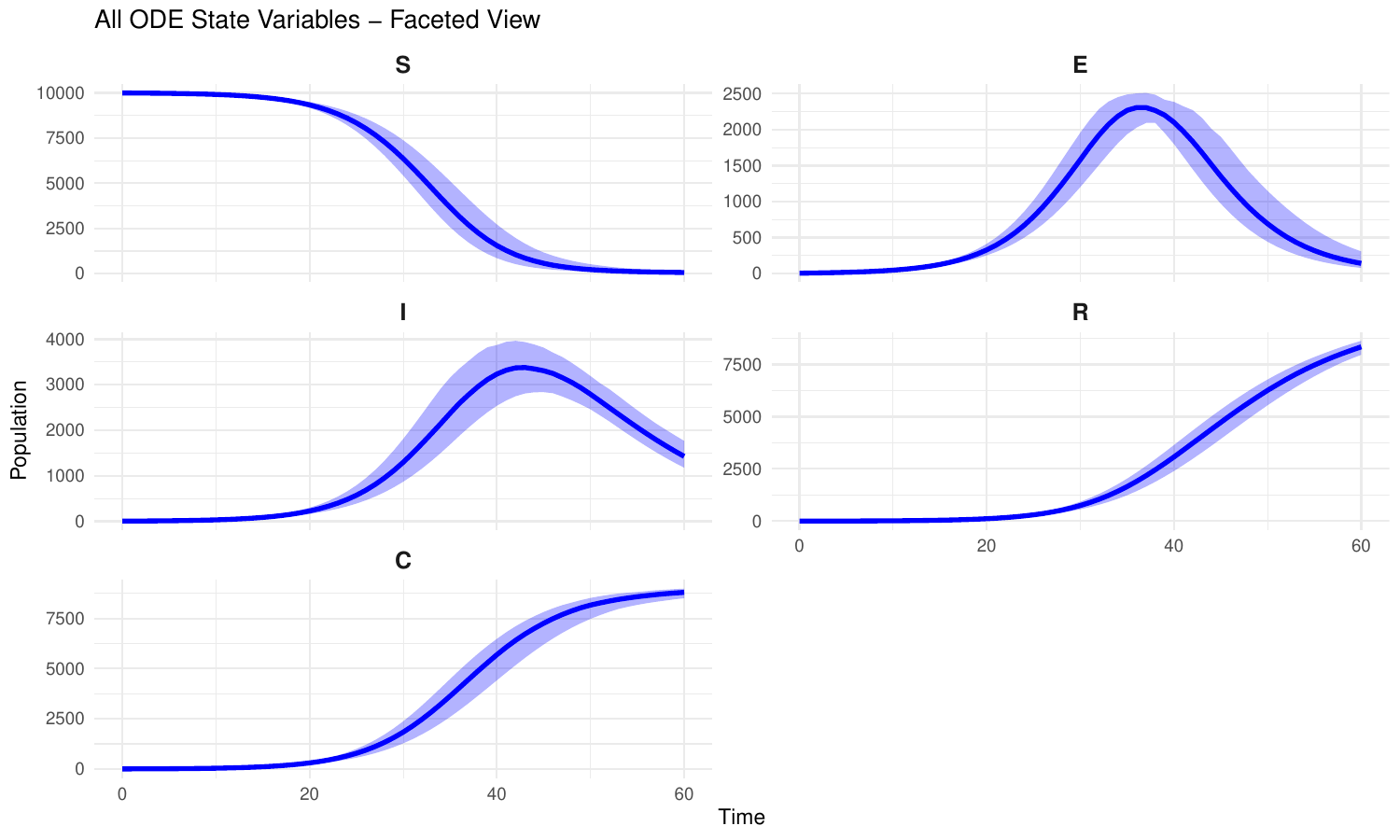}
\end{minipage}
\caption{\footnotesize Prior predictive simulation results using strong informative priors: (a) incidence cases and (b) all state variables.}
\label{fig:prior_pred}
\end{figure}

Figure~\ref{fig:prior_pred} shows prior uncertainty (95\% CI) for $\dfrac{dC}{dt}$ and the resulting ODE state variable trajectories. The tight prior distribution for the parameters yields low variability in the simulated trajectories, representing our strong initial beliefs before observing data.

\subsubsection{run\_MCMC}
This file is designed to fit the model to the data using user-defined parameters. As mentioned earlier, the user can configure the options in the options file and call it here. Therefore, it is crucial to update line 5 of this code to match the name of the configured options file.

\subsubsection{Options library}
To make the toolbox user-friendly, we have already prepared a library of option files in the main folder. These files allow users to leverage pre-configured case studies and serve as examples to learn how to define the parameters as explained in the previous section. This library includes analyses of simulated data as well as four real-world epidemics:

\begin{itemize}[left=2em] 
    \item San Francisco 1918 flu
    \item Cumberland 1918 flu
    \item Bombay plague 1896-97
    \item Switzerland COVID-19
\end{itemize}

For each dataset, we have provided two prior distributions for ODE parameters $\beta$, $\gamma$, $\kappa$: the normal and the uniform distribution. In total, 10 option files are already included in the toolbox for use. Their names are as follows:

\begin{itemize}[left=2em] 
    \item options\_SEIR\_simulated\_normalprior
    \item options\_SEIR\_simulated\_uniformprior
    \item options\_SEIR\_sanfrancisco\_normalprior
    \item options\_SEIR\_sanfrancisco\_uniformprior
    \item options\_SEIR\_cumberland\_normalprior
    \item options\_SEIR\_cumberland\_uniformprior
    \item options\_SEIRD\_plague\_normalprior
    \item options\_SEIRD\_plague\_uniformprior
    \item options\_SEIURC\_covid\_normalprior
    \item options\_SEIURC\_covid\_uniformprior
\end{itemize}
To select any of these option files, update line 5 of \texttt{"run\_MCMC"}. For example, to analyze the San Francisco 1918 flu with normal prior distributions, modify line 5 to:

\begin{Verbatim}

            source("options_SEIR_sanfrancisco_normalprior.R")
    
\end{Verbatim}
However, if you prefer to use your option file, configure the \texttt{"options\_generaluse.R"} file according to the explanation in the previous section, and then modify line 5 to:
\begin{Verbatim}

                  source("options_generaluse.R")
    
\end{Verbatim}
Note that the name of the option file can be anything, but what is important is the format of the option file, and keeping and setting all parameters as they appear in \texttt{"options\_generaluse.R"}.

\subsubsection{Generating the Stan file}
One of the key features of this toolbox is that the user does not need to interact directly with a Stan file. Typically, in the Bayesian framework, a Stan file is essential for performing all Bayesian analyses. However, with this toolbox, once you set the options file and run \texttt{"run\_MCMC"}, the Stan file is automatically generated and placed in the toolbox folder, enabling Bayesian inference. If you're interested in examining the Stan file, you can find it as \texttt{"ode\_model.stan"} in the same directory.

\subsubsection{Saved data}
After running \texttt{"run\_MCMC"}, the samples will be generated, and an \texttt{".Rdata"} file will be saved in the same directory (toolbox folder) for data analysis. For each calibration period in the \texttt{"calibrationperiods"} array, there will be a corresponding \texttt{".Rdata"} file. The generated file name is based on the following setting parameters:

\begin{itemize}[left=2em] 
    \item ``model\_name"
    \item ``calibrationperiod"
    \item ``forecastinghorizon"
    \item ``errstrc"
    \item ``caddisease"
\end{itemize}
Therefore, if two option files have identical setting parameters as described above, the second file will overwrite the first one. It is recommended to change the \texttt{model\_name} to avoid overwriting.

\subsubsection{run\_analyzeResults}{\label{sec:analyzeResults}}
By running this file, the user can generate and analyze the results. It is important to keep the option file the same as when running \texttt{"run\_MCMC"}(except the forecasting horizon to achieve a zoomed version as already explained). Additionally, line 13 of this file needs to be modified to include the name of the option file you are using. After running this code, several results will be generated in the output folder. The files are formatted as follows:

\begin{Verbatim}
model_name + caddisease + errstrc + calibrationperiod + 
                                                     forecastinghorizon 
\end{Verbatim}
The folder will include four Excel files as follows:

\begin{itemize}[left=2em] 
    \item \textbf{Convergence}: This file includes the convergence parameter \texttt{"Rhat"}. It is recommended to use the results obtained by Bayesian inference when the parameter \texttt{"Rhat"} is less than 1.1. To achieve this goal, you can increase \texttt{"niter"}, the number of iterations, or \texttt{"num\_chain"}, the number of chains. To illustrate the format of this file, consider the example: \texttt{convergence-Bayesian-weak-sanfrancisco-normal-cal-10-fcst-10}.
    \item \textbf{Forecast}: This file includes five columns: Date, Data, Median, Lower Bound and Upper Bound of the 95\% PIs. It can be used to plot the fit easily using the given values. To illustrate the format of this file, consider the example: \texttt{"forecast-Bayesian-weak-sanfrancisco-normal-cal-10-fcst-10"}.
    \item \textbf{Parameters}: This file provides an analysis of the parameter estimation by giving the median, mean, lower bound and upper bound of the 95\% CIs. In the case of using a normal error structure, the parameter $\sigma$ and using a negative binomial error structure, the parameter $\phi$ will also be estimated here. To illustrate the format of this file, consider the example: \texttt{"parameters-Bayesian-weak-sanfrancisco-normal-cal-10-fcst-10"}.
    \item \textbf{Performance Metrics}: This file provides four performance metrics for both calibration and forecasting periods: MAE, MSE, WIS, and Coverage of the 95\% PIs. It is important to note that if data is unavailable for the forecasting period, the performance metrics will show NA, which is not applicable for the forecasting part. To illustrate the format of this file, consider the example: \texttt{"performance metrics-Bayesian-weak-sanfrancisco-normal-cal-10-fcst-10"}.
\end{itemize}

In addition to the Excel files, some PDF files will be generated in the folder:

\begin{itemize}[left=2em] 
    \item \textbf{Histograms}: The histograms of all estimated parameters will be saved as PDF files. In the case of a normal error structure, the histogram of the parameter $\sigma$, and in the case of a negative binomial error structure, the histogram of the parameter $\phi$ can be found in the folder. If there are composite expressions, they will also be included here. To illustrate the format of this file, consider the example: \texttt{"beta-histogram-Bayesian-weak-\\sanfrancisco-normal-cal-10-fcst-10"}.
    \item \textbf{Forecast}: The plot for fitting the model and forecasting will be found here. It shows the median fit, the margins showing the 95\%PI, the available data points, and the calibration line. To illustrate the format of this file, consider the example: \texttt{"Forecast-Bayesian-weak-sanfrancisco-normal-\\cal-10-fcst-10"}.
    \item \textbf{Trace plot}: A trace plot is a graphical representation of the sampled values of a parameter over iterations of the MCMC algorithm. It shows how the parameter values change from one iteration to the next, allowing you to visually inspect the behavior of the sampler. To illustrate the format of this file, consider the example: \texttt{"traceplot-Bayesian-weak-sanfrancisco-normal-cal-10-fcst-10"}.
\end{itemize}

\section{Results and Discussion}
Here we will present some real examples of running a simulation. We follow a workflow similar to that described in \cite{chowell2024parameter}. Also, one example of the simulated data for fitting the SEIR model~(\ref{eq:SEIR}) to multiple data series simultaneously will be presented. 

\subsection{1918 influenza pandemic in San Francisco with Poisson error structure}\label{sec:examplePoisson}
In this example, the SEIR model (\ref{eq:SEIR}) with the parameter $\rho = 1$ and the SF 1918 flu dataset is considered. We begin by fitting the model using a Poisson error structure. To do this, we need to modify the options file to achieve the desired results and enable a fair comparison. Specifically, we modify the \texttt{"options\_SEIR\_sanfrancisco\_normalprior"} file, though not all sections need to be filled out. For instance, we can make a copy of this file and name it \texttt{"options\_SEIR\_sanfrancisco\_Ex1.R"}.  Modify the parameters as below.

\begin{Verbatim}

    calibrationperiods <- c(17)
    forecastinghorizon <- 10
    vars <- c("S", "E", "I", "R", "C")
    ode_system <- '
      diff_var1 = -params1 * vars3 * vars1 / params5
      diff_var2 = params1 * vars3 * vars1 / params5 - params3 * vars2
      diff_var3 = params3 * vars2 - params2 * vars3
      diff_var4 = params2 * vars3
      diff_var5 = params4 * params3 * vars2'
    paramsfix <- c(0,1,1,1,1)
    fitting_index <- 5
    fitting_diff <- 1
    errstrc <- 3
    params1_prior <- "uniform(0, 10)"
    params2_prior <- 1/4.1
    params3_prior <- 1/1.9
    params4_prior <- 1
    params5_prior <- 550000
    niter <- 1000
    
\end{Verbatim}
We retain all other parameters from the existing file. Special attention is given to the \texttt{paramsfix} option, where $\rho$ is fixed, and the prior is set accordingly. Since \texttt{params4\_prior = 1} implies $\rho = 1$, there is no need to exclude this parameter from the model.

\begin{figure}[t]
    \centering
    \begin{minipage}[b]{0.45\textwidth}
        \centering
        \includegraphics[width=\linewidth]{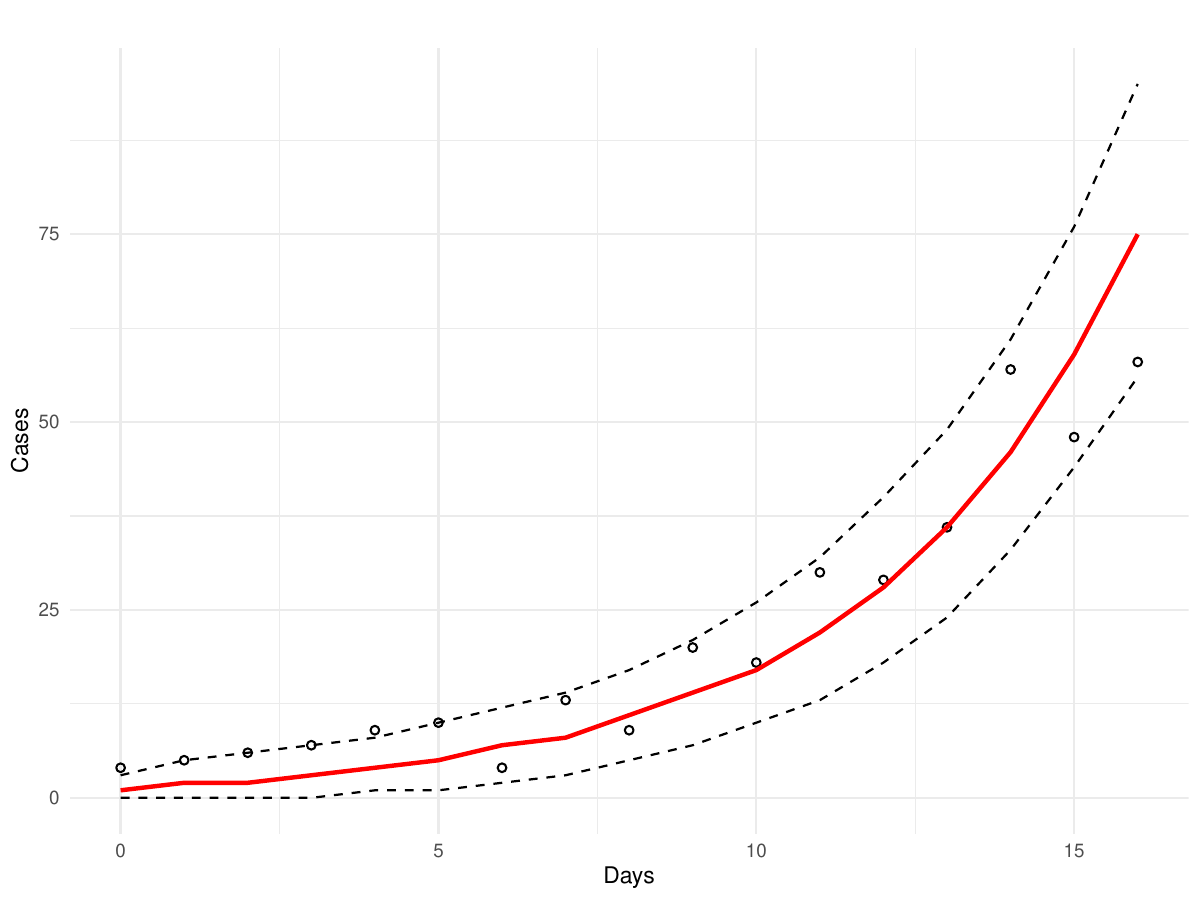}
    \end{minipage}
    \\
    \begin{minipage}[b]{0.45\textwidth}
        \centering
        \includegraphics[width=\linewidth]{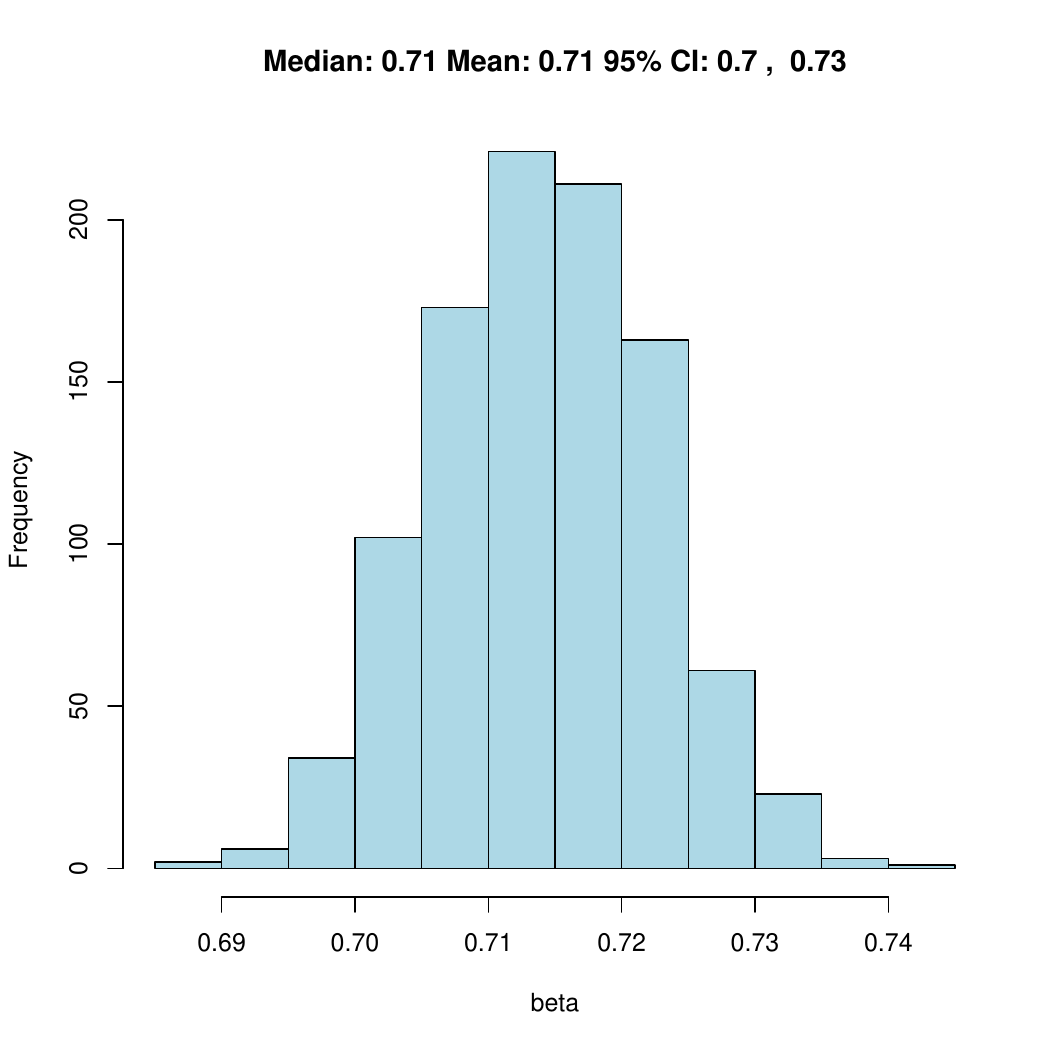}
    \end{minipage}
    \hfill
    \begin{minipage}[b]{0.45\textwidth}
        \centering
        \includegraphics[width=\linewidth]{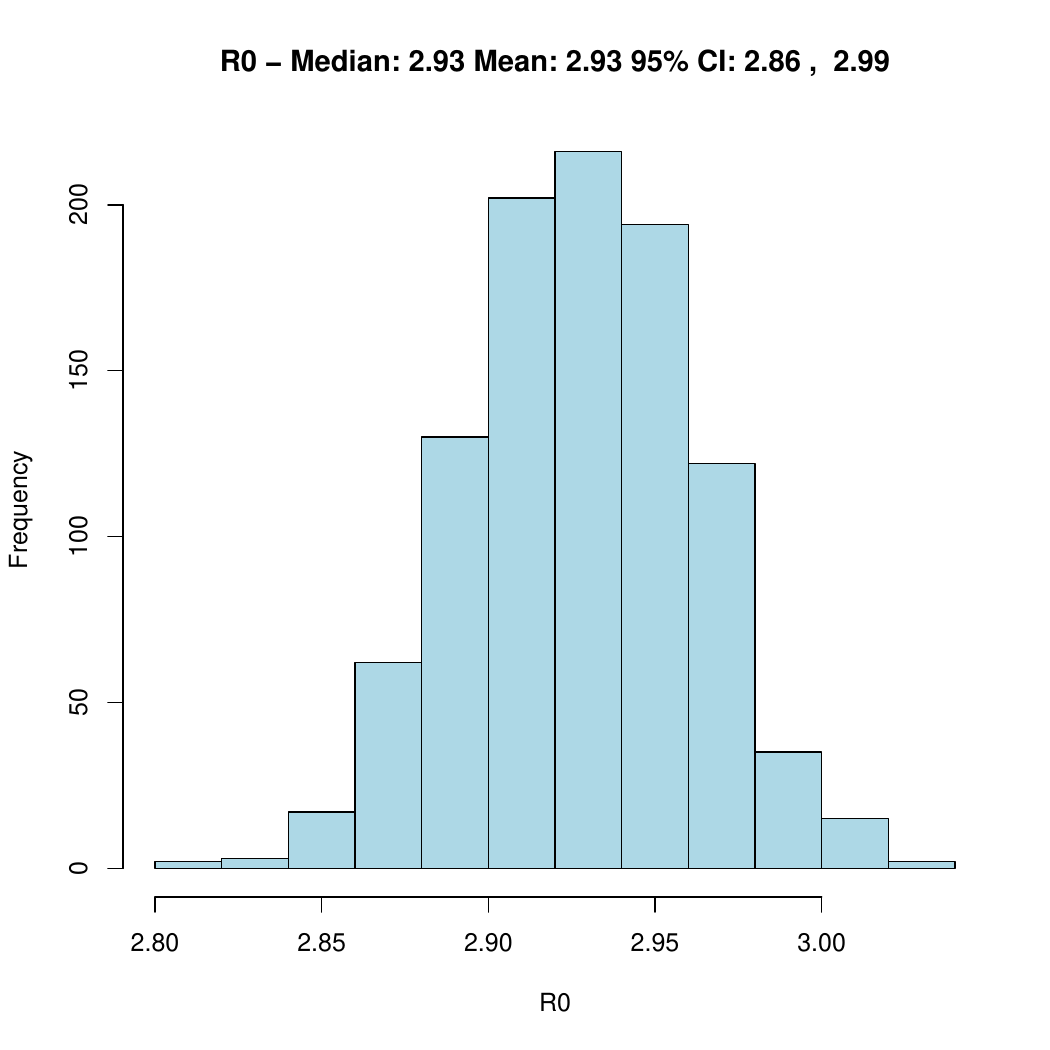}
    \end{minipage}\caption{\footnotesize Bayesian fitting of the SEIR model to the 1918 influenza dataset from San Francisco was performed using a Poisson error structure, calibrated over 17 days. The model captures the dynamics of newly infected individuals, represented by $\frac{dC}{dt}$. The parameters $\kappa = \frac{1}{1.9}$, $\gamma = \frac{1}{4.1}$, and $\rho = 1$ are held constant, while the prior distribution for $\beta$ is uniform(0,10). All parameters were constrained to have non-negative values, with a total population size fixed at $550,000$. Initial conditions were set as (549996, 0, 4, 0, 4), based on the first recorded data point. The MCMC algorithm was executed with 1,000 iterations across two independent chains to ensure robust convergence. In the top panel, the red curve represents the model’s best fit to the data, while the black dashed lines delineate the $95\%$ prediction interval (PI). The circles represent the observed data points for comparison.The bottom panel presents a histogram of the posterior distributions for the transmission rate parameter $\beta$, and the reproduction number $R\_0$, derived from the frequency of MCMC iterations. The median, mean, and $95\%$ credible intervals (CI) are highlighted, giving a comprehensive summary of the parameter estimates and their uncertainties.}
    \label{fig:Bayesian-uniform-sanfrancisco-poisson-cal-17-fcst-0} 
\end{figure}

Next, open \texttt{run\_MCMC} to execute the MCMC algorithm. Begin by modifying line 5 in \texttt{run\_MCMC} to reflect the new options file: \texttt{options\_SEIR\_sanfrancisco\_Ex1.R}. Select all lines of code, then click `Run' in RStudio. This process may take some time to complete. Once finished, an \texttt{.Rdata} file will be generated in the same directory, indicating that the data is ready for analysis. Proceed by opening the \texttt{run\_analyzeResults.R} script to generate the results. Ensure that you modify the name of the options file here as well. Once this section is executed, the results will be produced. Relevant figures and tables are shown in Figure~\ref{fig:Bayesian-uniform-sanfrancisco-poisson-cal-17-fcst-0} and Table \ref{tab:performance-Bayesian-uniform-sanfrancisco-poisson-cal-17-fcst-0} and \ref{tab:convergence-performance-Bayesian-uniform-sanfrancisco-poisson-cal-17-fcst-0}.

\begin{table}[t]
\centering
\caption{The performance metrics resulting from Bayesian fitting of the SEIR model to the 1918 influenza dataset from San Francisco, using a Poisson error structure and 17 days of calibration data.} 
\label{tab:performance-Bayesian-uniform-sanfrancisco-poisson-cal-17-fcst-0}
\begin{tabular}{|l|l|l|}
\hline 
& \textbf{Calibration} & \textbf{Forecasting} \\ \hline
\textbf{MAE} & 5.24& NA \\
\textbf{MSE} & 45.35 & NA \\
\textbf{WIS} & 3.33 & NA \\
\textbf{Coverage of 95\% CI} & 88.24 & NA \\ \hline
\end{tabular}
\footnotetext{The model fits the data to the number of newly infected individuals, represented by \(\frac{dC}{dt}\). The parameters \(\kappa = \frac{1}{1.9}\), \(\gamma = \frac{1}{4.1}\), and \(\rho = 1\) are held constant, with a uniform prior distribution \((0, 10)\) for \(\beta\). All parameters are constrained to have non-negative values. The population size is fixed at 550,000, with initial conditions set to (549996, 0, 4, 0, 4), based on the first recorded data point. The MCMC algorithm was executed with 1,000 iterations across two chains to ensure robust convergence.
}
\end{table}

\begin{table}[t]
\centering
\caption{Posterior inference and convergence analysis for the example applying the SEIR model to the San Francisco 1918 influenza dataset with a Poisson error structure.}
\begin{tabular}{|l|l|l|l|l|l|l|}
\hline
\textbf{Calibration} & \textbf{Parameter} & \textbf{Mean} & \textbf{Median} & \textbf{CI\_95} & \textbf{N\_eff} & \textbf{Rhat} \\ \hline
17                   & $\beta$            & 0.71          & 0.71            & ( 0.7 , 0.73 )  & 316.35          & 1          \\ \hline
\end{tabular}
\footnotetext{The model is fitted to the number of newly infected individuals, represented by $\frac{dC}{dt}$. The parameters $\kappa = \frac{1}{1.9}$, $\gamma = \frac{1}{4.1}$, and $\rho = 1$ are treated as constants, while the prior distribution for $\beta$ is uniform(0,10). All parameters are bounded below by zero to ensure non-negative values. The population size is set at $550,000$, with initial conditions of (549996, 0, 4, 0, 4) based on the first recorded data point in the dataset. The MCMC algorithm was executed for 1,000 iterations across two chains, ensuring robust posterior estimates.}
\label{tab:convergence-performance-Bayesian-uniform-sanfrancisco-poisson-cal-17-fcst-0}
\end{table}

\subsection{Comparison of Calibration Performance: Negative Binomial vs. Poisson Error Structures}
\begin{figure}[t]
    \centering
    % First row: Time series plot
    \begin{minipage}[b]{0.5\textwidth}
        \centering
        \includegraphics[width=\linewidth]{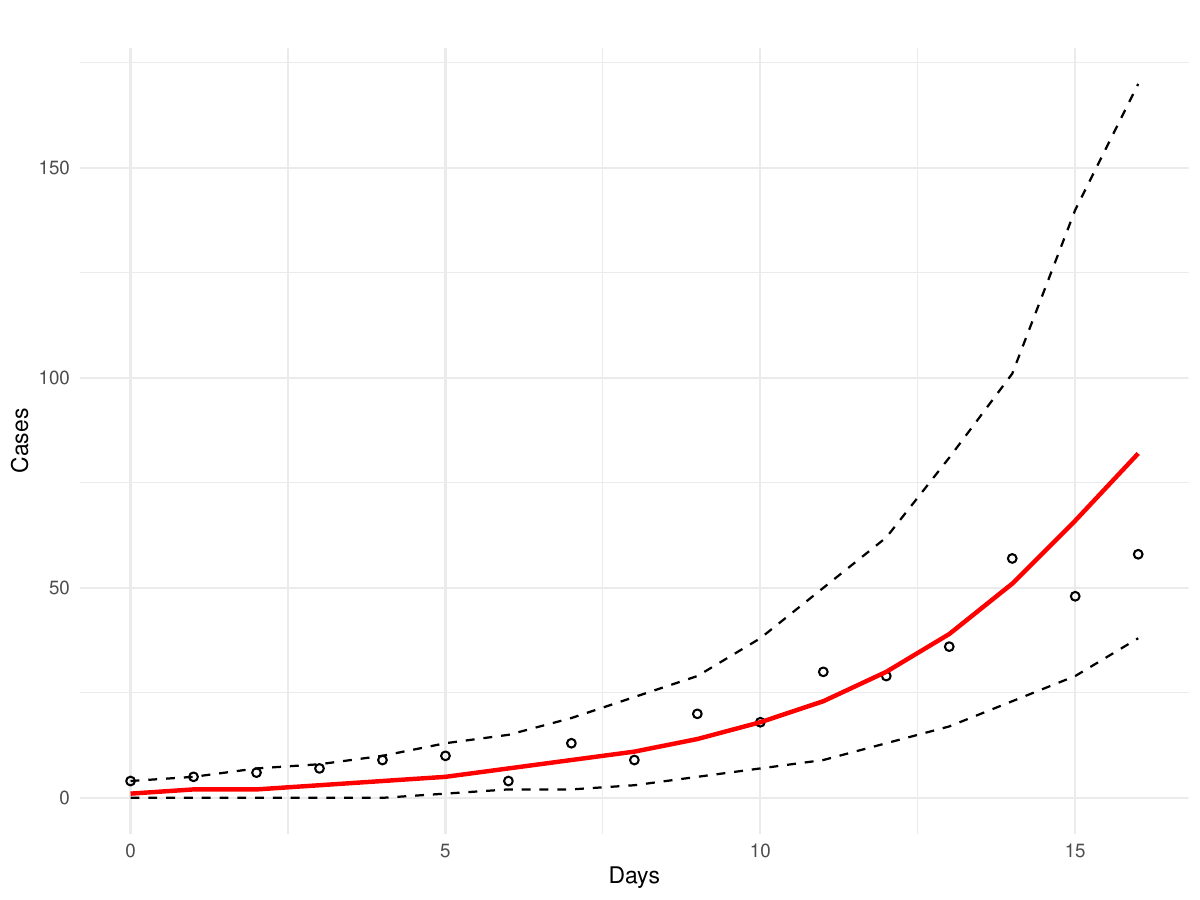}
    \end{minipage}
    
    % Second row: Histograms
    \vspace{0.5cm} % Add some space between rows
    \begin{minipage}[b]{0.3\textwidth}
        \centering
        \includegraphics[width=\linewidth]{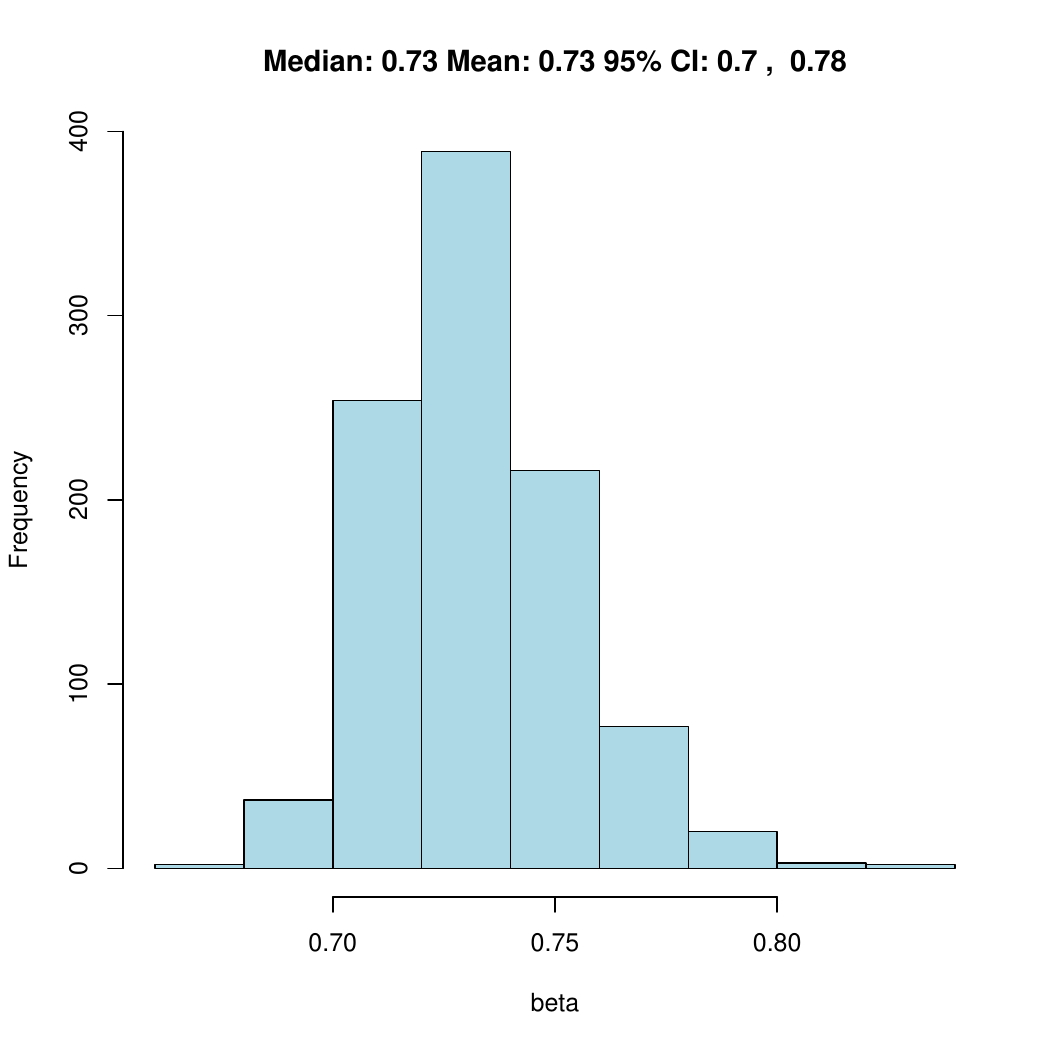}
    \end{minipage}
    \hfill
    \begin{minipage}[b]{0.3\textwidth}
        \centering
        \includegraphics[width=\linewidth]{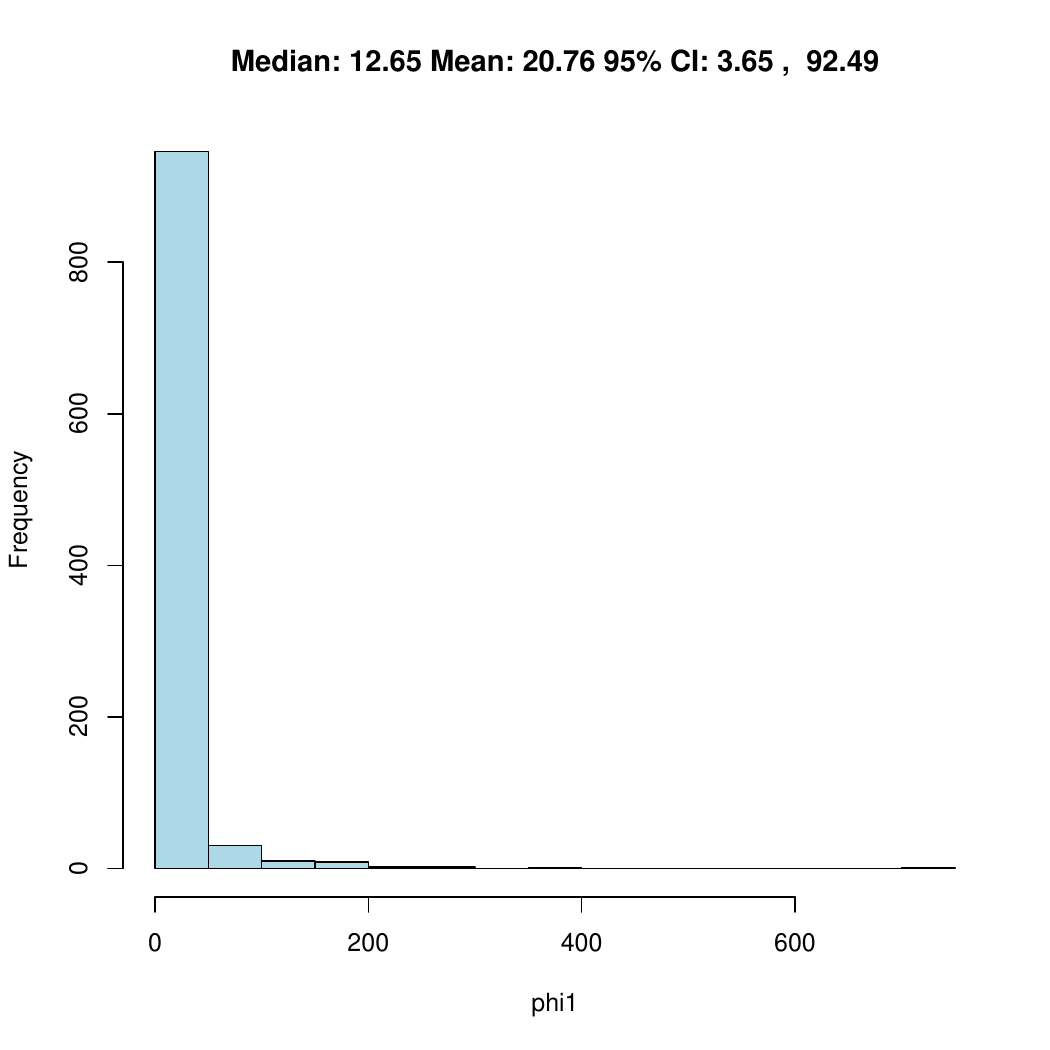}
    \end{minipage}
    \hfill
    \begin{minipage}[b]{0.3\textwidth}
        \centering
        \includegraphics[width=\linewidth]{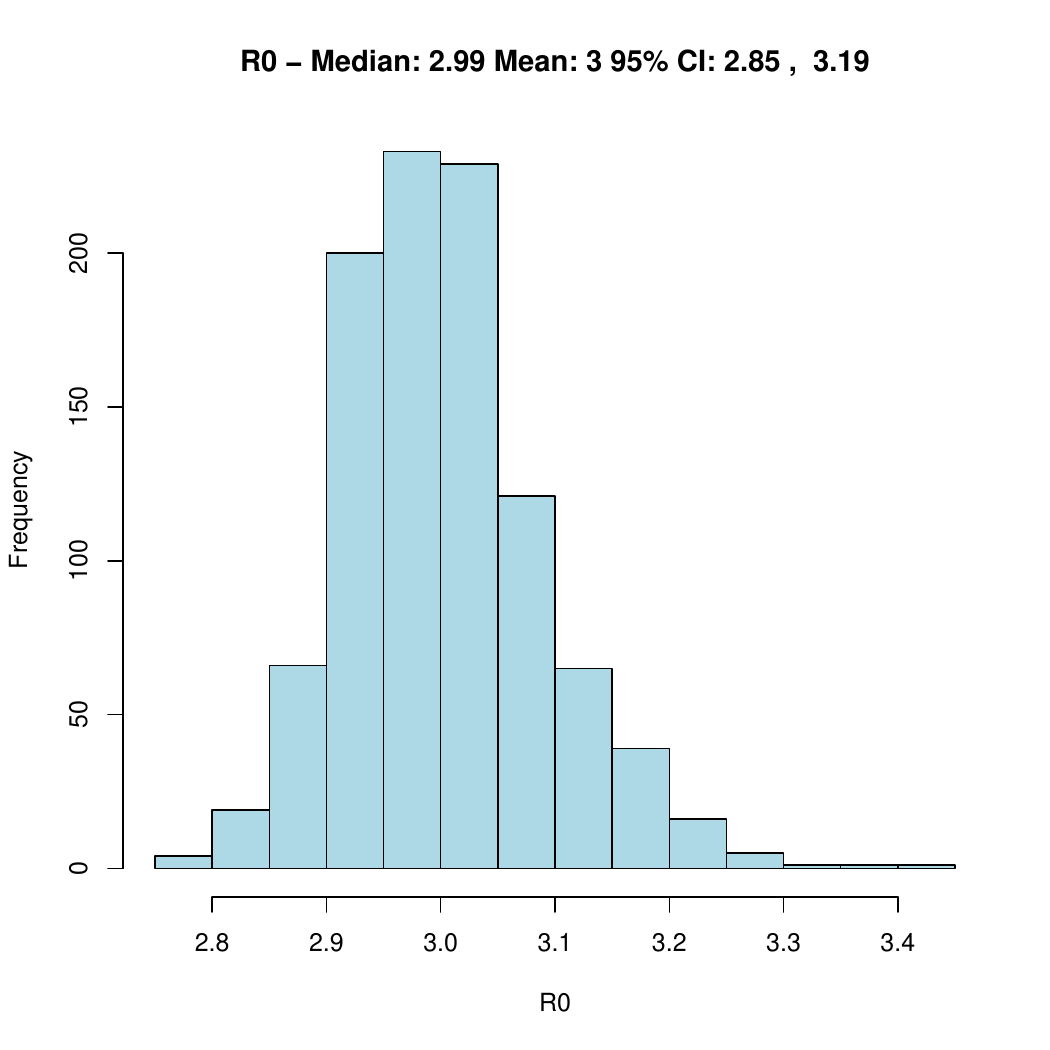}
    \end{minipage}
    
    \caption{\footnotesize The Bayesian fitting of the SEIR model to the first 17 days of the 1918 influenza pandemic in San Francisco was performed using a negative binomial error structure.  The model is fitted to the newly infected individuals, represented by $\frac{dC}{dt}$. The parameters $\kappa = \frac{1}{1.9}$, $\gamma = \frac{1}{4.1}$, and $\rho = 1$ are constants, while the prior distribution for $\beta$ is uniform(0,10) and for $\phi$ is exponential(5), with all parameters having a lower bound of zero. The population size is 550,000, with an initial condition of (549996, 0, 4, 0, 4) based on the first recorded data point. The MCMC algorithm was run with 1,000 iterations across two chains. The top panel shows the model fit to the data, with the red curve representing the median of fitted curves and black dashed lines indicating the 95\% PI, while the circles depict the actual data points. The bottom panels show the histograms of $\beta$ (left), $\phi$ (center), and $R_0$ (right).}
    \label{fig:Combined-Bayesian-uniform-sanfrancisco-negativebinomial-cal-17}
\end{figure}

In this example, we compare the performance metrics of the SEIR model using the SF 1918 flu dataset with two error structures: negative binomial and Poisson. Specifically, we replicate the results of TABLE 2 in \cite{chowell2024parameter} using a Bayesian approach. In Figure~\ref{fig:Combined-Bayesian-uniform-sanfrancisco-negativebinomial-cal-17}, the top panel illustrates the model fit, while the bottom panels display histograms of the estimated parameters: $\beta$, $\phi$, and $R_0$. Moreover, Table~\ref{tab:performance metrics-Bayesian-uniform-sanfrancisco-negativebinomial-cal-17-fcst-0} represents the performance metrics, and Table~\ref{tab:convergence-Bayesian-uniform-sanfrancisco-negativebinomial-cal-17-fcst-0} provides the convergence analysis. The comparison of this practice example using the negative binomial and Poisson error structure can be found in Table~\ref{tab:seir_model_comparison_Poisson_negativebinomial} based on their performance metrics. You can find the related option file in the folder under the name \texttt{"options\_SEIR\_sanfrancisco\_Ex2"}.

\begin{table}[t]
\caption{Performance metrics for the SEIR model with a negative binomial error structure, based on the San Francisco 1918 influenza dataset, calibrated over 17 days.}
\centering
\begin{tabular}{|c|c|c|}
\hline
\textbf{} & \textbf{Calibration} & \textbf{Forecasting} \\ \hline
\textbf{MAE} & 5.76 & NA \\ 
\textbf{MSE} & 68.24 & NA \\ 
\textbf{WIS} & 3.57 & NA \\ 
\textbf{Coverage of 95\% PI} & 100.00 & NA \\ \hline
\end{tabular}
\footnotetext{The model is fitted over 17-days to newly infected individuals, represented by $\frac{dC}{dt}$. The parameters $\kappa = \frac{1}{1.9}$, $\gamma = \frac{1}{4.1}$, and $\rho = 1$ are constants, while the prior distribution for $\beta$ is uniform(0,10) and for $\phi$ is exponential(5). All parameters have a lower bound of zero. The population size is 550,000, with an initial condition of (549996, 0, 4, 0, 4) based on the first recorded data point. The MCMC algorithm was run with 1,000 iterations across two chains.}
\label{tab:performance metrics-Bayesian-uniform-sanfrancisco-negativebinomial-cal-17-fcst-0}
\end{table}

\begin{table}[t]
\caption{Convergence and other statistics for the SEIR model applied to the San Francisco 1918 influenza dataset with a negative binomial error structure.}
\centering
\begin{tabular}{|l|l|l|l|l|l|}
\hline
\textbf{Parameter} & \textbf{Mean} & \textbf{Median} & \textbf{CI\_95} & \textbf{N\_eff} & \textbf{Rhat} \\ \hline
$\beta$            & 0.73          & 0.73            & ( 0.7 , 0.78 )  & 356.83          & 1          \\ \hline
$\phi$            & 20.76          & 12.65            & ( 3.65 , 92.49 )  & 395.63          & 1          \\ \hline
\end{tabular}
\footnotetext{The model is fitted over 17-days to newly infected individuals, represented by $\frac{dC}{dt}$. The parameters $\kappa = \frac{1}{1.9}$, $\gamma = \frac{1}{4.1}$, and $\rho = 1$ are constants, while the prior distribution for $\beta$ is uniform(0,10) and for $\phi$ is exponential(5). All parameters have a lower bound of zero. The population size is 550,000, with an initial condition of (549996, 0, 4, 0, 4) based on the first recorded data point. The MCMC algorithm was run with 1,000 iterations across two chains.}
\label{tab:convergence-Bayesian-uniform-sanfrancisco-negativebinomial-cal-17-fcst-0}
\end{table}

\begin{table}[t]
\caption{Comparison of performance metrics between the Poisson and negative binomial error structures for the SEIR model, based on the San Francisco 1918 influenza dataset.}
\centering
\begin{tabular}{|>{\centering\arraybackslash}p{4cm}|c|c|c|c|}
\hline
\textbf{Model} & \textbf{MAE} & \textbf{MSE} & \textbf{Coverage 95\% PI} & \textbf{WIS} \\ \hline
\centering Bayesian, SF 1918 flu data set, Poisson error structure & 15.24 & 45.35 & 88.24 & 3.33 \\ \hline
\centering Bayesian, SF 1918 flu data set, negative binomial error structure & 5.76 & 68.24 & 100 & 3.57 \\ \hline
\end{tabular}
\footnotetext{The model is fitted over 17-days to newly infected individuals, represented by $\frac{dC}{dt}$. The parameters $\kappa = \frac{1}{1.9}$, $\gamma = \frac{1}{4.1}$, and $\rho = 1$ are constants, while the prior distribution for $\beta$ is uniform(0,10) and for $\phi$ is exponential(5). All parameters have a lower bound of zero. The population size is 550,000, with an initial condition of (549996, 0, 4, 0, 4) based on the first recorded data point. The MCMC algorithm was run with 1,000 iterations across two chains. The red curve represents the median of a series of fitted curves, while the black dashed lines indicate the 95\% confidence interval. The circles depict the actual data points.}
\label{tab:seir_model_comparison_Poisson_negativebinomial}
\end{table}

\subsection{Generating Forecasts with the Best Model}
\begin{figure}[t]
    \centering
    \begin{minipage}{0.48\linewidth}
        \centering
        \includegraphics[width=\linewidth]{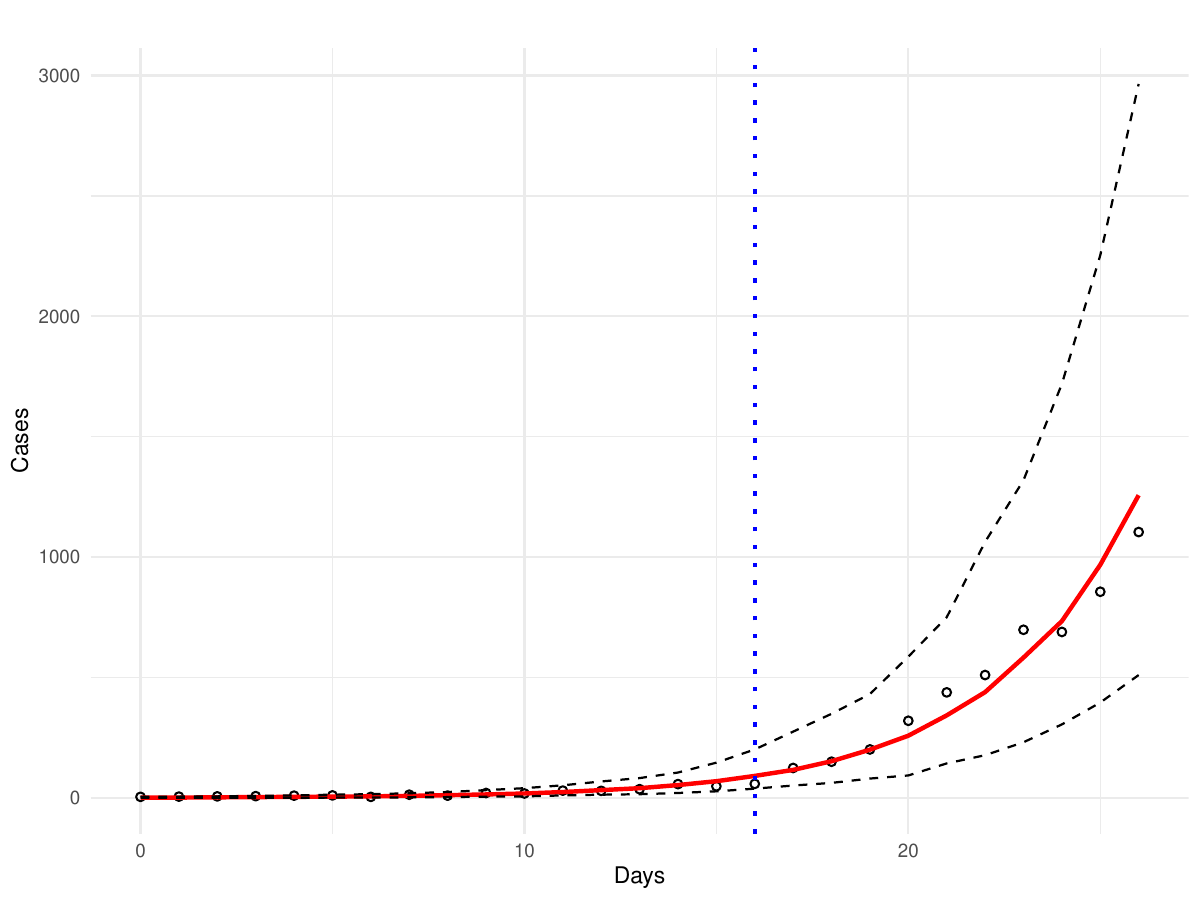}
        
    \end{minipage}
    \hfill
    \begin{minipage}{0.48\linewidth}
        \centering
        \includegraphics[width=\linewidth]{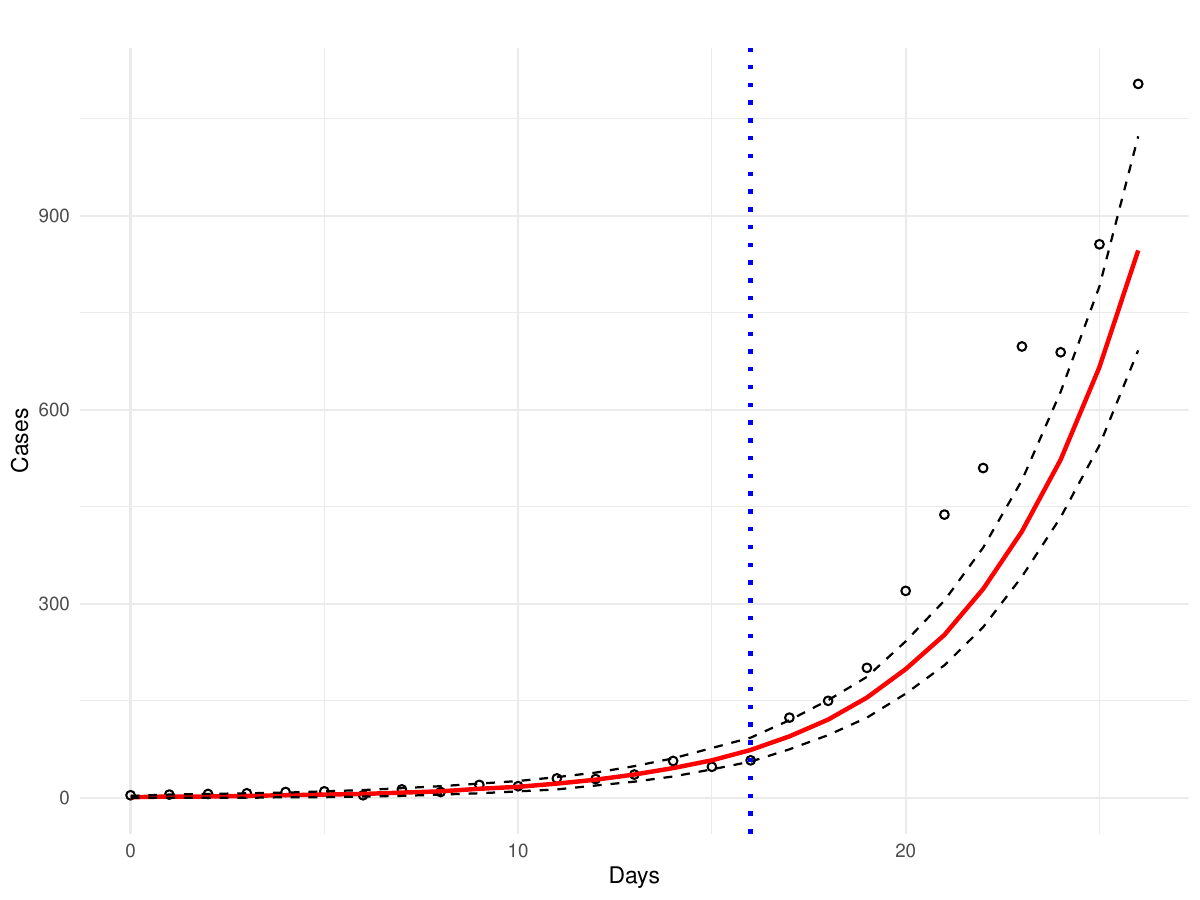}
    \end{minipage}
    \caption{\footnotesize Forecasting results for the SEIR model, calibrated over 17 days with a 10-day forecast horizon. This figure allows for a direct comparison, highlighting differences in forecast accuracy and uncertainty between the two error structures. The model is fitted to the newly infected individuals, represented by $\frac{dC}{dt}$. The parameters $\kappa = \frac{1}{1.9}$, $\gamma = \frac{1}{4.1}$, and $\rho = 1$ are constants, while the prior distribution for $\beta$ is uniform(0,10) and for $\phi$ is exponential(5). All parameters have a lower bound of zero. The population size is 550,000, with an initial condition of (549996, 0, 4, 0, 4) based on the first recorded data point. The MCMC algorithm was run with 1,000 iterations across two chains. The red curve represents the median of a series of fitted curves, while the black dashed lines indicate the 95\% PI. The circles depict the actual data points. The left panel uses a negative binomial error structure while the right one has a Poisson error structure.}
    \label{fig:Forecast-Bayesian-uniform-sanfrancisco-cal-17-fcst-10}
\end{figure}

In this example, we present the best model for forecasting 10 days after calibrating 17 days of the SF 1918 flu dataset. During the calibration period, the Poisson error structure performs better in terms of MAE, MSE, and WIS metrics, while the negative binomial model excels in $95\%$ PI coverage. However, the negative binomial model demonstrates superior performance across all metrics in the forecasting period. Figure \ref{fig:Forecast-Bayesian-uniform-sanfrancisco-cal-17-fcst-10} provide a forecasting for both Poisson and negative binomial error structure. Also, a comparison of the performance metrics is given in Table~\ref{tab:seir_model_comparison_Poisson_negativebinomial_forecast}. The related option files for this practice example are "\texttt{options\_SEIR\_sanfrancisco\_Ex3\_Negbin}" and "\texttt{options\_SEIR\_sanfrancisco\_Ex3\_Poisson}". 

\begin{table}[t]
\centering
\caption{A comparison of forecasting and model fit performance metrics between the Poisson and negative binomial error structures.}
\begin{tabular}{llcccc}
\rowcolor[HTML]{C0C0C0} 
\textbf{Error structure} & \textbf{MAE}  & \textbf{MSE} &\textbf{Coverage 95\% PI}  &\textbf{WIS}  \\
\midrule
\multicolumn{5}{c}{\textbf{Calibration Performance}} \\
\midrule
\textbf{Negative Binomial} & 5.85 & 68.25 & 100.0 & 3.62 \\
\textbf{Poisson}& 5.00 & 41.71 & 88.24 & 3.28 \\
\midrule
\multicolumn{5}{c}{\textbf{Forecasting Performance}} \\
\midrule
\textbf{Negative Binomial} & 92.10 & 12325.90 & 100.0 & 56.84 \\
 \textbf{Poisson} & 149.75 & 29976.22 & 10.0 & 122.28 \\
\bottomrule
\end{tabular}
\footnotetext{The table highlights differences in MAE, MSE, WIS, and $95\%$ PI coverage, emphasizing the superiority of the negative binomial model in forecasting performance and uncertainty quantification over a 10-day horizon, using a 17-day calibration period. The models are fitted to the newly infected individuals, represented by $\frac{dC}{dt}$. The parameters $\kappa = \frac{1}{1.9}$, $\gamma = \frac{1}{4.1}$, and $\rho = 1$ are constants, while the prior distribution for $\beta$ is uniform(0,10) and for $\phi$ is exponential(5). All parameters have a lower bound of zero. The population size is 550,000, with an initial condition of (549996, 0, 4, 0, 4) based on the first recorded data point. The MCMC algorithm was run with 1,000 iterations across two chains.}
\label{tab:seir_model_comparison_Poisson_negativebinomial_forecast}
\end{table}

\subsection{Comparison of the SEIR Model with the Exponential Growth Model}
\begin{figure}[t]
    \centering
    \begin{minipage}[b]{0.55\textwidth}
        \centering
        \includegraphics[width=\linewidth]{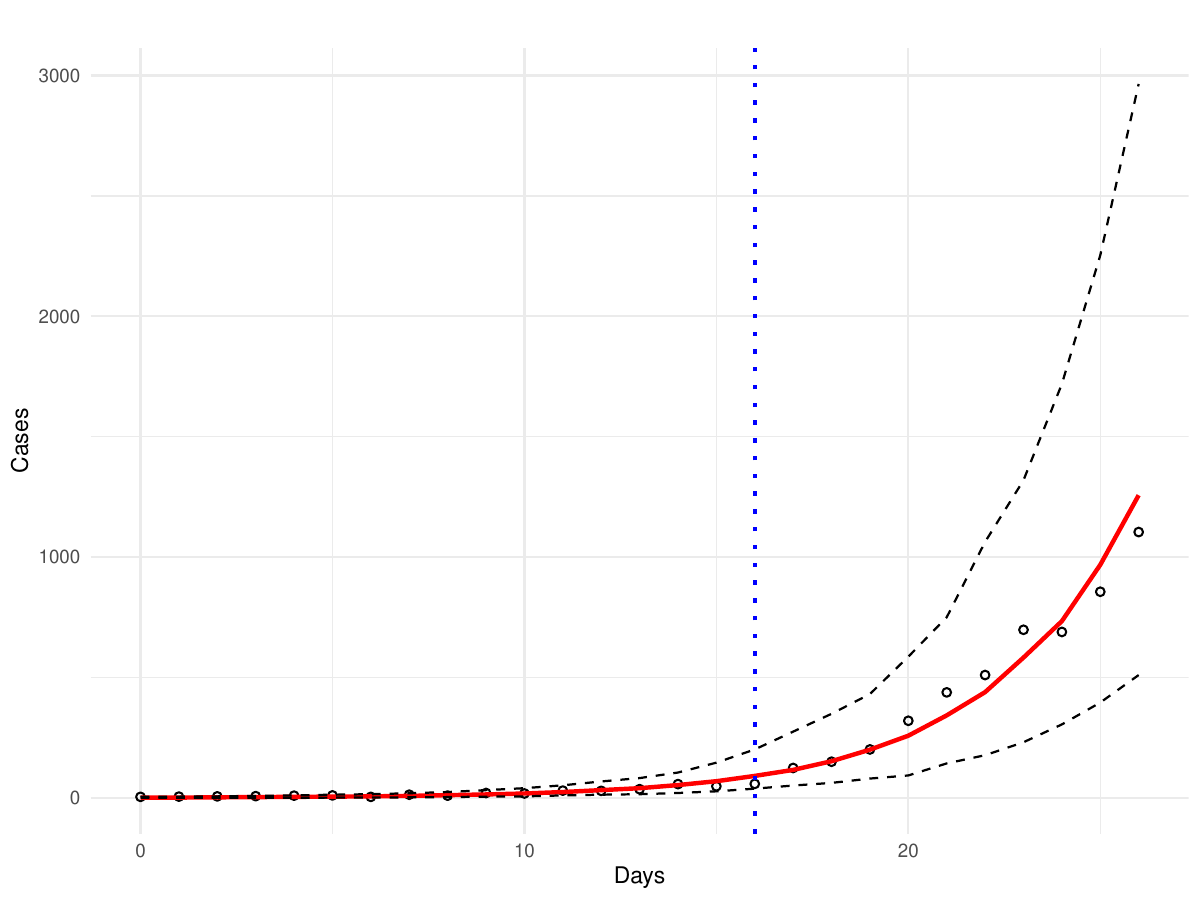}
    \end{minipage}
    \hfill
    \begin{minipage}[b]{0.4\textwidth}
        \centering
        \includegraphics[width=\linewidth]{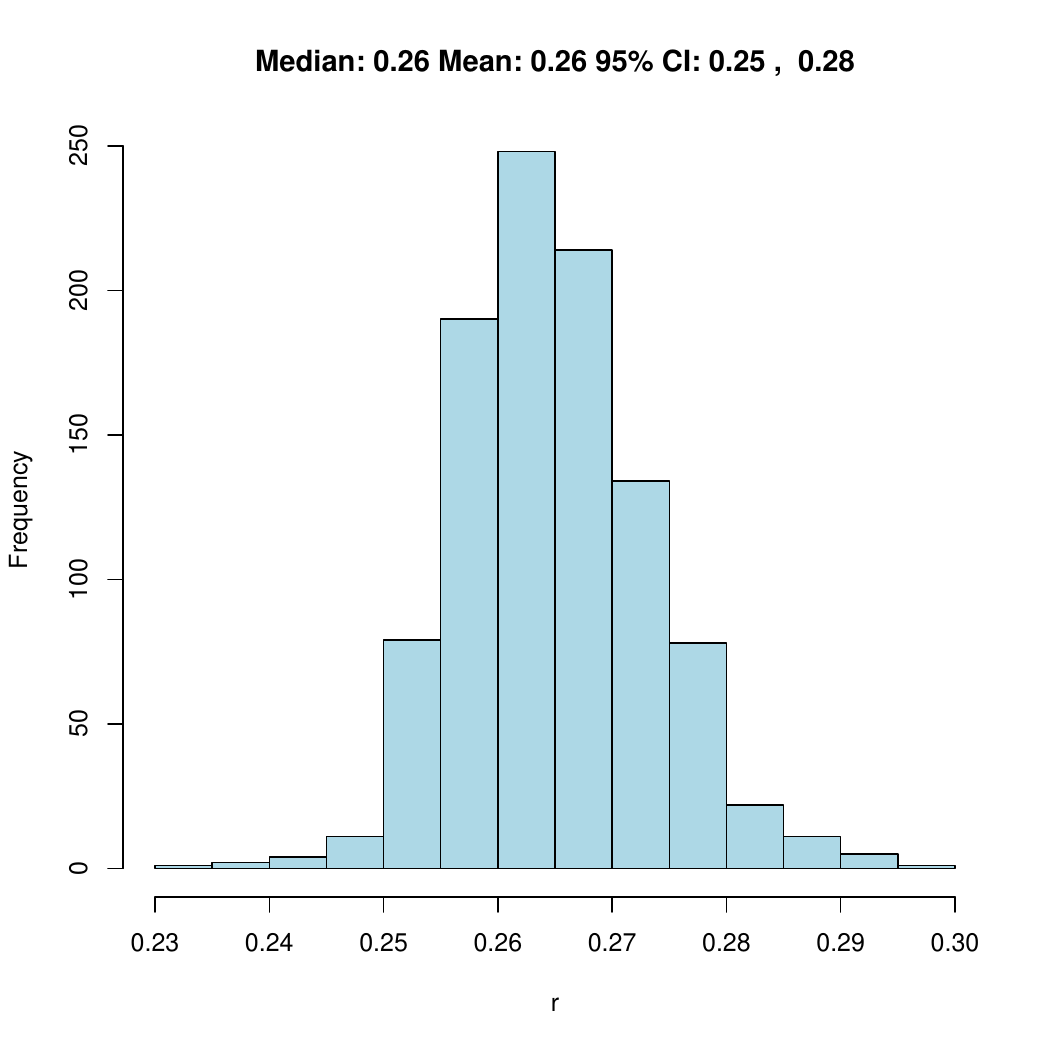}
    \end{minipage}
    \caption{\footnotesize Bayesian fitting of the exponential growth model to the San Francisco 1918 flu dataset with negative binomial error structure, using 17 days of calibration data with a forecasting horizon 10. The model is fitted to the newly infected individuals, represented by $\frac{dC}{dt}$. The prior distribution for $r$ is uniform(0,10), and for $\phi$ is exponential(5). All parameters have a lower bound of zero. The population size is 550,000, with an initial condition of (4) based on the first recorded data point. The MCMC algorithm was run with 1,000 iterations across two chains. The red curve represents the median of a series of fitted curves, while the black dashed lines indicate the 95\% PI. The circles depict the actual data points.}
    \label{fig:Bayesian-exp-sanfrancisco-negativebinomial-cal-17-fcst-10}
\end{figure}

In this practice example, we compare the SEIR model to the exponential growth model using SF 1918 flu data, negative binomial error structure, calibrated for 17 days with a forecast of 10 days. Check the option file \texttt{"options\_EXP\_sanfrancisco\_Ex4.R"} for more details about this model. Figure~\ref{fig:Bayesian-exp-sanfrancisco-negativebinomial-cal-17-fcst-10} provides forecasting as well as the histogram of the parameter $r$ using the exponential growth model. Table~\ref{tab:comparison_exp_SEIR} indicates that the SEIR model outperforms in calibration performance across all metrics, while the exponential growth model outcompetes in forecasting for all metrics.

\begin{table}[t]
\centering
\caption{A comparison of performance metrics between the SEIR model and the exponential growth model using Bayesian fitting to the San Francisco 1918 flu dataset with negative binomial error structure, using 17 days of calibration data with a forecasting horizon 10.}
\begin{tabular}{llcccc}
\rowcolor[HTML]{C0C0C0} 
\textbf{Model} & \textbf{MAE}  & \textbf{MSE} &\textbf{Coverage 95\% PI}  &\textbf{WIS}  \\
\midrule
\multicolumn{5}{c}{\textbf{Calibration Performance}} \\
\midrule
\textbf{SEIR model} & 5.85 & 68.25 & 100.0 & 3.62 \\
 \textbf{Exponential growth} & 6.53 & 104.88 & 100.0 & 4.11 \\
\midrule
\multicolumn{5}{c}{\textbf{Forecasting Performance}} \\
\midrule
\textbf{SEIR model} & 92.10 & 12325.90 & 100.0 & 56.84 \\
 \textbf{Exponential growth} & 66.50 & 6939.35 & 100.0 & 53.98 \\
\bottomrule
\end{tabular}
\footnotetext{The model is fitted to the newly infected individuals, represented by $\frac{dC}{dt}$. In the SEIR model, the parameters $\kappa = \frac{1}{1.9}$, $\gamma = \frac{1}{4.1}$, and $\rho = 1$ are constants. The prior distribution for $r$ and $\beta$ is uniform(0,10), and for $\phi$ is exponential(5). All parameters have a lower bound of zero. The population size is 550,000, with an initial condition of (549996, 0, 4, 0, 4) for the SEIR model, and (4) for the exponential growth model based on the first recorded data point. The MCMC algorithm was run with 1,000 iterations across two chains.}
\label{tab:comparison_exp_SEIR}
\end{table}

\subsection{The SEIR Model with Estimating the Initial number of infected people}

\begin{figure}[t]
    \centering
    % First row: Time series plot
    \begin{minipage}[b]{0.5\textwidth}
        \centering
        \includegraphics[width=\linewidth]{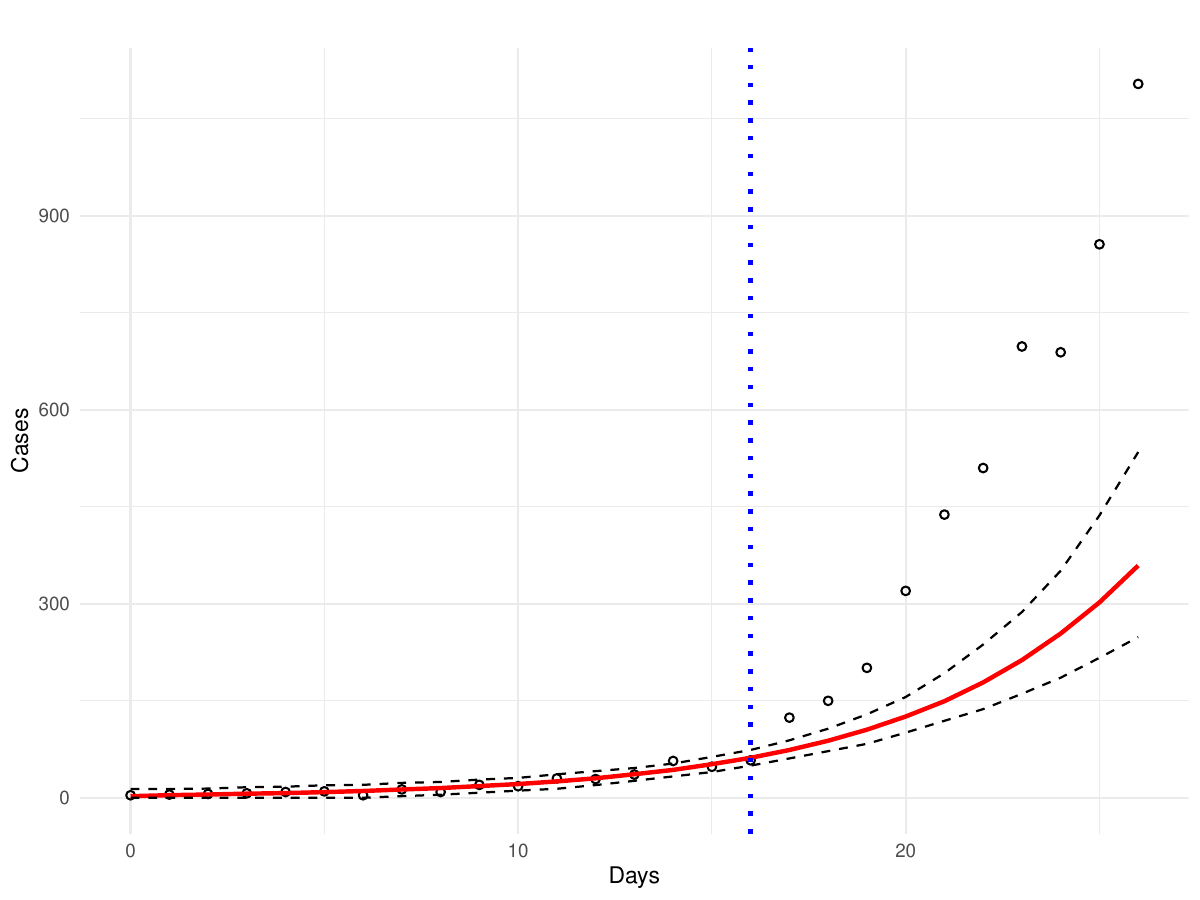}
    \end{minipage}
    
    % Second row: Histograms
    \vspace{0.5cm} % Add some space between rows
    \begin{minipage}[b]{0.3\textwidth}
        \centering
        \includegraphics[width=\linewidth]{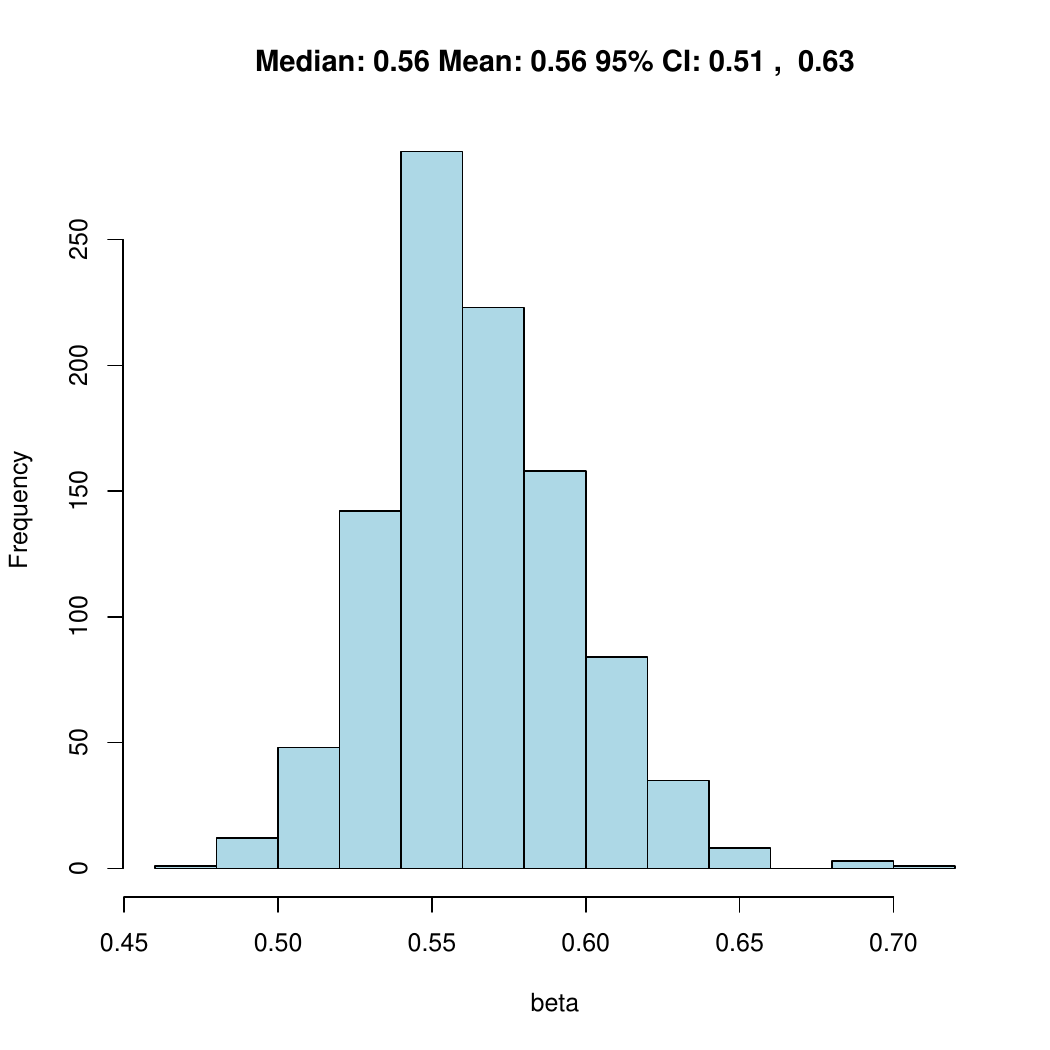}
    \end{minipage}
    \hfill
    \begin{minipage}[b]{0.3\textwidth}
        \centering
        \includegraphics[width=\linewidth]{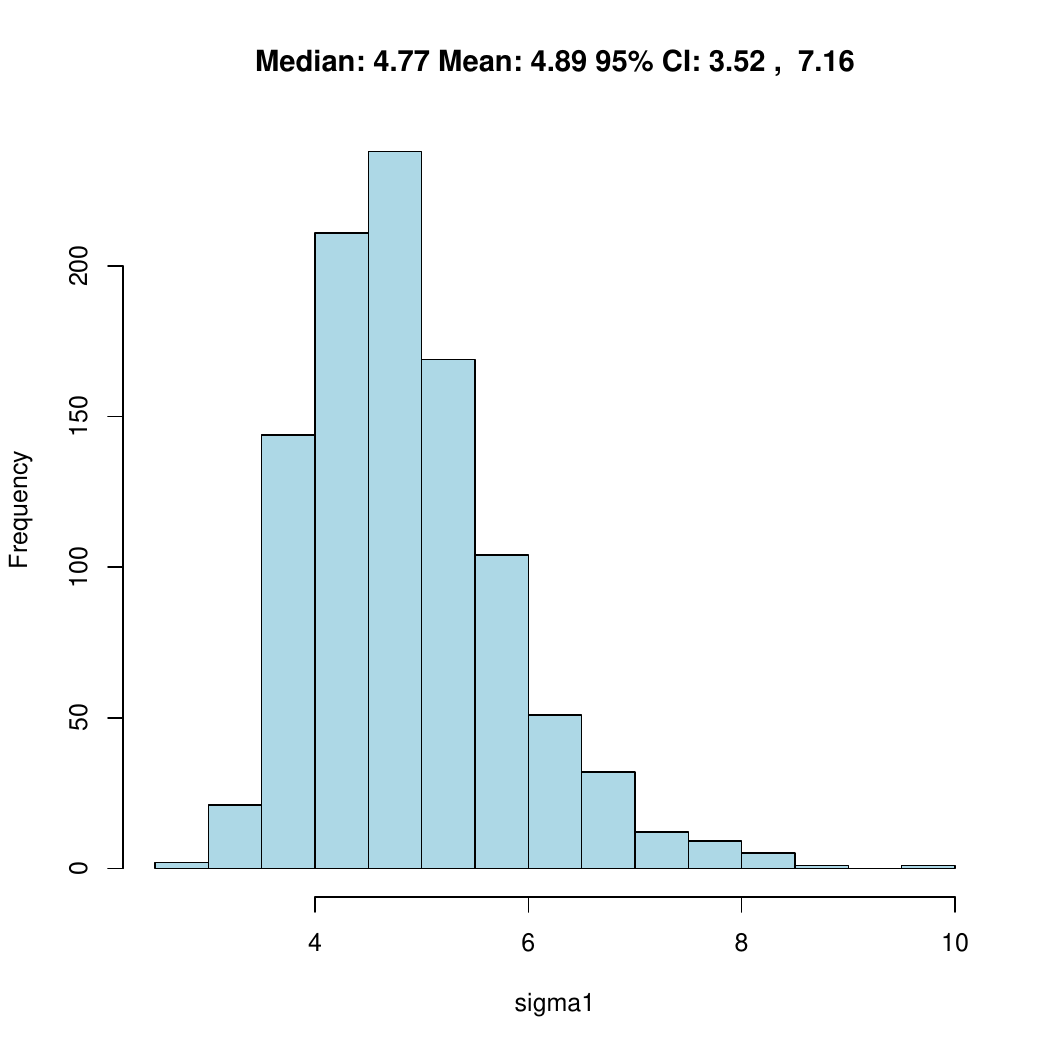}
    \end{minipage}
    \hfill
    \begin{minipage}[b]{0.3\textwidth}
        \centering
        \includegraphics[width=\linewidth]{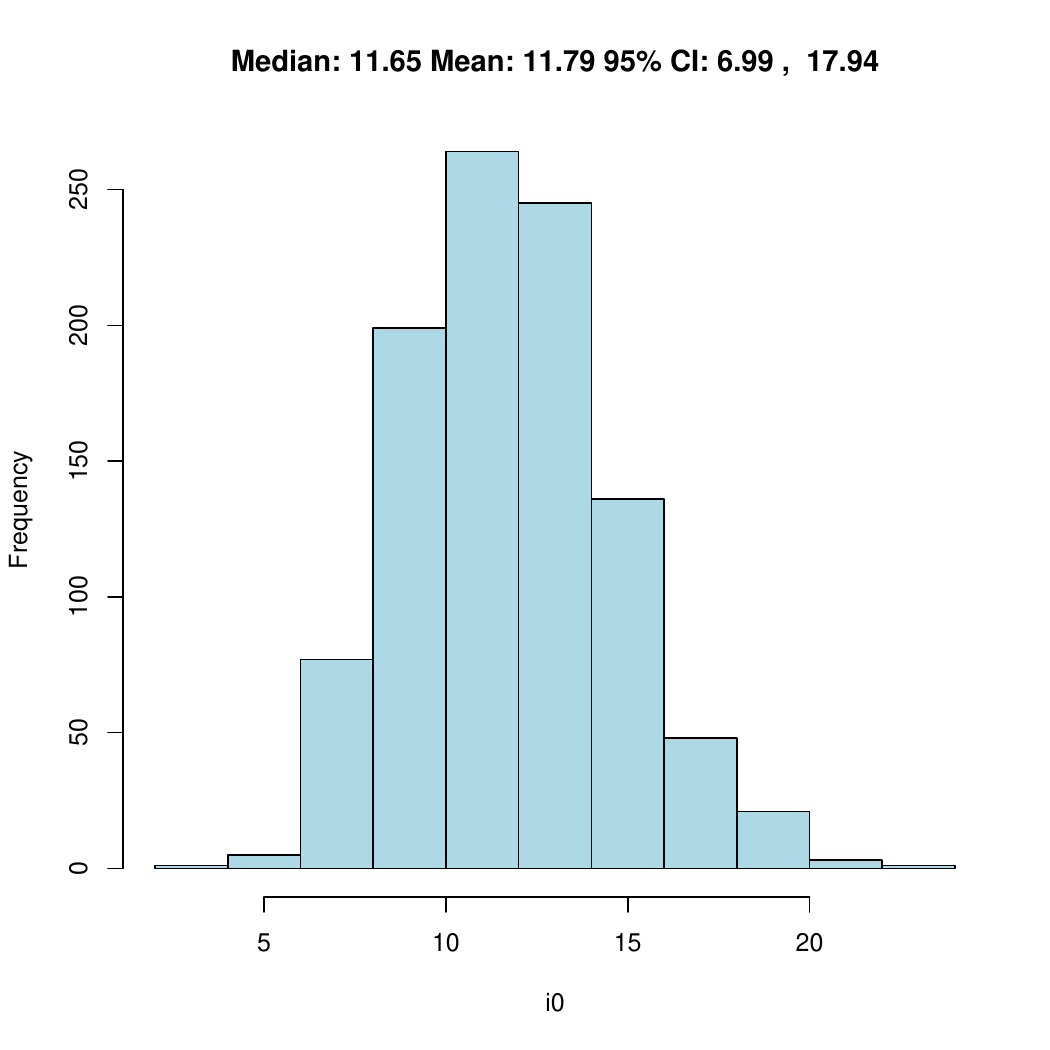}
    \end{minipage}
    
    \caption{\footnotesize Bayesian fitting of the SEIR model to the initial 17 days of the 1918 influenza pandemic in San Francisco using a normal error structure. The model is fitted to the newly infected individuals, represented by $\frac{dC}{dt}$. The parameters $\kappa = \frac{1}{1.9}$, $\gamma = \frac{1}{4.1}$, and $\rho = 1$ are constants, while the prior distribution for $\beta$ is uniform(0,10) and for $\sigma$ is cauchy(0, 2.5), with all parameters having a lower bound of zero. The population size is 550,000, with an initial condition of (N-$i_0$, 0, $i_0$, 0, $i_0$) with the purpose of estimating the first recorded data point $i_0$. The prior distribution for $i_0$ is normal(0,10) by assuming an upper bound of 100. The MCMC algorithm was run with 1,000 iterations across two chains. The top panel shows the model fit to the data, with the red curve representing the median of fitted curves and black dashed lines indicating the 95\% PI, while the circles depict the actual data points. The bottom panels show the histograms of $\beta$ (left), $\sigma$ (center), and $i_0$ (right).}
    \label{fig:Combined-Forecast-Bayesian-uniform-sanfrancisco-normal-cal-17-fcst-10}
\end{figure}

In this example, we use data from the 1918 influenza pandemic in San Francisco to forecast the number of infected individuals, employing a normal error structure. The model is calibrated over a 17-day period, followed by a 10-day forecast.  In this scenario, we also estimate the parameter $i_0$, which denotes the initial number of infected people. For more detailed instructions on how to configure this example, please refer to the option file \texttt{options\_SEIR\_sanfrancisco\_Ex5.R}. Figure \ref{fig:Combined-Forecast-Bayesian-uniform-sanfrancisco-normal-cal-17-fcst-10} illustrates both the model fitting and the estimation of $i_0$, which is approximately $11$.

\subsection{Fitting the SEIR model to multiple data sets using the simulated data}

\begin{figure}[t]
    \centering
    \begin{minipage}{0.32\textwidth}
        \centering
        \includegraphics[width=\linewidth]{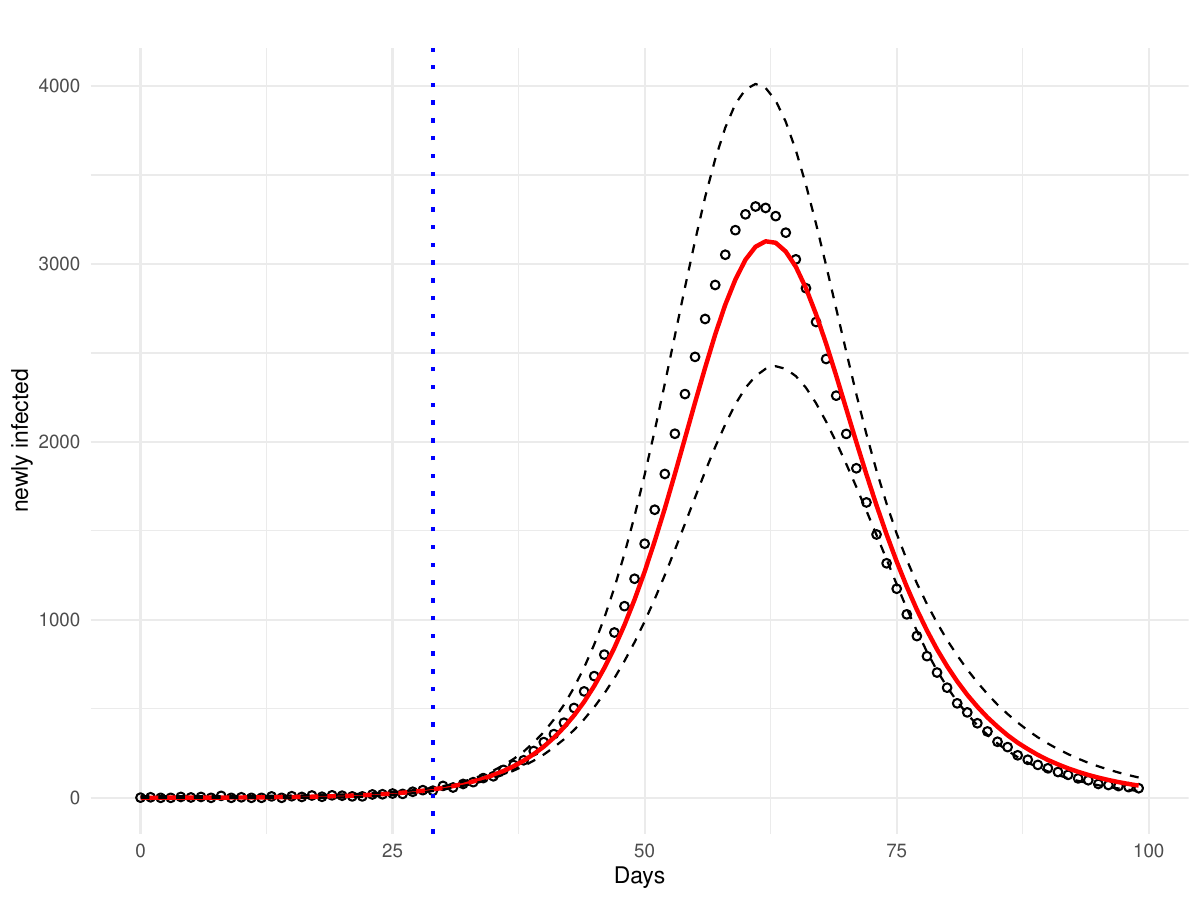}
    \end{minipage}%
    \hfill
    \begin{minipage}{0.32\textwidth}
        \centering
        \includegraphics[width=\linewidth]{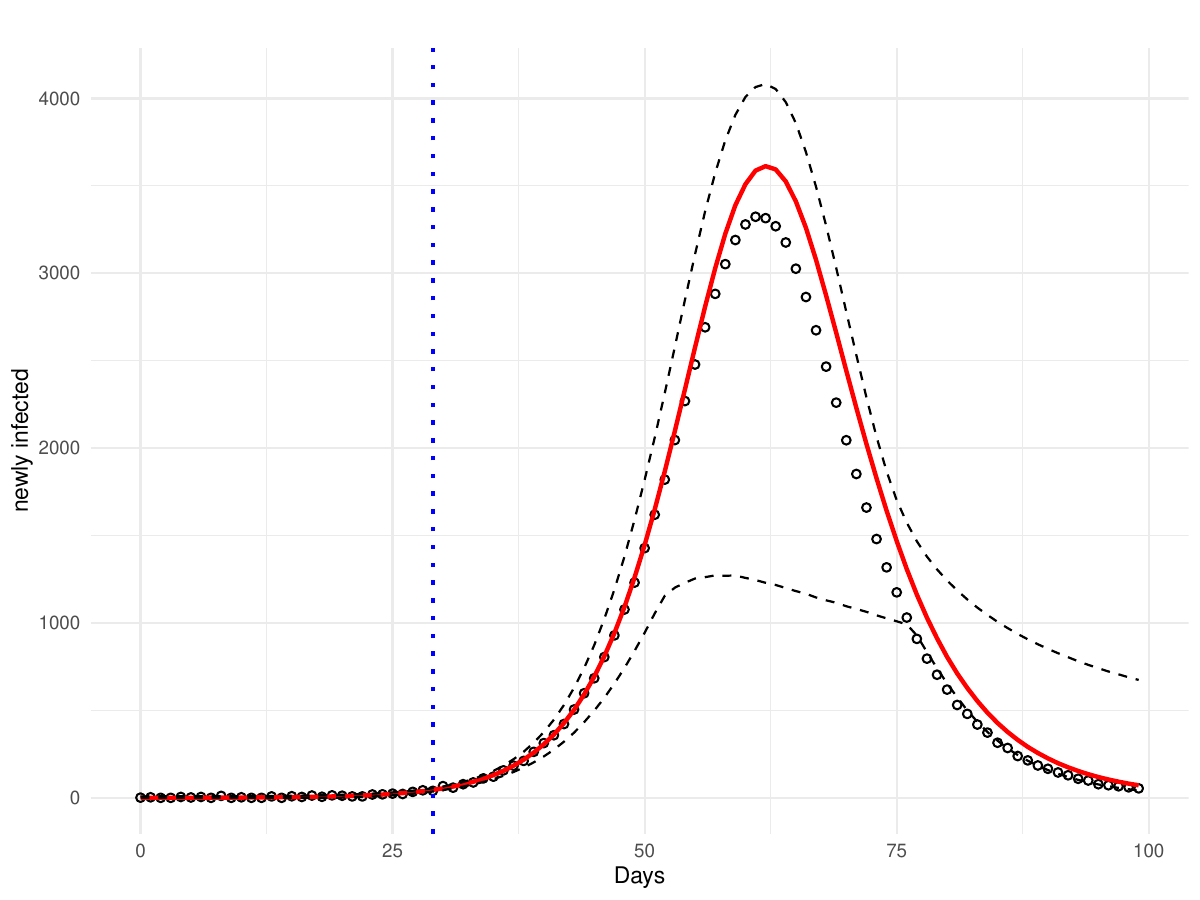}
    \end{minipage}%
    \hfill
    \begin{minipage}{0.32\textwidth}
        \centering
        \includegraphics[width=\linewidth]{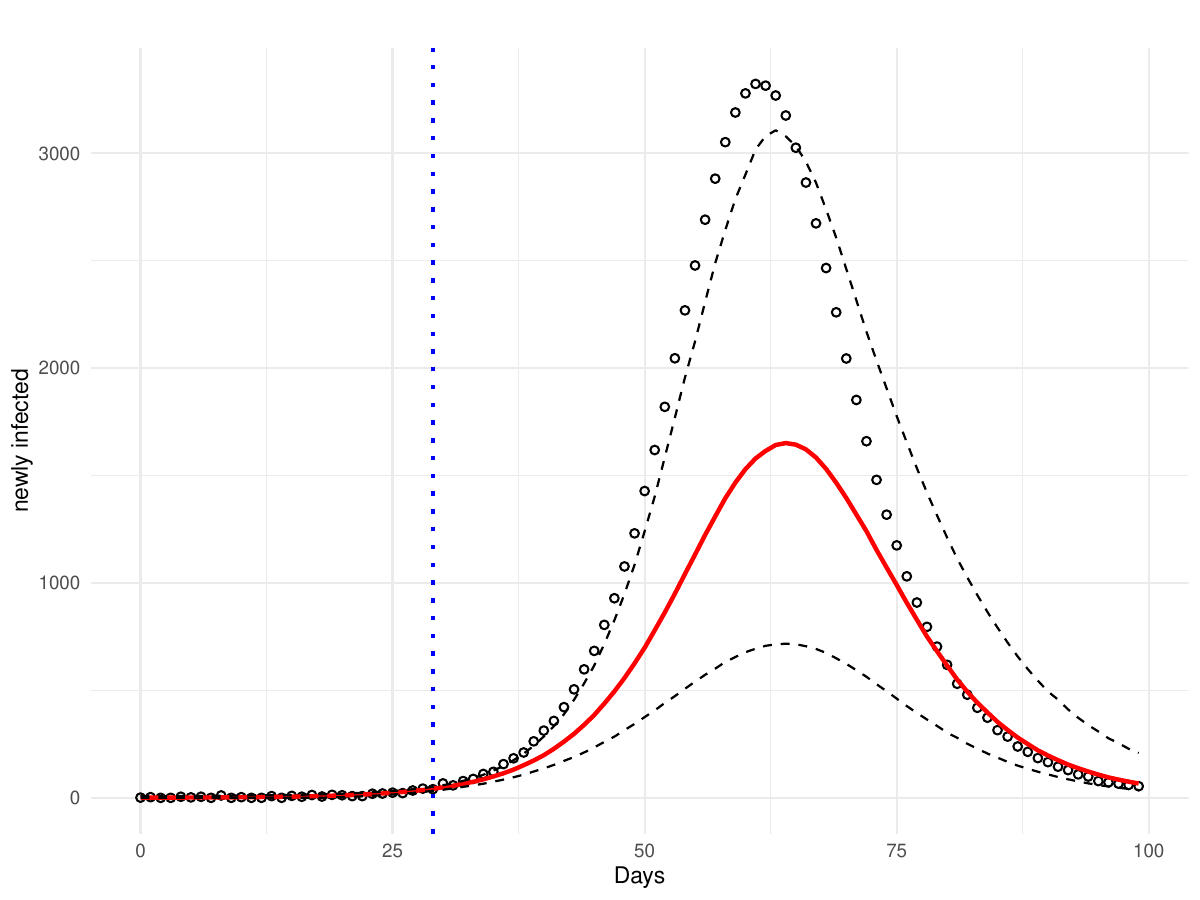}
    \end{minipage}
    
    \caption{\footnotesize We fit the SEIR model to the simulated data generated by the forward solution of the SEIR model ~(\ref{eq:SEIR}) using the parameters $\beta=0.5$, $\gamma = 0.25$, $\kappa = 1$, $\rho = 1$, and $N = 100{,}000$, with added noise sampled from $N(0, 25)$, using three different scenarios: (i) left panel---fitting the SEIR model to all the data, (ii) middle panel---fitting the SEIR model to only the first $I$ and the third column $\frac{dC}{dt}$, and (iii) right panel---fitting the SEIR model to only the third column $\frac{dC}{dt}$. In all scenarios, the figures display the time series of newly infected individuals, represented by $\frac{dC}{dt}$. }%The left panel corresponds to the first scenario, which utilizes all the available data. The middle panel shows the second scenario, incorporating both $I$ and $\frac{dC}{dt}$, and the right panel represents the third scenario, where only $\frac{dC}{dt}$ was used for fitting.}
    \label{fig:Forecast-SEIR-all-simulated-normal-cal-newly_infected}
\end{figure}

In this final practice example, we simulate data using the forward solution of the SEIR model~(\ref{eq:SEIR}) with the parameters $\beta=0.5$, $\gamma = 0.25$, $\kappa = 1$, $\rho = 1$, and $N = 100{,}000$ and adding normally distributed noise with $SD=5$. The Excel file containing the simulated data can be found in the toolbox with the name \texttt{"curve-SEIR\_plain\_I-dRdt-dCdt-dist1-0-factor1-5"}. In this file, the first column represents the days, the second column represents the number of infectious individuals ($I$), the third column represents the derivative of recovered individuals ($\frac{dR}{dt}$), and the fourth column represents the derivative of cumulative cases ($\frac{dC}{dt}$). We consider three different modeling scenarios: (i) fitting the SEIR model to all the data, (ii) fitting the model to only the infectious individuals ($I$) and the derivative of cumulative cases ($\frac{dC}{dt}$), and (iii) fitting the model to only the derivative of cumulative cases ($\frac{dC}{dt}$). A calibration period of 70 days and a forecasting horizon of 30 days were used.  Figure~\ref{fig:Forecast-SEIR-all-simulated-normal-cal-newly_infected} shows the fitted curves of the newly incidence together with the prediction bands from the three scenarios.  As expected, the results from the first scenario, where all available data is used, outperform those from the second scenario, which only uses the $I$ and $\frac{dC}{dt}$ columns. The second scenario, in turn, performs better than the third scenario, which relies solely on $\frac{dC}{dt}$. This is because including more data provides more information for model fitting. % and improves model accuracy. %, as demonstrated in the first scenario where all data points are included (see Figure~\ref{fig:Forecast-SEIR-all-simulated-normal-cal-newly_infected}). 
Table~\ref{tab:Performance-SEIR-all-simulated-normal-cal-newly_infected} highlights an interesting observation: even though the third scenario performs better in the calibration phase, it underperforms in the forecasting phase. The parameter estimates from the third scenario are greatly different from the true values, and the first scenario provides more accurate parameter estimates (Table~\ref{tab:Parameter-SEIR-all-simulated-normal-cal-newly_infected}). This implies that there might be an overfitting to the calibration period in the third scenario which has less data, and this leads to poor forecasting. Using multiple states provides more information and better avoids the overfitting. % suggests that fitting the model to more datasets may result in a less accurate calibration, but it allows for more precise parameter estimation, leading to improved forecasting performance (see Table~\ref{tab:Parameter-SEIR-all-simulated-normal-cal-newly_infected}). 
The option files used for each of these scenarios are:\texttt{"options\_SEIR\_simulated\_Ex6.R"}, \texttt{"options\_SEIR\_simulated\_Ex6\_I\_dcdt.R"}, and \texttt{"options\_SEIR\_simulated\_Ex6\_\\dcdt.R"}, respectively.

\begin{table}[t]
\caption{Performance metrics for the rate of new cases, $\frac{dC}{dt}$, obtained by fitting the SEIR model to simulated data generated using the forward solution of the SEIR model~(\ref{eq:SEIR}).}
\centering
\begin{tabular}{llcccc}
\rowcolor[HTML]{C0C0C0} 
\textbf{Scenario} & \textbf{Period} & \textbf{MAE} & \textbf{MSE} & \textbf{Coverage 95\% PI} & \textbf{WIS} \\
\midrule
\multicolumn{6}{c}{\textbf{Calibration Performance}} \\
\midrule
(i)\textbf{$I$, $\frac{dR}{dt}$, $\frac{dC}{dt}$} & 30 & 3.08 & 15.65 & 96.67 & 1.99 \\
(ii)\textbf{$I$, $\frac{dC}{dt}$} & 30 & 3.23 & 16.11 & 93.33 & 2.03 \\
(iii)\textbf{$\frac{dC}{dt}$} & 30 & 3.01 & 13.98 & 96.67 & 1.89 \\
\midrule
\multicolumn{6}{c}{\textbf{Forecasting Performance}} \\
\midrule
(i)\textbf{$I$, $\frac{dR}{dt}$, $\frac{dC}{dt}$} & 70 & 98.86 & 16487.18 & 87.14 & 65.43 \\
(ii)\textbf{$I$, $\frac{dC}{dt}$} & 70 & 131.26 & 35876.48 & 84.29 & 92.26 \\
(iii)\textbf{$\frac{dC}{dt}$} & 70 & 471.14 & 564219.77 & 55.71 & 314.63 \\
\bottomrule
\end{tabular}
\footnotetext{The model utilized the following parameters: transmission rate $\beta=0.5$, recovery rate $\gamma = 0.25$, progression rate $\kappa = 1$, reporting rate $\rho = 1$, and a total population size of $N = 100{,}000$. Three distinct scenarios were analyzed: (i) fitting the SEIR model to the entire dataset, (ii) fitting the model to both the infected population $I$ and the third column of data, $\frac{dC}{dt}$, and (iii) fitting the SEIR model solely to the third column, $\frac{dC}{dt}$.}
\label{tab:Performance-SEIR-all-simulated-normal-cal-newly_infected}
\end{table}

\begin{table}[t]
\caption{Parameter estimates obtained by fitting the SEIR model to 30-days of simulated data.}
\centering
\begin{tabular}{lccccc}
\textbf{Scenario} & \textbf{Calibration} & $\beta$ & $\gamma$ & $\kappa$ \\
\midrule
(i) \textbf{$I$, $\frac{dR}{dt}$}, $\frac{dC}{dt}$  & 30 & 0.48 (0.4, 0.83) & 0.24 (0.21,0.27) & 1.08 (0.19,8.84) \\
(ii)\textbf{$I$, $\frac{dC}{dt}$} & 30 & 0.56 (0.38,4.88) & 0.21 (0.16,0.25) & 0.38 (0.02,7.82) \\
(iii)\textbf{$\frac{dC}{dt}$} & 30 & 2.05 (0.34,9.72) & 1.52 (0.16,7.85) & 0.8 (0.16,9.15) \\
\bottomrule
\end{tabular}
\footnotetext{Parameter estimation was performed by fitting the SEIR model to simulated data generated using the forward solution of the SEIR model ~(\ref{eq:SEIR}) with the following parameters: $\beta=0.5$, $\gamma = 0.25$, $\kappa = 1$, $\rho = 1$, and a population size of $N = 100{,}000$. Three different scenarios were considered: (i) fitting the SEIR model to the entire dataset, (ii) fitting the SEIR model to only the first column ($I$) and the third column $\frac{dC}{dt}$, and (iii) fitting the model solely to the third column $\frac{dC}{dt}$.}
\label{tab:Parameter-SEIR-all-simulated-normal-cal-newly_infected}
\end{table}

\subsection{Fitting the SEIR model with a time-dependent parameter to the simulated data\label{sec:time-dependent}}

In this practice example, we simulate data using the forward solution of the SEIR model~(\ref{eq:SEIR}) with the time-dependent parameter $\beta$, defined as below:

\begin{equation}
\label{eq:time_dependent}
\beta = \beta(t) =
\begin{cases}
\beta_0, & \text{if } t < t_{\text{int}} \\
\beta_1 + (\beta_0 - \beta_1) e^{-q (t - t_{\text{int}})}, & \text{if } t \geq t_{\text{int}}
\end{cases}
\end{equation}

This piecewise function describes a transition in $\beta(t)$ from an initial value $\beta_0$ to a new value $\beta_1$ after a specific intervention time $t_{\text{int}}$. For times $t$ earlier than $t_{\text{int}}$, the function remains constant at $\beta_0$. After $t_{\text{int}}$, the function decays exponentially towards $\beta_1$, with the rate of decay governed by the parameter $q$.

We generate the forward solution of this model over the time period $[0,100]$, using the following parameters: $\beta_0 = 0.5$, $\beta_1 = 0.1$, $q = 0.1$, $t_{\text{int}} = 10$, $\gamma = \frac{1}{7}$, $\kappa = \frac{1}{5}$, $\rho = 0.6$, and $N = 1{,}000{,}000$. The Excel file containing the simulated data is located in the toolbox under the name \texttt{"seir\_simulated"}. In this file, the first column represents the days, and the second column contains the derivative of the cumulative cases, denoted by $\frac{dC}{dt}$. The option file related to this practice example is under the name \texttt{"options\_seir\_timedep\_Ex7.R"} located in the toolbox directory.

\begin{figure}[H]
    \centering
    \includegraphics[width=0.5\linewidth]{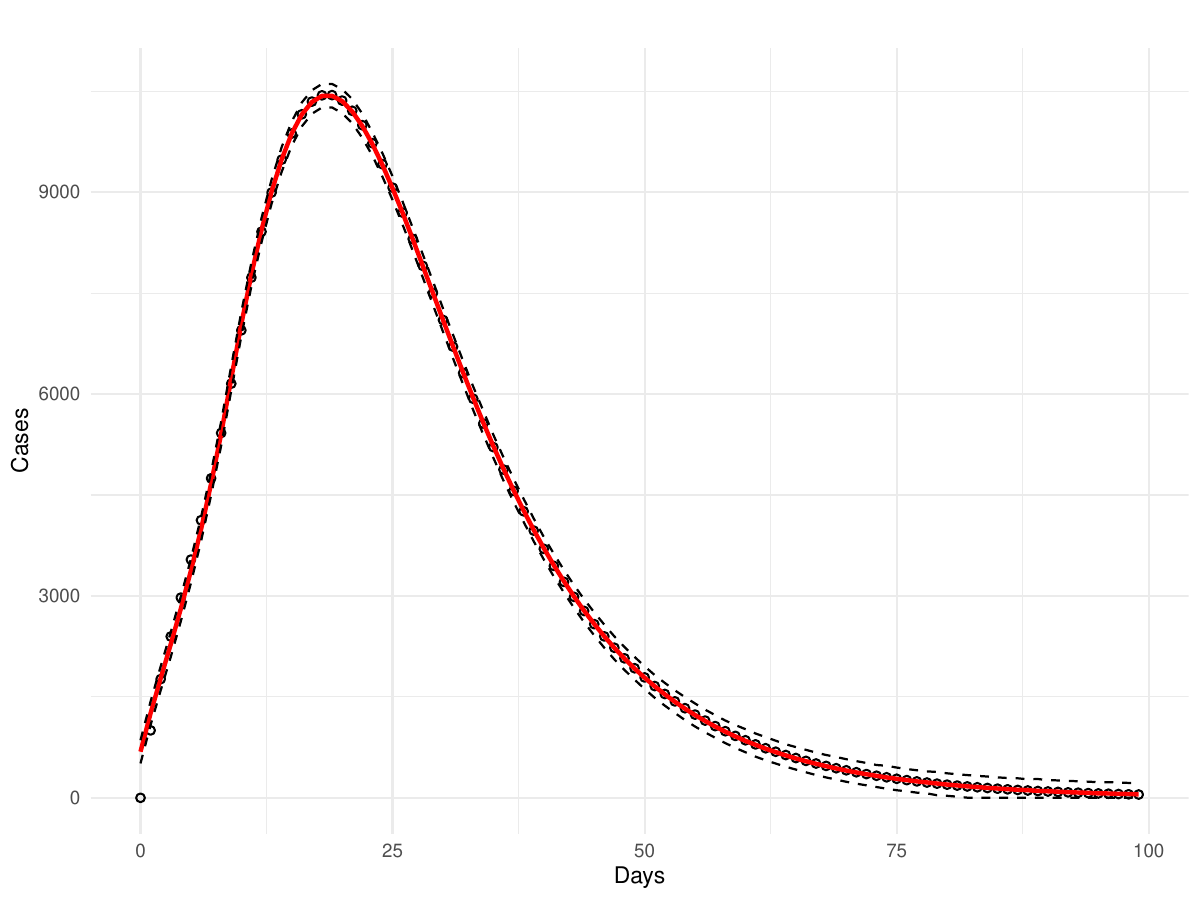}
    \caption{Fitting of the SEIR model~(\ref{eq:SEIR}) with a time-dependent transmission rate~(\ref{eq:time_dependent}) to synthetic data generated from the forward simulation of the same model, using parameters $\beta_0 = 0.5$, $\beta_1 = 0.1$, $q = 0.1$, $t_{\text{int}} = 10$, $\gamma = \frac{1}{7}$, $\kappa = \frac{1}{5}$, $\rho = 0.6$, and $N = 1{,}000{,}000$.}
    \label{fig:Forecast-SEIR-tdp-simulated-normal-cal-Cases-fcst-100}
\end{figure}

Figure~\ref{fig:Forecast-SEIR-tdp-simulated-normal-cal-Cases-fcst-100} illustrates an excellent fit of the SEIR model with a time-dependent transmission rate to the synthetic data. Notably, the impact of the time-dependent transmission rate is evident, as the epidemic curve is asymmetric with respect to the peak, reflecting the dynamic nature of the transmission over time.

Table~\ref{tab:convergence-Bayesian-SEIR-simulated} shows that the estimated parameters do not exactly match the true values, but they are reasonably close. This suggests that while parameter estimates may not be fully reliable when using a time-dependent parameter in the model, the model can still provide a good fit to the data. 

\begin{table}[t]
\caption{Convergence and posterior statistics for the SEIR model with a time-dependent transmission rate fitted to synthetic data.}
\centering
\begin{tabular}{lccccc}
\toprule
\textbf{Parameter} & \textbf{Mean} & \textbf{Median} & \textbf{CI\textsubscript{95}} & \textbf{N\textsubscript{eff}} & \textbf{\^R} \\
\midrule
$\beta_0$ & 0.91 & 0.67 & (0.55, 1.62) & 26.08 & 1.01 \\
$\beta_1$ & 0.34 & 0.06 & (0.02, 1.20) & 25.77 & 1.01 \\
$q$       & 0.16 & 0.10 & (0.10, 0.45) & 33.43 & 1.00 \\
$\kappa$  & 0.11 & 0.12 & (0.08, 0.14) & 30.95 & 1.01 \\
$\gamma$  & 0.14 & 0.11 & (0.07, 0.25) & 28.97 & 1.00 \\
$\rho$    & 0.45 & 0.48 & (0.33, 0.55) & 29.37 & 1.00 \\
\bottomrule
\end{tabular}
\footnotetext{The model was fitted to synthetic epidemic data generated from the SEIR model with a time-dependent transmission rate. Parameters $\beta_0$, $\beta_1$, $q$, $\kappa$, $\gamma$, and $\rho$ were estimated using MCMC, with 10{,}000 iterations across two chains. The effective sample size ($N_{\text{eff}}$) and potential scale reduction factor ($\hat{R}$) are reported to assess convergence.}
\label{tab:convergence-Bayesian-SEIR-simulated}
\end{table}

\subsection{Optimizing Computational Efficiency with SLURM and Parallelism in \texttt{rstan}}

Running multiple MCMC chains is standard practice in Bayesian modeling with Stan to ensure reliable and robust results. However, these chains can be computationally intensive, especially when dealing with complex models or large datasets. To mitigate this issue, parallel execution of MCMC chains can significantly reduce computation time, increasing the efficiency of the Bayesian estimation process.

Within the R environment, the \texttt{rstan} package utilizes R’s \texttt{parallel} package to achieve parallelism. By configuring the \texttt{mc.cores} option, users can specify the number of cores to be utilized for running the chains in parallel. For example, setting \texttt{options(mc.cores = parallel::detectCores())} will automatically detect and use the maximum number of cores available on the system, enabling each chain to run on a separate core. This setup accelerates computation and promotes faster convergence of the MCMC chains.

For users working on high-performance computing (HPC) clusters, integrating SLURM (Simple Linux Utility for Resource Management) can further optimize computational efficiency. SLURM is a robust workload manager that allows users to allocate resources effectively across multiple nodes in an HPC environment. By submitting jobs through SLURM, users can run MCMC chains across several nodes, each with multiple cores, thereby enabling large-scale parallelism. For example, if running four chains, you would request four CPU cores with the directive \texttt{\#SBATCH --cpus-per-task=4}. This allows \texttt{rstan} to distribute the chains across the allocated cores, reducing the total computation time compared to running chains sequentially.

Moreover, when conducting simulations for multiple calibration periods, SLURM's job array feature can be utilized to parallelize the process further. By using the \texttt{\#SBATCH --array=} directive, you can run separate jobs for each calibration period in parallel. For example, if the calibration periods are \texttt{c(10, 15, 20, 25, 30)}, setting \texttt{\#SBATCH --array=10,15,20,25,30} in the SLURM script will launch independent jobs for each period simultaneously.This approach greatly enhances simulation speed by fully utilizing the cluster's parallel processing capabilities.

To implement this approach, the \texttt{run\_MCMC.R} script should be modified to retrieve the current calibration period from the environment variable \texttt{SLURM\_ARRAY\_TASK\_ID}. This can be achieved with the following R code:

\begin{Verbatim}

            calibrationperiod = Sys.getenv('SLURM_ARRAY_TASK_ID')
            print("calibrationperiod")
            print(calibrationperiod)
    
\end{Verbatim}

By incorporating these modifications, both the MCMC chains and calibration periods can run in parallel, thus maximizing the use of HPC resources. However, as the number of chains and calibration periods increases, it is essential to ensure sufficient memory allocation using the \texttt{--mem=} directive, as each additional task will require more RAM.

In summary, leveraging parallelism through multiple CPU cores and SLURM job arrays can substantially reduce the overall runtime of Stan simulations, optimizing the cluster's computational resources and improving the efficiency of Bayesian analysis.

\subsection{Usage of the \texttt{BayesianFitForecast} Library}

The \texttt{BayesianFitForecast} library is available on the Comprehensive R Archive Network (CRAN)~\cite{BayesianFitForecast}. Users can install and load the library in R as shown below:

\begin{Verbatim}

                    # Install the library from CRAN
                    install.packages("BayesianFitForecast")

                    # Load the library
                    library(BayesianFitForecast)

\end{Verbatim}

Once loaded, the library provides access to two primary functions: \texttt{Run\_MCMC} and \texttt{Run\_analyzeResults}. These functions allow users to perform Bayesian fitting and forecasting of time series data. Below, we explain the usage of \texttt{Run\_MCMC}, the function responsible for generating MCMC data.

\subsubsection{\texttt{Run\_MCMC}}

The \texttt{Run\_MCMC} function generates data that can be analyzed further using \texttt{Run\_analyzeResults}. This function requires three inputs:

\begin{enumerate}
    \item \textbf{Option file path}: A string specifying the path to the option file, which must be an \texttt{.R} file. The format and settings of the option file are detailed in Section~\ref{sec:option}.
    \item \textbf{Excel file path}: A string specifying the path to the Excel file containing the time series data. The file should be in \texttt{.xlsx} format, with its structure described in Section~\ref{sec:Excel}.
    \item \textbf{Output folder path}: A string specifying the path to the folder where the output will be saved. The output is an \texttt{.Rdata} file containing the results of the MCMC run.
\end{enumerate}

The \texttt{Run\_MCMC} function can be executed as follows:

\begin{Verbatim}

                    # Example usage of Run_MCMC
                    Run_MCMC(
                        option_file_path = "path/to/option_file.R",
                        Excel_file_path = "path/to/data_file.xlsx",
                        output_folder_path = "path/to/output_folder"
                    )

\end{Verbatim}

The output \texttt{.Rdata} file serves as the input for the \texttt{Run\_analyzeResults} function, which facilitates further analysis and will be described in detail later.

\subsubsection{\texttt{Run\_analyzeResults}}

The \texttt{Run\_analyzeResults} function is used to process and analyze the results generated by the \texttt{Run\_MCMC} function, as described in Section~\ref{sec:analyzeResults}. This function requires the following inputs:

\begin{enumerate}
    \item \textbf{\texttt{data\_file\_location}}:  
    The location of the data file generated after running \texttt{Run\_MCMC}. This file is required for performing the analysis.

    \item \textbf{\texttt{option\_file}}:  
    The name of the option file (e.g., \texttt{"option1.R"}) located in the current working directory. This file contains the necessary settings for the analysis.

    \item \textbf{\texttt{Excel\_file}}:  
    The path to an Excel file containing the data. This parameter is required and must be a valid path to an existing file.

    \item \textbf{\texttt{output\_path}}:  
    The directory where the output will be saved. If set to \texttt{NULL}, a temporary directory will be created in the system's default temporary location (e.g., as determined by \texttt{tempdir()} in R).

\end{enumerate}

An example of how to use \texttt{Run\_analyzeResults} is provided below:

\begin{Verbatim}

            # Example usage of Run_analyzeResults
            Run_analyzeResults(
                data_file_location = "path/to/folder_with_data_file",
                option_file = "option1.R",
                Excel_file = "path/to/additional_data.xlsx",
                output_path = "path/to/output_directory"
            )

\end{Verbatim}

The output of this function includes detailed analysis and processed results, saved in the specified \texttt{output\_path} (or the temporary directory, if \texttt{output\_path = NULL}). These results provide valuable insights and can be used for further interpretation and reporting.

By leveraging the \texttt{BayesianFitForecast} library through CRAN, users can easily integrate Bayesian fitting and forecasting workflows into their R environment. Additionally, we provide an R Shiny App interface that offers a graphical dashboard for the BayesianFitForecast workflow. This application enables users to upload data, configure model parameters, specify priors and ODE structures, and view diagnostic and forecasting outputs without writing code. The Shiny app mirrors the core functionality of the R scripts, allowing consistent and reproducible results through a more accessible interface.

\subsection{The \texttt{BayesianFitForecast} Shiny App}

\begin{figure}[t]
    \centering
        \centering
        \includegraphics[width=\linewidth]{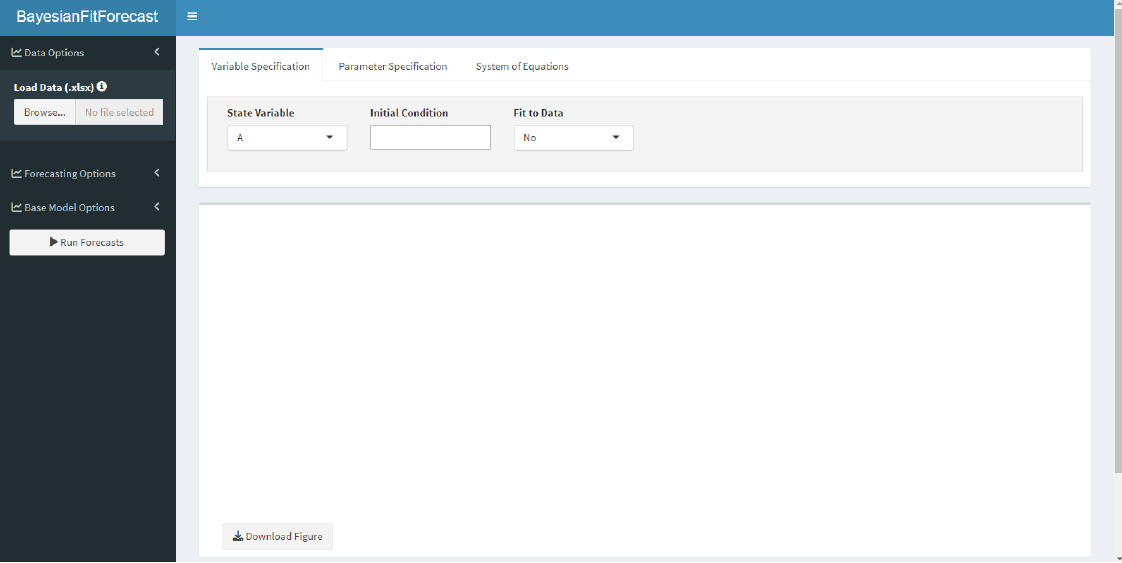}
    \caption{\footnotesize A screenshot of the \texttt{BayesianFitForecast} R-Shiny app user interface. This figure provides a snapshot of the user interface that first appears when the app is loaded. Users can enter data, forecast, model, and MCMC specifications on the sidebar of the page, and details regarding the state variables, parameter specifications, and system of ODEs in the top panel of the body of the dashboard. The remaining sections auto-populate with the associated data files and figures produced during the analysis process.}
    \label{fig:Figure-App-Structure}
\end{figure}

To further facilitate the use of \texttt{BayesianFitForecast}, we provide an associated R-Shiny app interface designed to simplify model calibration and forecasting workflows for users with limited coding experience. It provides interactive controls for uploading data, configuring model settings, and viewing outputs generated by the \texttt{BayesianFitForecast} library (Fig. \ref{fig:Figure-App-Structure}). The Shiny app interface utilizes the functionality discussed above and is built around the \texttt{BayesianFitForecast} library \cite{BayesianFitForecast}. However, it provides a graphical interface for users to specify the details previously included in the \texttt{options.R} file and obtain all of the output presented in the earlier examples. Below we supply a brief overview of the dashboard's structure, and illustrate its utility employing the example presented in Section \ref{sec:examplePoisson} in Appendix \ref{appendix:A} and a tutorial video (\url{https://www.youtube.com/watch?v=3ZiLxmMVfyc}).

% General App Structure 

\subsubsection{Installing and loading the \texttt{BayesianFitForecast} App}

Prior to running the \texttt{BayesianFitForecast} R-Shiny application, the most recent version of \texttt{R} and \texttt{RStudio} must be downloaded. A system-specific package compiler must also be loaded (i.e., \texttt{Rtools}, \texttt{XCode}). Additional details can be found in the "package\_builder\_instructions" document available within the Shiny App source folder on the GitHub page (\url{https://github.com/gchowell/BayesianFitForecast/tree/main/Shiny%20app%20(Beta%20version)}). Once the necessary software is installed, users can then:

\begin{itemize}
    \item Download the folder containing the app and required functions from \url{https://github.com/gchowell/BayesianFitForecast/tree/main/Shiny%20app%20(Beta%20version)}.
    \item Load the \texttt{BayesianFitForecast} project file into RStudio.
    \item Open the \texttt{app.R} file and click \textit{Run App}.
\end{itemize}

Table \ref{tab:app_meta} contains the R packages and compilation requirements required to load the Shiny application successfully, as well as the web address for the host repository. While the packages must be manually loaded when running the source code discussed above, they should be downloaded automatically upon the first rendering of the shiny application. 

\begin{table}[ht]
\caption{\texttt{BayesianFitForecast} Shiny app metadata. The table below includes details regarding the required software, versions, and packages needed to successfully launch the \texttt{BayesianFitForecast} R-Shiny application. Additionally, it provides the link to the permanent repository containing all needed functions, the user interface files, and tutorial materials.}
\centering
\begin{tabular}{|p{3.5cm}|p{8cm}|}
\hline
\textbf{Required Software} & R (>= 4.3), R-Studio (2024.09.0 Build 375), System-Specific Package Compiler\\
\hline
\textbf{Compilation Requirements} & \texttt{shiny, shinydashboard, shinyWidgets, ggplot2, plotly, DT, mathjaxr,
readxl, shinyjqui, devtools, bayesplot, xlsx, openxlsx, rstan,
shinyjs, BayesianFitForecast, stringr, shinybusy, eList,
shinyalert, glue, remotes} \\
\hline
\textbf{Permanent link to repository} & \url{https://github.com/gchowell/BayesianFitForecast/tree/main/Shiny%20app%20(Beta%20version)} \\
\hline
\end{tabular}
\footnotetext{Upon initial compilation of the dashboard, it checks if the required packages are downloaded. If they are not, R will proceed to install the packages for the given session. During this process, pop-up messages may appear asking if the user would like to compile the package. To successfully utilize all features of the dashboard, the user must select “yes”. }
\label{tab:app_meta}
\end{table}

\subsubsection{Dashboard Structure}\label{sec:loading-data-app}

The dashboard utilizes the same input data structure and naming scheme described in Section \ref{sec:Excel}. To load the data into the dashboard, users first must navigate to the "Data Options" menu on the sidebar, select the \textit{Browse} button, and identify the file location within their personal computer. Once loaded, additional data specification parameters related to the process of interest included in the data (ex., "Influenza"), its temporal resolution (ex., "Days"), and the type of data included (ex., "Cases"), will become available. 

Once all parameters related to the input data have been specified, users can then proceed with providing the desired calibration periods, forecasting horizon, the system of ODEs, and its related state variables and parameters. While the dashboard utilizes most of the parameters specified within the \texttt{options.R} file's structure described above, the parameters specification process has been streamlined to remove any programming requirements. Appendix \ref{appendix:A} contains a side-by-side comparison of the user parameters included within the \texttt{options.R} file and their corresponding label within the dashboard, as well as a description of their specification within the app. Additionally, it presents a step-by-step walk-through of all of the dashboard's features, including the full workflow to obtain the desired output. 

\begin{figure}[t]
    \centering
    \includegraphics[width=\linewidth]{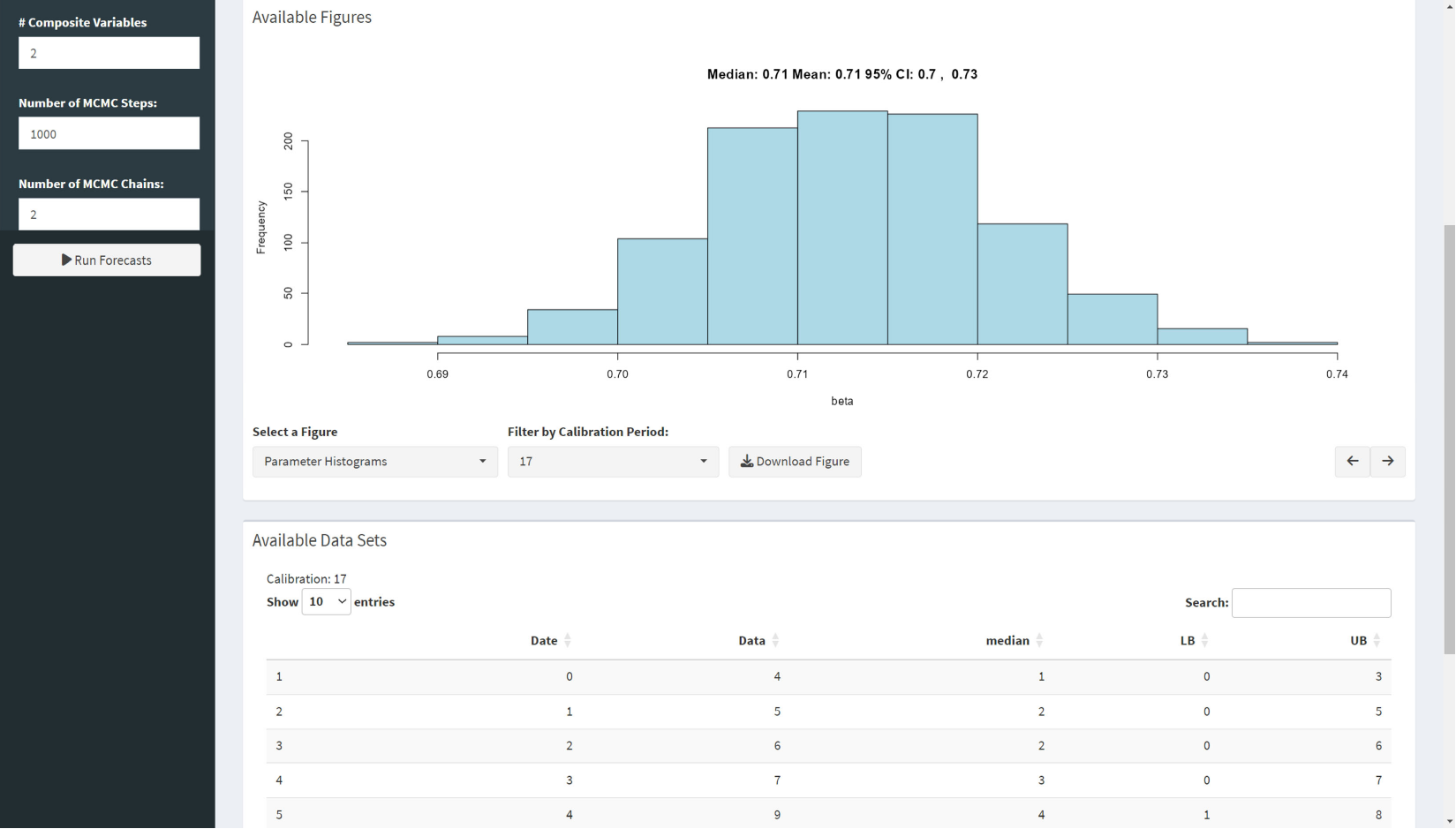}
    \caption{\footnotesize A screenshot of the dashboard with output loaded. This figure shows a screenshot of the dashboard after all fitting and forecasting processes have been completed. The “Available Figures” box contains all available figures, and the “Available Data Sets” box contains available data sets. Users can filter the output by type and calibration period, and download all of the output to the folder of their choosing.}
    \label{fig:Fig-All-Output-Shiny}
\end{figure}

The \textit{BayesianFitForecast} R-Shiny interface produces the same output as presented in the above sections, which the users will see appear in the body of the dashboard (Fig. \ref{fig:Fig-All-Output-Shiny}). Using the provided drop-down menus, users can filter the viewable figures and data sets, as well as navigate through all output produced using the arrows located in the bottom-right corner of the figure and data sections. Additionally, the entire set of provided figures, data, and ODE R-Stan file are available to download via their respective \textit{Download} buttons. While the figure and data download buttons are available in the main body of the dashboard, the user will presented with an opportunity to download the R-STAN file immediately after its creation and prior to the fitting of the model, and production of subsequent forecasts. Appendix \ref{appendix:A} describes the process of obtaining all available output from the dashboard, and a full tutorial illustrating the application of the dashboard to forecasting the 1918 influenza pandemic in San Francisco can be found at: \url{https://www.youtube.com/watch?v=3ZiLxmMVfyc}. 

The Shiny app fully replicates the functionality of the core scripts \texttt{Run\_MCMC} and \texttt{run\_analyzeResults.R}, ensuring identical results and outputs regardless of whether the toolbox is used through the GUI or script interface.

\section{Conclusions}

In this tutorial paper, we have introduced \texttt{BayesianFitForecast}, an advanced R toolbox designed to facilitate Bayesian parameter estimation and forecasting with quantified uncertainty for ODE models, such as those frequently used for infectious disease modeling and forecasting. The strength of \texttt{BayesianFitForecast} lies in its ability to seamlessly integrate prior knowledge with observed data, producing posterior distributions that reflect both empirical evidence and expert judgment. This is particularly advantageous in infectious disease modeling, where data may be scarce or noisy, and the ability to rigorously quantify uncertainty is critical for making informed decisions. 

One of the significant strengths of \texttt{BayesianFitForecast} is its flexibility. Users can define a wide array of ODE models and apply various error structures, such as Poisson or negative binomial, depending on the nature of the data and the specific requirements of the study. This adaptability ensures that the toolbox can be employed across diverse scenarios, enhancing its utility for researchers working in various fields.

However, it is important to acknowledge some limitations of \texttt{BayesianFitForecast}. First, the toolbox is primarily tailored for deterministic ODE models. Although these models are powerful tools for understanding disease dynamics, they do not account for the random fluctuations inherent in real-world dynamical processes. As a result, stochastic ODE models, which incorporate these random variations, are becoming increasingly crucial in epidemic modeling and other fields. 

Another limitation is the computational cost associated with running complex models, especially when using large datasets or performing extensive sensitivity analyses.  Although \texttt{BayesianFitForecast} leverages advanced sampling algorithms such as Hamiltonian Monte Carlo (HMC) and the No-U-Turn Sampler (NUTS) to enhance efficiency, the significant computational resources required may pose a barrier to its use in some settings. To mitigate this, future updates could explore integrating parallel computing techniques, reducing computational time and improving accessibility for large-scale applications.

Additionally, the toolbox currently assumes that users possess a foundational understanding of Bayesian methods and ODE modeling.  While the tutorial provided helps bridge this gap, incorporating more user-friendly interfaces or guided workflows could make the toolbox even more accessible to a broader audience, including those with limited statistical or programming backgrounds.

Regardless of the parameter estimation methods employed—whether Bayesian, frequentist, or otherwise—ensuring the structural and practical identifiability of model parameters is crucial. Structural identifiability refers to the theoretical ability to uniquely estimate model parameters from perfect data, given the model structure (\cite{dong2023differential,bellu2007daisy,chowell2023structural}). On the other hand, practical identifiability concerns the feasibility of estimating these parameters from real, noisy data (\cite{wieland2021structural,raue2009structural,saucedo2024comparative,roosa2019assessing}). Failure to address parameter identifiability can lead to erroneous conclusions, so modelers should assess both structural and practical identifiability during the model development process. This involves determining whether the available data and model structure provide sufficient information to uniquely estimate parameters, while also considering factors such as data quality, measurement error, and model complexity.

Looking forward, we anticipate that the adoption of \texttt{BayesianFitForecast} will facilitate more accurate and transparent modeling efforts across various scientific disciplines. Future developments could focus on expanding the library of pre-configured models and error structures, further increasing the toolbox’s utility. Additionally, as Bayesian methods continue to evolve, integrating more advanced techniques such as hierarchical modeling and machine learning-based priors could further strengthen the toolbox’s capabilities.

In summary, \texttt{BayesianFitForecast} represents a substantial step forward in the practical application of Bayesian inference to ODE models. Its combination of user-friendly design and rigorous statistical foundation makes it an invaluable tool for researchers aiming to conduct reproducible modeling studies based on ODE systems. We encourage researchers to explore the capabilities of \texttt{BayesianFitForecast} and contribute to its ongoing development and refinement.

\section*{Availability and requirements}

\begin{unenumerate}
    \item \textbf{Project name:} \texttt{BayesianFitForecast}
    \item \textbf{Project home page:} https://github.com/gchowell/BayesianFitForecast
    \item \textbf{Operating system(s):} Platform independent
    \item \textbf{Programming language: } R
    \item \textbf{Other requirements:} R, RStudio, Java, and Rtools (or equivalent package builder)
    \item \textbf{License:} This project is licensed under the Creative Commons Attribution-NonCommercial-ShareAlike 4.0 International (CC BY-NC-SA 4.0).
\end{unenumerate}

\section*{List of abbreviations}

\begin{unenumerate}
    \item \textbf{ODEs:} Ordinary Differential Equations 
    \item \textbf{MCMC:} Markov chain Monte Carlo
    \item \textbf{HMC:} Hamiltonian Monte Carlo
    \item \textbf{NUTS:} No U-Turn Sampler
    \item \textbf{DIC:} Deviance Information Criterion
    \item \textbf{AIC:} Akaike Information Criterion
    \item \textbf{PI:} Prediction Interval
    \item \textbf{CI:} Confidence Interval
    \item \textbf{MAE:} Mean Absolute Error
    \item \textbf{MSE:} Mean Squared Error
    \item \textbf{WIS:} Weighted Interval Scores
    \item \textbf{IS:} Interval Scores
    \item \textbf{SEIR:} Susceptible-Exposed-Infected-Recovered model
    \item \textbf{SEIUR:} Susceptible-Exposed-Infected-Under-reported model
    \item \textbf{SEIRD:} Susceptible-Exposed-Infected-Recovered-Death model
\end{unenumerate}

\section*{Declarations}

\begin{unenumerate}
    \item \textbf{Ethics approval and consent to participate:} Not applicable. 
    \item \textbf{Consent for publication:} Not applicable. 
    
    \item \textbf{Availability of data and materials:} This program and all associated files and data mentioned within the manuscript are publicly available at https://github.com/gchowell/BayesianFitForecast.
    
    \item \textbf{Competing interests: } Not applicable
    \item \textbf{Funding:} HK and AB are supported by a 2CI Fellowship from Georgia State University. G.C. is partially supported by NSF grants 2125246 and 2026797. The funding bodies played no role in the design of the study and collection, analysis, and interpretation of data and in writing the manuscript.
    \item \textbf{Authors' contributions:} H.K., R.L., and G.C. conceived and developed the first version of the toolbox, with H.K. responsible for coding. H.K., R.L., and G.C. wrote the first draft of the tutorial. H.K., A.B., R.L., and G.C. contributed to writing and reviewing subsequent drafts of the tutorial. A.B. produced the tutorial video. A.B., H.K., and G.C. generated the Shiny App, with A.B. responsible for the UI and related coding. A.B. wrote the first draft related to the Shiny App and Appendix, and H.K. and G.C. contributed to subsequent drafts.
     \item \textbf{Acknowledgements:} Not applicable
\end{unenumerate}

\newpage

% Setting up the references 
\bibliographystyle{sn-vancouver}
\bibliography{reference}

\begin{thebibliography}{10}
\providecommand{\doi}[1]{\url{https://doi.org/#1}}
\bibcommenthead

\bibitem[\protect\citeauthoryear{Strogatz}{2024}]{strogatz2024nonlinear}
Strogatz SH.
\newblock Nonlinear dynamics and chaos: with applications to physics, biology, chemistry, and engineering.
\newblock Chapman and Hall/CRC; 2024.

\bibitem[\protect\citeauthoryear{Brauer and Castillo-Chavez}{2012}]{brauer2012mathematical}
Brauer F, Castillo-Chavez C.
\newblock Mathematical models in population biology and epidemiology. vol.~40.
\newblock Springer; 2012.

\bibitem[\protect\citeauthoryear{Yan and Chowell}{2019}]{yan2019quantitative}
Yan P, Chowell G.
\newblock Quantitative methods for investigating infectious disease outbreaks. vol.~70.
\newblock Springer; 2019.

\bibitem[\protect\citeauthoryear{Grinsztajn et~al.}{2021}]{grinsztajn2021bayesian}
Grinsztajn L, Semenova E, Margossian CC, Riou J.
\newblock Bayesian workflow for disease transmission modeling in Stan.
\newblock Statistics in medicine. 2021;40(27):6209--6234.

\bibitem[\protect\citeauthoryear{Bouman et~al.}{2024}]{bouman2024bayesian}
Bouman JA, Hauser A, Grimm SL, Wohlfender M, Bhatt S, Semenova E, et~al.
\newblock Bayesian workflow for time-varying transmission in stratified compartmental infectious disease transmission models.
\newblock PLoS computational biology. 2024;20(4):e1011575.

\bibitem[\protect\citeauthoryear{Gelman et~al.}{2020}]{gelman2020bayesian}
Gelman A, Vehtari A, Simpson D, Margossian CC, Carpenter B, Yao Y, et~al.
\newblock Bayesian workflow.
\newblock arXiv preprint arXiv:201101808. 2020;.

\bibitem[\protect\citeauthoryear{Belasso et~al.}{2023}]{belasso2023bayesian}
Belasso CJ, Cai Z, Bezgin G, Pascoal T, Stevenson J, Rahmouni N, et~al.
\newblock Bayesian workflow for the investigation of hierarchical classification models from tau-PET and structural MRI data across the Alzheimer’s disease spectrum.
\newblock Frontiers in Aging Neuroscience. 2023;15:1225816.

\bibitem[\protect\citeauthoryear{Girolami}{2008}]{girolami2008bayesian}
Girolami M.
\newblock Bayesian inference for differential equations.
\newblock Theoretical Computer Science. 2008;408(1):4--16.

\bibitem[\protect\citeauthoryear{Kypraios et~al.}{2017}]{kypraios2017tutorial}
Kypraios T, Neal P, Prangle D.
\newblock A tutorial introduction to Bayesian inference for stochastic epidemic models using Approximate Bayesian Computation.
\newblock Mathematical biosciences. 2017;287:42--53.

\bibitem[\protect\citeauthoryear{McKinley et~al.}{2014}]{mckinley2014simulation}
McKinley TJ, Ross JV, Deardon R, Cook AR.
\newblock Simulation-based Bayesian inference for epidemic models.
\newblock Computational Statistics \& Data Analysis. 2014;71:434--447.

\bibitem[\protect\citeauthoryear{Gelman et~al.}{1995}]{gelman1995bayesian}
Gelman A, Carlin JB, Stern HS, Rubin DB.
\newblock Bayesian data analysis.
\newblock Chapman and Hall/CRC; 1995.

\bibitem[\protect\citeauthoryear{Alahmadi et~al.}{2020}]{alahmadi2020influencing}
Alahmadi A, Belet S, Black A, Cromer D, Flegg JA, House T, et~al.
\newblock Influencing public health policy with data-informed mathematical models of infectious diseases: Recent developments and new challenges.
\newblock Epidemics. 2020;32:100393.

\bibitem[\protect\citeauthoryear{O’Neill and Roberts}{1999}]{o1999bayesian}
O’Neill PD, Roberts GO.
\newblock Bayesian inference for partially observed stochastic epidemics.
\newblock Journal of the Royal Statistical Society Series A: Statistics in Society. 1999;162(1):121--129.

\bibitem[\protect\citeauthoryear{Jackson et~al.}{2015}]{jackson2015calibration}
Jackson CH, Jit M, Sharples LD, De~Angelis D.
\newblock Calibration of complex models through Bayesian evidence synthesis: a demonstration and tutorial.
\newblock Medical decision making. 2015;35(2):148--161.

\bibitem[\protect\citeauthoryear{Trejo and Hengartner}{2022}]{trejo2022modified}
Trejo I, Hengartner NW.
\newblock A modified Susceptible-Infected-Recovered model for observed under-reported incidence data.
\newblock PloS one. 2022;17(2):e0263047.

\bibitem[\protect\citeauthoryear{Team}{2023}]{stan2023}
Team SD. Team SD, editor.: Stan User's Guide, Version 2.35.
\newblock Stan Development Team.
\newblock Available from: \url{https://mc-stan.org/users/documentation/}.

\bibitem[\protect\citeauthoryear{Annis et~al.}{2017}]{annis2017bayesian}
Annis J, Miller BJ, Palmeri TJ.
\newblock Bayesian inference with Stan: A tutorial on adding custom distributions.
\newblock Behavior research methods. 2017;49:863--886.

\bibitem[\protect\citeauthoryear{Kelter}{2020}]{kelter2020bayesian}
Kelter R.
\newblock Bayesian survival analysis in STAN for improved measuring of uncertainty in parameter estimates.
\newblock Measurement: Interdisciplinary Research and Perspectives. 2020;18(2):101--109.

\bibitem[\protect\citeauthoryear{Sennhenn-Reulen}{2018}]{sennhenn2018bayesian}
Sennhenn-Reulen H.
\newblock Bayesian regression for a Dirichlet distributed response using Stan.
\newblock arXiv preprint arXiv:180806399. 2018;.

\bibitem[\protect\citeauthoryear{Sorensen and Vasishth}{2015}]{sorensen2015bayesian}
Sorensen T, Vasishth S.
\newblock Bayesian linear mixed models using Stan: A tutorial for psychologists, linguists, and cognitive scientists.
\newblock arXiv preprint arXiv:150606201. 2015;.

\bibitem[\protect\citeauthoryear{Monnahan et~al.}{2017}]{monnahan2017faster}
Monnahan CC, Thorson JT, Branch TA.
\newblock Faster estimation of Bayesian models in ecology using Hamiltonian Monte Carlo.
\newblock Methods in Ecology and Evolution. 2017;8(3):339--348.

\bibitem[\protect\citeauthoryear{B{\"u}rkner}{2017}]{burkner2017brms}
B{\"u}rkner PC.
\newblock brms: An R package for Bayesian multilevel models using Stan.
\newblock Journal of statistical software. 2017;80:1--28.

\bibitem[\protect\citeauthoryear{Carpenter et~al.}{2017}]{carpenter2017stan}
Carpenter B, Gelman A, Hoffman MD, Lee D, Goodrich B, Betancourt M, et~al.
\newblock Stan: A probabilistic programming language.
\newblock Journal of statistical software. 2017;76.

\bibitem[\protect\citeauthoryear{Betancourt}{2017}]{betancourt2017conceptual}
Betancourt M.
\newblock A conceptual introduction to Hamiltonian Monte Carlo.
\newblock arXiv preprint arXiv:170102434. 2017;.

\bibitem[\protect\citeauthoryear{Spiegelhalter et~al.}{2002}]{spiegelhalter2002bayesian}
Spiegelhalter DJ, Best NG, Carlin BP, Van Der~Linde A.
\newblock Bayesian measures of model complexity and fit.
\newblock Journal of the royal statistical society: Series b (statistical methodology). 2002;64(4):583--639.

\bibitem[\protect\citeauthoryear{Watanabe}{2010}]{watanabe2010asymptotic}
Watanabe S.
\newblock Asymptotic equivalence of Bayes cross validation and widely applicable information criterion in singular learning theory.
\newblock Journal of machine learning research. 2010;11(12).

\bibitem[\protect\citeauthoryear{Vehtari et~al.}{2017}]{vehtari2017practical}
Vehtari A, Gelman A, Gabry J.
\newblock Practical Bayesian model evaluation using leave-one-out cross-validation and WAIC.
\newblock Statistics and computing. 2017;27(5):1413--1432.

\bibitem[\protect\citeauthoryear{Gneiting and Raftery}{2007}]{gneiting2007strictly}
Gneiting T, Raftery AE.
\newblock Strictly proper scoring rules, prediction, and estimation.
\newblock Journal of the American statistical Association. 2007;102(477):359--378.

\bibitem[\protect\citeauthoryear{Kuhn et~al.}{2013}]{kuhn2013applied}
Kuhn M, Johnson K, et~al.
\newblock Applied predictive modeling. vol.~26.
\newblock Springer; 2013.

\bibitem[\protect\citeauthoryear{Bracher et~al.}{2021}]{bracher2021evaluating}
Bracher J, Ray EL, Gneiting T, Reich NG.
\newblock Evaluating epidemic forecasts in an interval format.
\newblock PLoS computational biology. 2021;17(2):e1008618.

\bibitem[\protect\citeauthoryear{Cramer et~al.}{2022}]{cramer2022evaluation}
Cramer EY, Ray EL, Lopez VK, Bracher J, Brennen A, Castro~Rivadeneira AJ, et~al.
\newblock Evaluation of individual and ensemble probabilistic forecasts of COVID-19 mortality in the United States.
\newblock Proceedings of the National Academy of Sciences. 2022;119(15):e2113561119.

\bibitem[\protect\citeauthoryear{Murray}{2003}]{Murray2003}
Murray JD.
\newblock Mathematical Biology II: Spatial Models and Biomedical Applications. vol.~18 of Interdisciplinary Applied Mathematics.
\newblock 3rd ed. New York: Springer; 2003.

\bibitem[\protect\citeauthoryear{Chowell et~al.}{2024}]{chowell2024parameter}
Chowell G, Bleichrodt A, Luo R.
\newblock Parameter estimation and forecasting with quantified uncertainty for ordinary differential equation models using QuantDiffForecast: A MATLAB toolbox and tutorial.
\newblock Statistics in Medicine. 2024;43(9):1826--1848.

\bibitem[\protect\citeauthoryear{Karami et~al.}{2024}]{BayesianFitForecast}
Karami H, Bleichrodt A, Luo R, Chowell G.: BayesianFitForecast: A library for Bayesian fitting and forecasting.
\newblock Comprehensive R Archive Network (CRAN).
\newblock R package version 1.0.0.
\newblock Available from: \url{https://CRAN.R-project.org/package=BayesianFitForecast}.

\bibitem[\protect\citeauthoryear{Dong et~al.}{2023}]{dong2023differential}
Dong R, Goodbrake C, Harrington HA, Pogudin G.
\newblock Differential elimination for dynamical models via projections with applications to structural identifiability.
\newblock SIAM Journal on Applied Algebra and Geometry. 2023;7(1):194--235.

\bibitem[\protect\citeauthoryear{Bellu et~al.}{2007}]{bellu2007daisy}
Bellu G, Saccomani MP, Audoly S, D’Angi{\`o} L.
\newblock DAISY: A new software tool to test global identifiability of biological and physiological systems.
\newblock Computer methods and programs in biomedicine. 2007;88(1):52--61.

\bibitem[\protect\citeauthoryear{Chowell et~al.}{2023}]{chowell2023structural}
Chowell G, Dahal S, Liyanage YR, Tariq A, Tuncer N.
\newblock Structural identifiability analysis of epidemic models based on differential equations: a tutorial-based primer.
\newblock Journal of Mathematical Biology. 2023;87(6):79.

\bibitem[\protect\citeauthoryear{Wieland et~al.}{2021}]{wieland2021structural}
Wieland FG, Hauber AL, Rosenblatt M, T{\"o}nsing C, Timmer J.
\newblock On structural and practical identifiability.
\newblock Current Opinion in Systems Biology. 2021;25:60--69.

\bibitem[\protect\citeauthoryear{Raue et~al.}{2009}]{raue2009structural}
Raue A, Kreutz C, Maiwald T, Bachmann J, Schilling M, Klingm{\"u}ller U, et~al.
\newblock Structural and practical identifiability analysis of partially observed dynamical models by exploiting the profile likelihood.
\newblock Bioinformatics. 2009;25(15):1923--1929.

\bibitem[\protect\citeauthoryear{Saucedo et~al.}{2024}]{saucedo2024comparative}
Saucedo O, Laubmeier A, Tang T, Levy B, Asik L, Pollington T, et~al.
\newblock Comparative analysis of practical identifiability methods for an seir model.
\newblock arXiv preprint arXiv:240115076. 2024;.

\bibitem[\protect\citeauthoryear{Roosa and Chowell}{2019}]{roosa2019assessing}
Roosa K, Chowell G.
\newblock Assessing parameter identifiability in compartmental dynamic models using a computational approach: application to infectious disease transmission models.
\newblock Theoretical Biology and Medical Modelling. 2019;16:1--15.

\end{thebibliography}

\newpage

\backmatter

\newpage

\begin{appendices}

\section{The \texttt{BayesianFitForecast} Shiny App} \label{appendix:A}

\subsection{Using the Dashboard Application}

\begin{figure}[H]
    \centering
    \includegraphics[width=\linewidth]{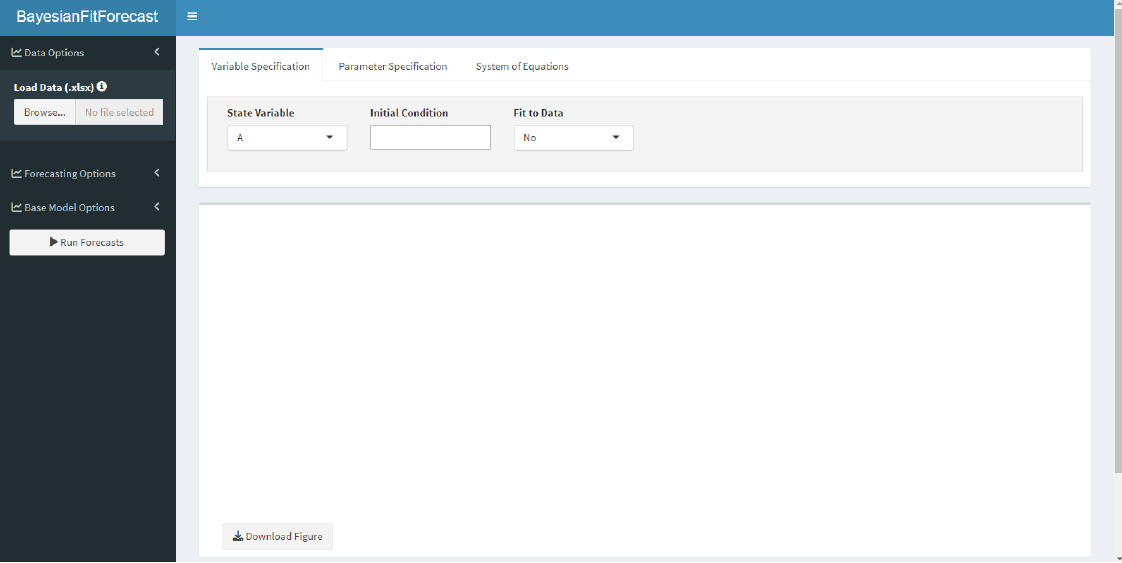}
    \caption{\footnotesize The general structure of the \texttt{BayesianFitForecast} R-Shiny application. Above shows the user-view of the \texttt{BayesianFitForecast} dashboard upon first loading the \texttt{app.R} file. The sidebar contains user inputs for the required data details, model calibration and forecasting specifications, and base model options. Users can specify their desired state variables, parameters, system of ordinary differential equations (ODEs) and associated details in the top panel of the main body. The second panel hosts the available figure outputs, and the third panel hosts the available data sets.}
    \label{fig:Fig1A}
\end{figure}

The \texttt{BayesianFitForecast} R-Shiny application consists of a single page, with a sidebar and three sections within its main body (Fig. \ref{fig:Fig1A}). While the dashboard utilizes many of the same user parameters presented in the main text, it streamlines their entry by removing any necessary programming requirements. Table \ref{tab:comparisonAPP} below provides a side-by-side comparison of the options required in the source \texttt{options.R} file and the user parameters required to run the dashboard. Once all required parameters have been specified and the \textit{Run Forecasts} button has been selected, the resulting model fits, forecasts, metrics, and visualizations will be displayed. 

\begin{longtable}{|P{3.6cm}|P{2.2cm}|P{6.25cm}|}
\caption{A side-by-side comparison, including descriptions, of the parameters needed for the \texttt{BayesianFitForecast} R-library and \texttt{BayesianFitForecast} R-Shiny dashboard.} \label{tab:comparisonAPP} \\
\hline
\textbf{Options Parameter} & \textbf{Dashboard\newline Parameter(s)} & \textbf{Description} \\
\hline
\endfirsthead
\multicolumn{3}{r}{\textit{(Continued on next page)}} \\
\endfoot
\endlastfoot

\texttt{calibrationperiods} & Calibration Period Length & A pre-filled drop-down menu containing the available calibration period lengths. The options range from the inclusion of a single piece of data to the entire length of the input time series. Multiple calibration period lengths can be selected. \\
\hline
\texttt{forecastinghorizon} & Forecasting Horizon & A numeric-based free-text input to specify the length of the forecasting horizon. Users can only enter numeric values ranging from 0 to $\infty$. Additionally, users can incrementally increase or decrease the forecasting horizon by selecting the up and down arrows within the input field. \\
\hline
\texttt{modelname} & NA & This parameter is not needed for the functionality of the Shiny application. Therefore, unlike the \texttt{options.R} file specification, users do not need to enter a model name within the dashboard. \\
\hline
\texttt{vars} & State Variable & A pre-filled drop-down menu containing the 26 letters of the English alphabet. The letter represents the \textit{State Variable} to be used within the specification of the system of ODEs. The order of state variables corresponds to the order of the ODE equation rows. \\
\hline
\texttt{params} & Parameter & A \LaTeX-input to specify the symbol of the parameter. This is the symbol that will be used when specifying the system of ODEs. To enter non-alphanumeric symbols such as $\beta$, users should first type "\texttt{\textbackslash}" followed by the name of their desired symbol. For example, to achieve the $\beta$ symbol, users should type "\texttt{\textbackslash beta}". \\
\hline
\texttt{odesystem} & Equation \# & A \LaTeX-input to specify the derivative equation of interest. The order of the equations corresponds to the order of the state variable specification. They can only contain state variables and parameters included within the "Variable" and "Parameter Specification" tabs. To enter non-alphanumeric symbols such as $\beta$, users should first type "\texttt{\textbackslash}" followed by the name of their desired symbol. For example, to achieve the $\beta$ symbol, users should type "\texttt{\textbackslash beta}". \\
\hline
\texttt{paramsfix} & Estimation & A drop-down menu to indicate the estimation status for a given parameter. Users can either choose a \texttt{<Fixed>} status or \texttt{<Estimate>}. If \texttt{<Estimate>} is selected, users must specify the prior and related bounds. \\
\hline
\texttt{compositeexpressions} & Variable \& Equation & A free-text (\textit{Variable}) and \LaTeX-input (\textit{Equation}) to specify the equation for the composite variable of interest. Only parameters included within the system of ODEs can be used in the \textit{Equation} input. To enter non-alphanumeric symbols such as $\beta$, users should first type "\texttt{\textbackslash}" followed by the name of their desired symbol. For example, to achieve the $\beta$ symbol, users should type "\texttt{\textbackslash beta}". \\
\hline
\texttt{fittingindex} & Fit to Data \& Time Series to Fit & A Boolean drop-down to indicate if the given compartment should be fit to the data, and a drop-down with labels for each time series. If \textit{Fit to Data} is \texttt{<Yes>}, the \textit{Time Series to } and \textit{Fit to Derivative} options will appear. The number of selected \texttt{<Yes>} responses must match the number of provided time series. \\
\hline
\texttt{fittingdiff} & Fit to Derivative & A drop-down Boolean menu to indicate if the derivative of the model’s fitting variable should be fit to the data. \\
\hline
\texttt{errstrc} & Error Structure & A drop-down menu with available error structures: Normal, Negative Binomial, and Poisson. The normal and negative binomial error structure options require additional parameters. Only one structure can be selected at a time. \\
\hline
\texttt{cadfilename1} & Load Data (.xlsx) & See Section~\ref{sec:loading-data-app}. \\
\hline
\texttt{caddisease} & Process of Interest & A free-text input to specify the process of interest (e.g., disease) included within the data. \\
\hline
\texttt{seriescases} & Time Series Type \# & A free-text input to specify the type of data (e.g., Cases, Deaths). Multiple inputs appear if multiple time series are included. \\
\hline
\texttt{datetype} & Temporal Resolution & A free-text input to specify the temporal resolution (e.g., Days, Weeks) of the input data. \\
\hline
\texttt{paramsprior} & Prior & A drop-down menu containing available prior distributions. Depending on the selection, the corresponding parameter options appear. Custom priors can be added. This input appears only when \textit{Estimation} is set to \texttt{<Estimate>}. \\
\hline
\texttt{LB and UB} & Lower Bound \& Upper Bound & Free-text numeric inputs to indicate the bounds of the parameter estimate. If no bound is needed, leave blank or enter NA. \\
\hline
\texttt{normalerrorxprior, negbinerrorprior} & Required Parameters & A drop-down for prior distributions for parameters. Available only if the error structure is \texttt{<Negative Binomial>} or \texttt{<Normal>}. Separate priors can be selected for each input time series. \\
\hline
\texttt{Ic} & Initial Condition & A \LaTeX-input for initial condition of each state variable. Accepts: (1) numeric, (2) numeric and parameter symbols, or (3) only parameter symbols. Mathematical equations can also be entered. To enter non-alphanumeric symbols such as $\beta$, users should first type "\texttt{\textbackslash}" followed by the name of their desired symbol. For example, to achieve the $\beta$ symbol, users should type "\texttt{\textbackslash beta}". \\
\hline
\texttt{vars.init} & NA & This parameter is derived from the initial condition, parameter, and equation specifications. Users do not set this explicitly in the dashboard. \\
\hline
\texttt{niter} & Number of MCMC Steps & Numeric input for the number of MCMC steps in the sampling process. \\
\hline
\texttt{numchain} & Number of MCMC Chains & Numeric input for the number of MCMC chains in the sampling process. \\
\hline
\end{longtable}

If any required options have not been specified and processed, the dashboard will return an error message prompting users to review their selections. The remainder of this text will focus on describing each of the dashboard's available features, utilizing the 1918 influenza pandemic in San Francisco data and example described in Section \ref{sec:examplePoisson} of the main text. 

\subsection{Sidebar Parameters}

\subsubsection{Data Options}

\begin{figure}[t]
    \centering
    \includegraphics[]{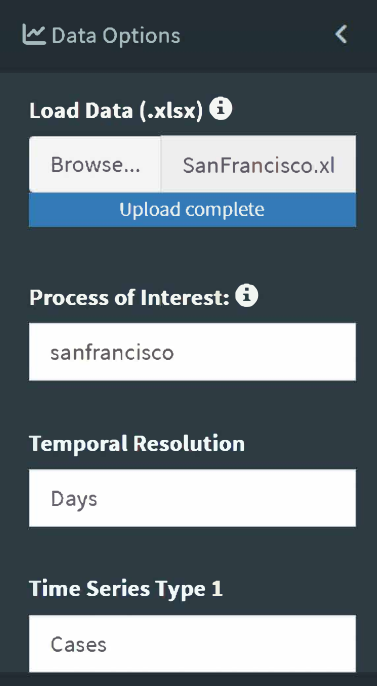}
    \caption{\footnotesize The "Data Options" available within the sidebar menu of the \texttt{BayesianFitForecast} dashboard. The first step in utilizing the dashboard includes loading the input data (ex. \texttt{SanFrancisco.xlsx}) and specifying the process of interest (ex. \texttt{<sanfranscio>}), temporal resolution of the data (ex. \texttt{<Days>}), and label indicating the type of included time series (ex. \texttt{<Cases>}). The \textit{Process of Interest}, \textit{Temporal Resolution}, and \textit{Time Series Type \#} parameters are free-text inputs that are not required to follow any specific format.}
    \label{fig:Fig2A}
\end{figure}

After uploading the data following the format described in the main text via the \textit{Load Data (.xlsx)} input option, additional parameters related to the process of interest, temporal resolution, and type of data included within each time series (i.e., column of data) become available (Fig. \ref{fig:Fig2A}). The \textit{Process of Interest} data parameter is a free text input that refers to the infectious disease or “process” measured within the time series data. It is a unique string indicator for the input data and does not require any specific formatting. For example, as related to the tutorial example, we set the parameter to \texttt{<sanfrancisco>}. The \textit{Temporal Resolution} data parameter is a free text input that refers to the temporal spacing between each time point, including in the data (i.e., “days”, “weeks”, years”, etc.) Like the \textit{Process of Interest} parameter, it is a unique string indicator and does not require any specific formatting. As we are working with daily data, we set the parameter to \texttt{<Days>}. Finally, we must provide labels for each included time series or non-temporal resolution column of data. The dashboard will determine the number of time series included within the input data and provide the corresponding number of \textit{Time Series Type \#} free text inputs. The \textit{Time Series Type \#} options order corresponds to the order of the time series columns. For example, as we are working with only one time series in this example, only one option will appear, \textit{Time Series Type 1} (Fig. \ref{fig:Fig2A}). As with the other two data parameters, the \textit{Time Series Type \#} parameter is a unique string indicator that does not require any specific formatting. Given our single time series contains case data, we used the label \texttt{<Cases>}. 

\subsubsection{Forecasting Options}

\begin{figure}[t]
    \centering
    \includegraphics[]{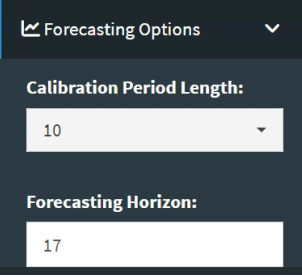}
    \caption{\footnotesize The "Forecasting Options" available within the sidebar menu of the \texttt{BayesianFitForecast} dashboard. The second step in utilizing the dashboard includes specifying the calibration period length and forecasting horizon. The \textit{Calibration Period Length} parameter is an auto-populated drop-down menu, and the \textit{Forecasting Horizon} parameter is a numeric input. For this tutorial, we set the calibration period length to 17 and the forecasting horizon to 10.}
    \label{fig:Fig3A}
\end{figure}

The dashboard supports both model-fitting and forecasting analysis; therefore, users must specify the length of the input data fed into the model, or calibration period length, and the desired forecasting horizon (Fig. \ref{fig:Fig3A}). The \textit{Calibration Period Length} parameter is drop-down menu input that is auto populated with all available calibration period lengths based upon the length of the inputted data. At a minimum, at least one data point must be fed into the model (i.e., \texttt{<Calibration Period Length: = 1>}) and, at a maximum, all the data can be utilized during the model fitting process. As with the \texttt{BayesianFitForecast} R library, multiple calibration periods can be selected, and outputs will be provided for each selected calibration period length. 

The \textit{Forecasting Horizon} option is a numeric input that refers to the number of time points in the future for which the user would like to produce forecasts after the forecast date. To obtain only the model fit, the \textit{Forecasting Horizon} should be set to \texttt{<0>}. To align with the example presented in Section \ref{sec:examplePoisson} of the main text, we set our calibration period to \texttt{<17>} and the forecasting horizon to \texttt{<10>} (Fig. \ref{fig:Fig3A}). 

\subsubsection{Base Model Options}

\begin{figure}[t]
    \centering
    \includegraphics[]{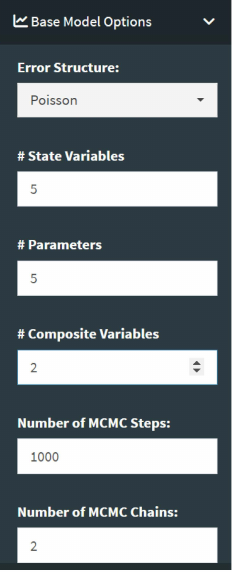}
    \caption{\footnotesize The "Base Model" options available within the last menu of the sidebar. The figure illustrates the base model options available within the dashboard’s sidebar. Prior to specifying the state variable, parameters, and system of ordinary differential equations (ODE), users must select the error structure of interest (\textit{Error Structure}) via the drop-down menu, and specify the number of desired state variables (\textit{\# State Variables}), parameters (\textit{\# Parameters}), composite variables (\textit{\# Composite Variables}), number of MCMC steps (\textit{Number of MCMC Steps}), and the number of chains to include (\textit{Number of MCMC Chains}) using the corresponding numeric input. For this tutorial, we indicated we wanted five state variables, four unique parameters, and two composite variables. We also employed 1000 MCMC steps and two MCMC chains.}
    \label{fig:Fig4A}
\end{figure}

The "Base Model Options" menu located within the sidebar contains the base controls for the state variable, parameter, and composite equation specifications (Fig. \ref{fig:Fig4A}). The \textit{Error Structure} option is a drop-down menu containing the three available error structures: (1) \texttt{Negative Binomial}, (2) \texttt{Normal}, and (3) \texttt{Poisson}. Only one error structure can be selected at a time. As discussed in the main text, additional parameters are required to be estimated when the \texttt{Negative Binomial} or \texttt{Normal} error structures are selected. To accommodate this requirement, when either error structure is selected, the corresponding symbol for the required parameters will appear under the “\textit{Required Parameters}” header on the “Parameter Specifications” tab. Additional details regarding prior selection for the required parameters can be found below. For our example, we selected the \texttt{<Poisson>} error structure option.

The next three parameters include the \textit{\# State Variables}, \textit{\# Parameters}, and \textit{\# Composite Variables} numeric inputs. Each option controls the number of state variable, parameter, and composite variable entries desired for the analysis. Given that the number of ODEs required in the system of equations corresponds to the number of state variables specified, the \textit{\# State Variables} parameter also controls the number of ODE entries available under the “System of Equations” tab. 

Finally, the \textit{Number of MCMC Steps} and \textit{Number of MCMC Chains} parameters are numeric inputs that control the number of MCMC steps taken in the analysis and the number of chains used, respectively. For this tutorial, we indicated we wanted five state variables, five unique parameters, and two composite variables. We also employed 1000 MCMC steps, and two MCMC chains (Fig. \ref{fig:Fig4A}). 

\subsection{Variable, Parameter, and Equation Specification}

\subsubsection{State Variable Specification}

\begin{figure}[t]
    \centering
    \includegraphics[width=\linewidth]{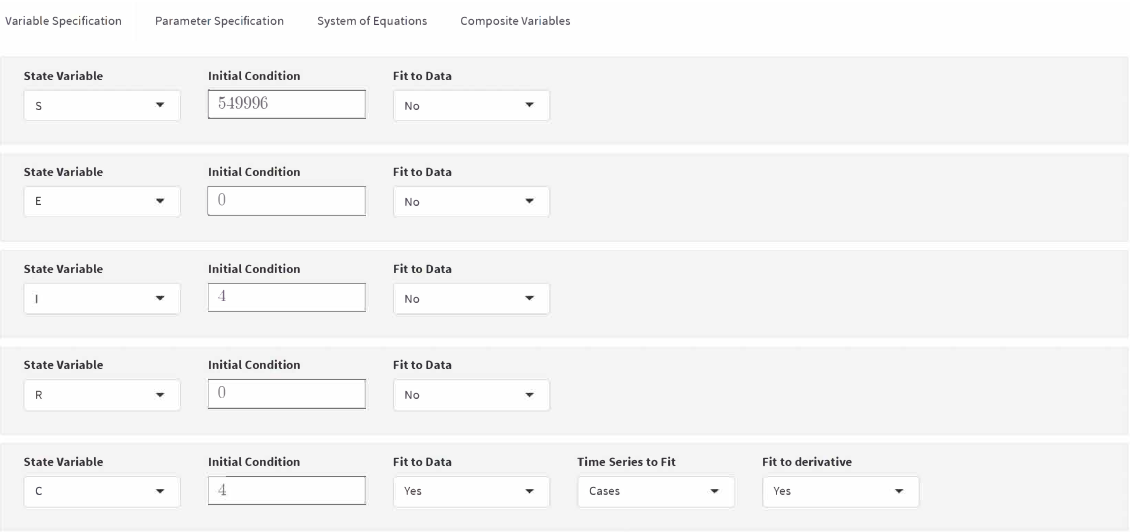}
    \caption{\footnotesize The “Variable Specification” tab available at the top of the main dashboard body. This figure illustrates the “Variable Specification” tab located at the top of the dashboard, where users can specify the required information about their state variables. After specifying all sidebar parameters, users can then specify their desired state variable labels (\textit{State Variable}), their initial conditions (\textit{Initial Condition}), fit status (\textit{Fit to Data}), and related parameters (\textit{Time Series to Fit} \& \textit{Fit to derivative}). The order in which state variables are entered must correspond to the desired order of the system of ordinary differential equations.}
    \label{fig:Fig5A}
\end{figure}

Once the sidebar options have been entered, users can proceed with specifying the desired state variables and associated parameters in the “Variable Specification” tab within the top panel of the main body of the dashboard (Fig. \ref{fig:Fig5A}). As discussed above, the number of state variable entries available corresponds to the number indicated within the \textit{\# State Variables} parameter. For example, as we set \texttt{<\# State Variable = 5>} we see five rows within the “Variable Specification” tab, each with three initial user-parameters to specify: (1) \textit{State Variable}, (2) \textit{Initial Condition}, and (3) \textit{Fit to Data}. First, users need to specify labels for the state variables of interest via the \textit{State Variable} drop-down menu for each row. For example, as we are utilizing an $SEIRC$ model, we select \texttt{<S>} for our first state variable, \texttt{<E>} for the second, and continue until we reach the end of the available rows. Once the state variables have been selected, users can then specify each of their initial conditions. 

\begin{figure}[t]
    \centering
    \includegraphics[width=\linewidth]{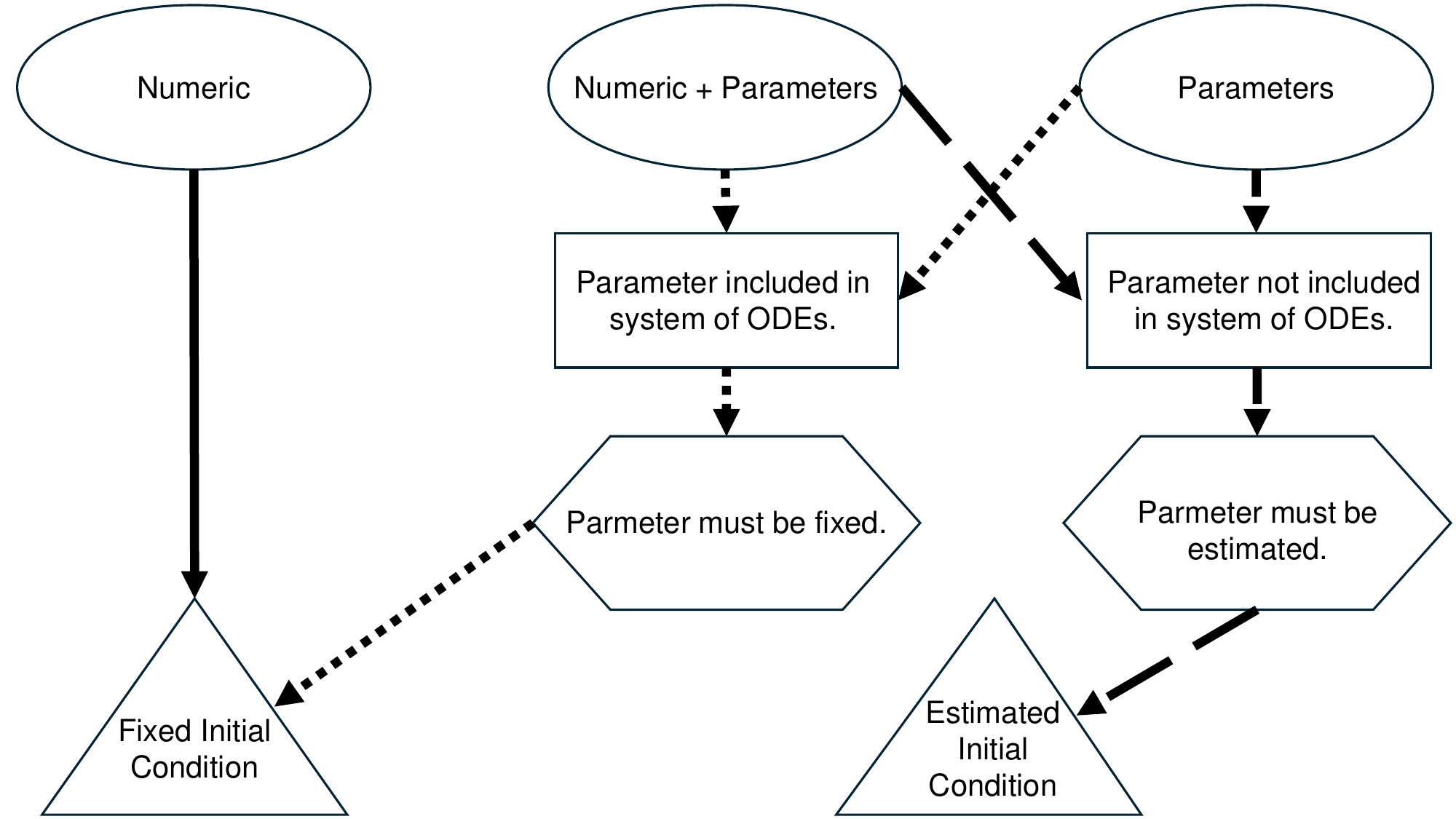}
    \caption{\footnotesize A flowchart illustrating the dashboard’s behavior and parameter fixed status dependent on the initial condition type. Corresponding arrow types (i.e., solid, dotted, dashed) correspond to possible front-facing and backend workflows based upon the entered initial condition. For example, if a single parameter is included within the initial conditions and not within the system of ODEs, the parameter must be estimated, and the dashboard will assume the initial condition of the state variable containing the corresponding parameter will also be estimated.}
    \label{fig:Fig6A}
\end{figure}

The \textit{Initial Conditions} are free-text, \LaTeX inputs that can take in numeric and mathematical symbols. As discussed in the main text, state variable initial conditions can be fixed or estimated, and the dashboard automatically detects the initial conditions’ fixed status based on the users’ specifications. Initial conditions within the dashboard can be entered as two types of objects: numeric and mathematical symbols. If working with only numeric values, users can either specify a single, fixed value as an initial condition (ex. \texttt{<10000>}) or a mathematical expression (ex. \texttt{<10000-6>}). Users can also use a combination of mathematical symbols and numbers (ex. \texttt{<N-6>}); the dashboard will assume that any mathematical symbols entered within the initial conditions are to be included as part of the parameters list. To enter non-alphanumeric symbols such as $\beta$ within the initial conditions, users should first type "\texttt{\textbackslash}" followed by the name of their desired symbol. For example, to achieve the $\beta$ symbol, users should type "\texttt{\textbackslash beta}". Finally, users can also enter only mathematical symbols as initial conditions (ex.\texttt{<N-i>}), and any such included symbols will be loaded as part of the parameters list. When any mathematical symbol is indicated, the dashboard will examine the specified system of ODEs to determine if the symbol is also included there. If it is not, the dashboard will treat the initial condition containing the parameter as it is to be estimated; otherwise, the initial condition is treated as fixed. Figure \ref{fig:Fig6A} presents a flow chart showing the dashboard’s behavior given the initial condition type specified. For our tutorial, we use only numeric values for initial conditions as shown in Figure \ref{fig:Fig5A}.

After specifying the initial conditions, users must indicate the state variable(s) the provided time series corresponds to. The number of state variables specified must match the number of time series included in the original data file. For example, as the tutorial input data contains a single time series of the number of infected persons, we could indicate that we want to fit our data to the \texttt{I} (infected state variable), or \texttt{C} (cumulative number of infected individuals). We do this first by changing the \textit{Fit to Data} drop-down menu input to \texttt{<Yes>} and then selecting the name of the time series of interest via the \textit{Time Series to Fit} drop-down menu input. A time series can be assigned only to one state variable at a time, and an error will occur if the time series is not assigned to any of the state variable. Finally, using the \textit{Fit to derivative} drop-down menu, we can indicate if we want the selected time series to be associated with the derivative of the variable. As shown in Figure \ref{fig:Fig5A}, we fit our single time series to the derivative of the \texttt{C} state variable. 

\subsubsection{Parameter Specification}

\begin{figure}[t]
    \centering
    \includegraphics[width=\linewidth]{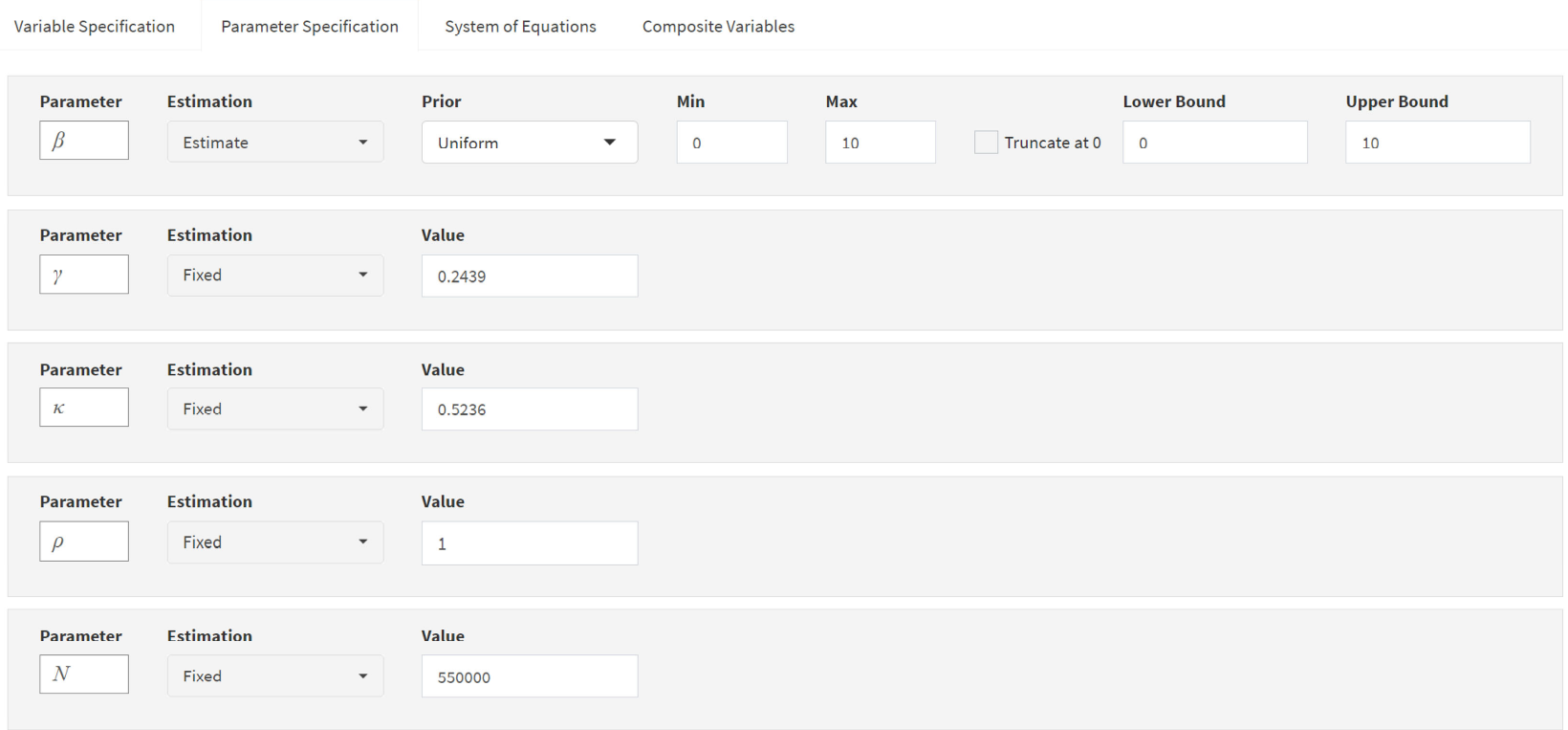}
    \caption{\footnotesize The “Parameter Specification” tab available at the top of the main dashboard body. This figure illustrates the “Parameter Specification” tab located at the top of the dashboard, where users can specify the required information about their parameters. After specifying all state variable entries, users can then indicate their desired parameter labels (\textit{Parameter}), their estimation status (\textit{Estimation}), their value (\textit{Value}) or prior distribution (\textit{Prior}), and associated prior parameters and bounds (\textit{Lower Bound} \& \textit{Upper Bound}).}
    \label{fig:Fig7A}
\end{figure}

After entering the state variables, the users can then proceed to the “Parameter Specification” tab, where they can indicate both required parameters (i.e., for error distribution) and the parameters of interest, their estimation status, and the values or prior distributions. The \textit{Parameter} options are \LaTeX inputs that can take single mathematical symbols; any parameters included within the state variable initial conditions will appear at the top of the parameter list (Fig. \ref{fig:Fig7A}). As we are working with a $SEIRC$ model, and do not include any parameters in the initial conditions, we need to specify five total parameters, $\beta$, $\kappa$, $\gamma$, $\rho$, and $N$. 

Once all parameters’ symbols have been indicated, users can then select their fixed status via the \textit{Estimation} drop-down menu. If \texttt{<Fixed>} is indicated, the \textit{Value} numeric input will become available for users to enter the value of the parameter. However, if \texttt{<Estimate>} is selected, the \textit{Prior} drop-down menu, associated parameters, option to truncate the estimate at zero (\textit{Truncate at 0}), and lower (\textit{Lower Bound}) and upper bound (\textit{Upper Bound}) options become available. The associated parameters that appear dependent on the prior selected are numeric inputs and will take only a single number for each entry. The \textit{Truncate at 0} is a checkbox to indicate if the user wants to truncate the estimate at zero. Finally, the \textit{Lower Bound} and \textit{Upper Bound} options are text inputs to indicate the lower and upper bounds of the parameter estimate. The input should always be numeric, albeit when there is no bound to be set. If no bound need be set, the user should either leave the input blank or enter \texttt{NA}. At least one parameter must always be estimated. 

Any parameters that are specified in the initial conditions follow a specific set of rules (Fig. \ref{fig:Fig6A}). For example, if any parameters are included within the state variable initial conditions and within the system of ODEs, the parameter must remain fixed (i.e., it cannot be estimated). However, if the parameter is not included within the system of ODEs, it must be estimated, and the dashboard will assume that the initial condition of the state variable, including the associated parameter, must be estimated. In addition to parameters included within the initial conditions, other parameters may be required, depending on the error distribution selected. For example, if instead of the \texttt{Poisson} distribution we selected the \texttt{Normal} error distribution, the “Required Parameters” selection will become available. Users will see the associated required parameter symbol, a drop-down menu to specify the desired prior (\textit{Prior}), and associated numeric inputs for the associated prior parameters. Multiple “Required Parameters” will appear if more than one time series is loaded into the dashboard; the order of the required parameters corresponds to the order of the time series in the original data set.

For this tutorial we are utilizing 5 parameters, $\beta$, $\kappa$, $\gamma$, $\rho$, and $N$, and estimating only the first parameter, $\beta$ (Fig. \ref{fig:Fig7A}). We select a uniform prior, with a minimum estimate of zero and a maximum of ten. Similarly, we set the lower bound to zero and upper bound of the parameter estimate to ten. 

\subsubsection{System of Equations}

\begin{figure}[t]
    \centering
    \includegraphics[width=\linewidth]{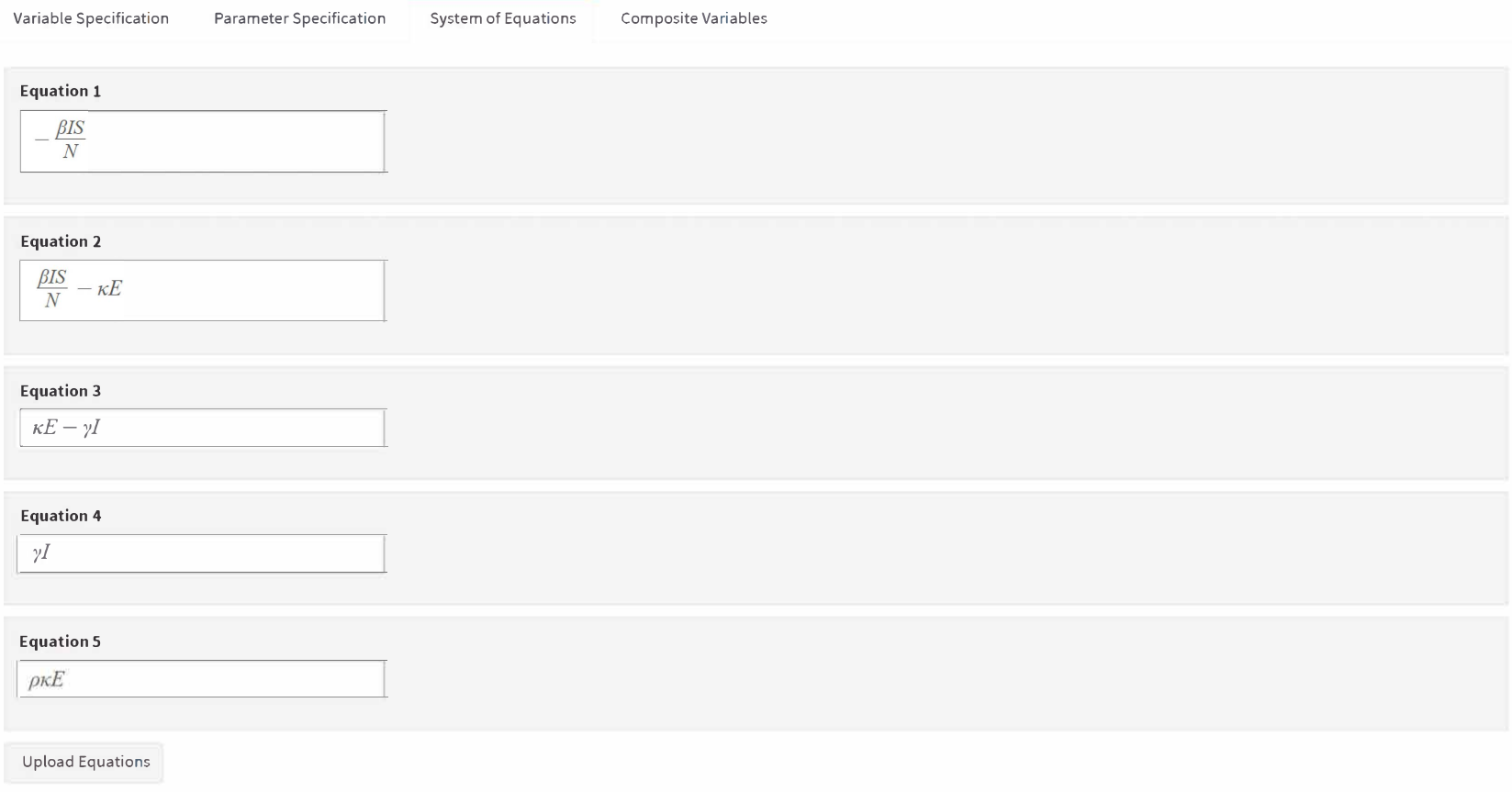}
    \caption{\footnotesize The “System of Equations” tab available at the top of the main dashboard body. This figure illustrates the “System of Equations” tab located at the top of the dashboard, where users can specify the system of ordinary differential equations (ODEs). Each row corresponds to a unique equation designated with \LaTeX syntax, and the order of the equations corresponds to the order of state variable specification. For example, given that we are using a $SEIRC$ model, “\texttt{Equation 1}” corresponds to the derivative of the “S” compartment. The ODEs must only contain state variables and parameters included within the “Variable Specification” and “Parameter Specification” tabs.}
    \label{fig:Fig8A}
\end{figure}

After specifying all desired and required parameters, the users can then proceed to entering the system of ODEs. Figure \ref{fig:Fig8A} illustrates the “System of Equations” tab, with the corresponding equations for the tutorial ($SEIRC$) specified. Each row corresponds to a unique equation and takes \LaTeX syntax. For example, to obtain \textit{Equation 1}, users should enter \texttt{<-\textbackslash beta*I*S/N>}. The order of the equations corresponds to the order of state variable specification, and the ODEs must only contain state variables and parameters included within the “Variable Specification” and “Parameter Specification” tabs. Once all equations are entered, the \textit{Upload Equations} button must be selected. If the equations were uploaded properly, the “Equations have been uploaded successfully” message will appear.

\subsubsection{Composite Equations}

\begin{figure}[t]
    \centering
    \includegraphics[width=\linewidth]{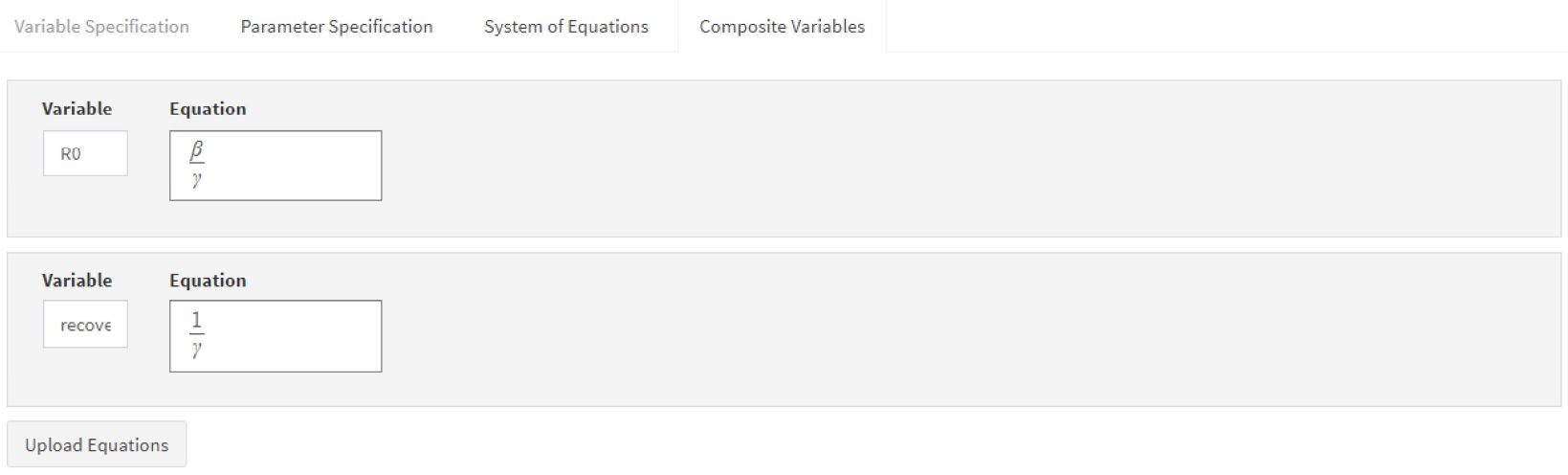}
    \caption{\footnotesize The “Composite Variables” tab available at the top of the main dashboard body. This figure illustrates the “Composite Variable” tab located at the top of the dashboard, where users can specify the left and right sides of each desired composite equation. Each row corresponds to a unique equation designated with text (\textit{Variable}) and \LaTeX (\textit{Equation}) syntax. The composite variables must only contain parameters included within the “System of Equations” tab.}
    \label{fig:Fig9A}
\end{figure}

Given that we indicated two composite equations in the sidebar of the dashboard (\texttt{<\# Composite Variables = 2>}), we must also specify both the left (\textit{Variable}) and right (\textit{Equation}) sides of each equation. The \textit{Variable} user parameter is a free-text input used to enter the label for the composite variable. For example, in our tutorial, we want to calculate both the R0 and recovery time; therefore, label the first composite equation \texttt{<R0>} and the second equation \texttt{<recovery\_time>}. The \textit{Equation} user parameter is a \LaTeX input to specify the equation to obtain the composite variable of interest. Specification of alphanumeric and mathematical symbols follows the same format used for entering the system of ODEs. Only parameters included within the system of ODEs can be included within the \textit{Equation} input. Once all labels and equations are entered, the \textit{Upload Equations} button must be selected. If the equations were uploaded properly, the “Composite equations have been uploaded successfully” message will appear. Figure \ref{fig:Fig9A} shows the “Composite Variables” tab, and required user-parameters, \textit{Variable} and \textit{Equation}. 

\subsection{Running the Dashboard}

\begin{figure}[t]
    \centering
    \includegraphics[width=\linewidth]{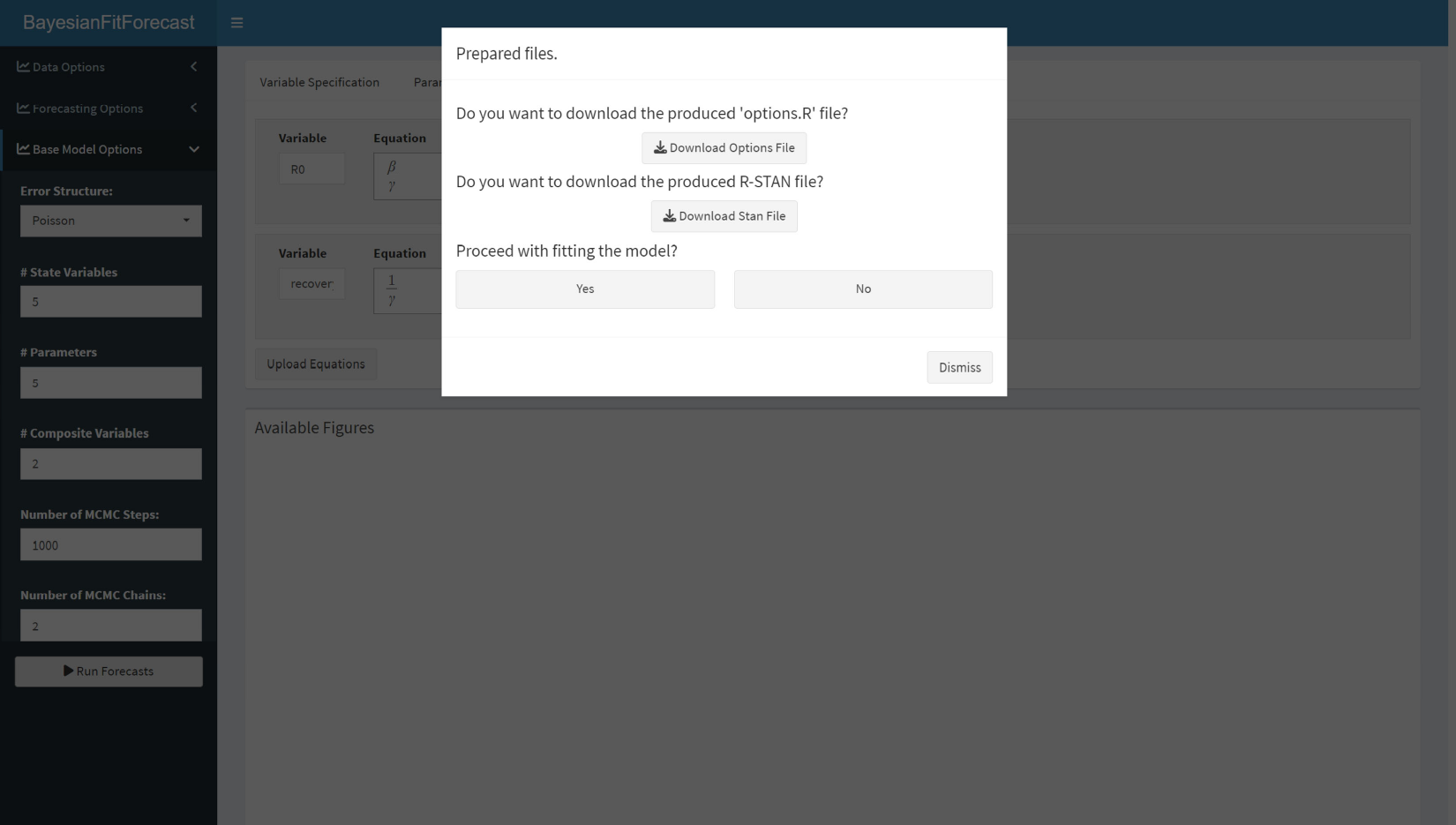}
    \caption{\footnotesize The pop-up menu available after the \textit{Run Forecasts} button has been selected. After all user inputs have been indicated, and the \textit{Run Forecasts} button has been clicked, the users will see a menu pop-up with options to download the produced options and R-Stan files, and to proceed with the model fitting process. The R-STAN file will match the one produced by the source package; the options file will be missing the \texttt{modelname}, \texttt{cadfilename1}, and \texttt{vars.init} user parameters. After the \textit{Yes} button is selected, the dashboard will proceed with fitting the model and producing subsequent forecasts.}
    \label{fig:Fig10A}
\end{figure}

Once all sidebar entries, state variables, parameters, system of equations, and composite variables have been entered and equations uploaded users can then proceed with running the dashboard (\textit{Run Forecasts}). After the \textit{Run Forecasts} button has been selected, users will see a menu (Fig. \ref{fig:Fig10A}) containing options to download the produced options and R-STAN files and must indicate if they wish to proceed with fitting the model. While the R-STAN file will match that which is produced by the source package, the options file will be missing the \texttt{modelname}, \texttt{cadfilename1}, and \texttt{vars.init} user parameters. After the \textit{Yes} button is selected, the dashboard will proceed with fitting the model and producing subsequent forecasts. 

\begin{figure}[t]
    \centering
    \includegraphics[width=\linewidth]{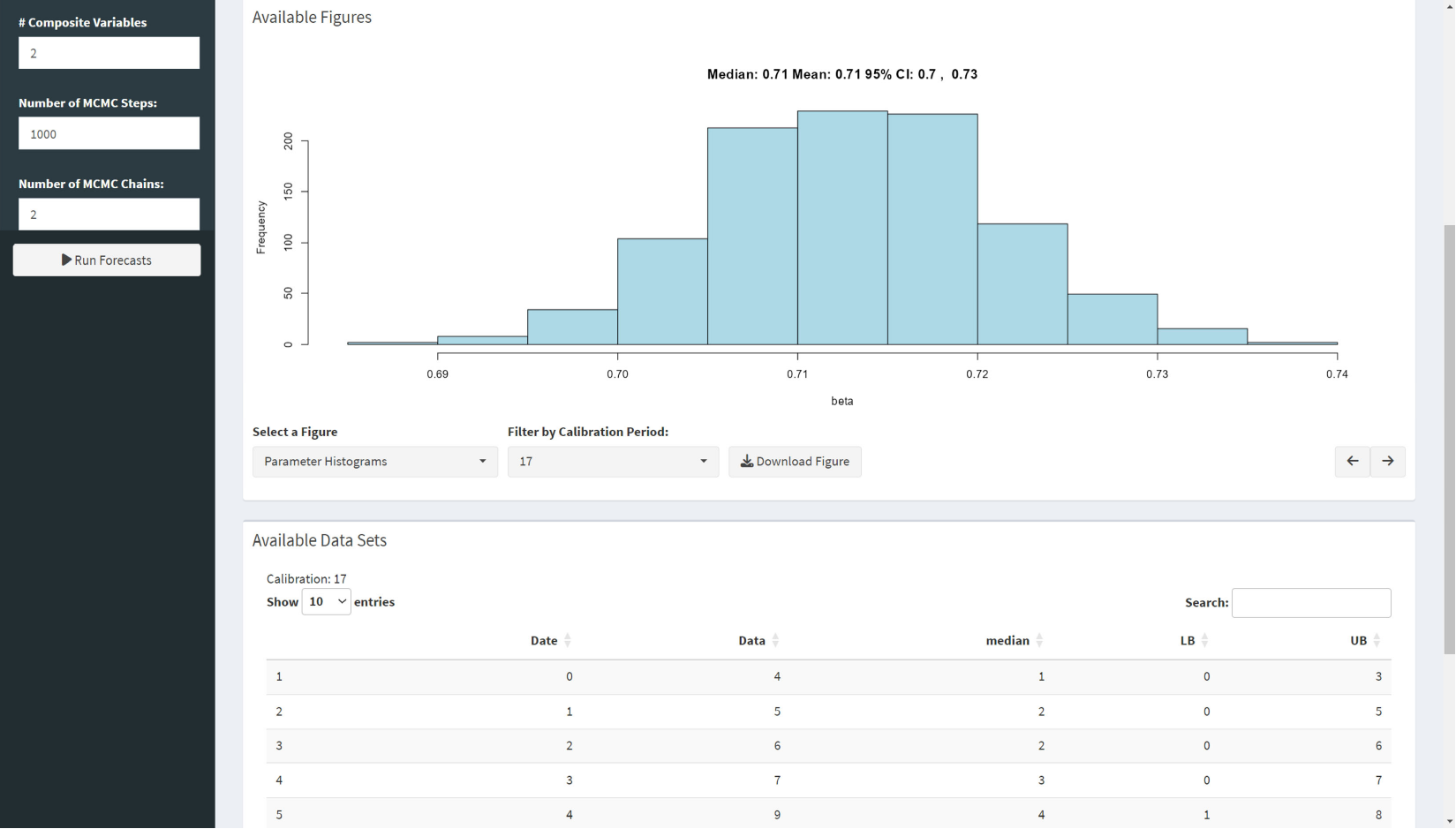}
    \caption{\footnotesize A screenshot of the dashboard with output loaded. This figure shows a screenshot of the dashboard after all fitting and forecasting processes have been completed. The “Available Figures” box contains all available figures, and the “Available Data Sets” box contains available data sets. Users can filter the output by type and calibration period, and download all of the output to the folder of their choosing.}
    \label{fig:Fig11A}
\end{figure}

The dashboard produces the same output (figures and data) as the source R package. However, it does not automatically download the figures and data sets; rather, it provides them within the main body of the dashboard (Fig. \ref{fig:Fig11A}). All figures and data sets are available to download to the folder of their choosing via their respective download buttons. 

\end{appendices}

\end{document}